\documentclass[preprint,authoryear,3p,12pt]{elsarticle}%
\usepackage{graphicx}
\usepackage{amssymb}
\usepackage{lineno}
\usepackage{amsmath}
\usepackage{color}
\usepackage{amsfonts}%
\providecommand{\U}[1]{\protect\rule{.1in}{.1in}}
\journal{Physics Reports}
\begin{document}
\begin{frontmatter}
\title{Many-electron dynamics of atomic processes
studied by photon-induced fluorescence spectroscopy}
\author[rndphys]{V.L.~Sukhorukov\corref{cor1}}
\ead{vlsu@sfedu.ru}
\author[rnd]{I.D.~Petrov}
\author[rnd]{B.M.~Lagutin}
\author[ksl]{A.~Ehresmann}
\ead{ehresmann@physik.uni-kassel.de}
\author[gs]{K.-H.~Schartner}
\author[kl]{H.~Schmoranzer\corref{cor2}}
\ead{schmoran@rhrk.uni-kl.de}
\cortext[cor1]{Corresponding author}
\cortext[cor2]{Principal corresponding author}
\address[rndphys]{Institute of Physics, Southern Federal University, 344090 Rostov-on-Don, Russia}
\address[rnd]{Rostov State Transport University, 344038 Rostov-on-Don, Russia}
\address[ksl]{Institut f\"{u}r Physik und Center for Interdisciplinary Nanostructure Science and Technology (CINSaT), Universit\"{a}t Kassel, D-34132, Kassel, Germany}
\address[gs]{I. Physikalisches Institut, Justus-Liebig-Universit\"at, D-35392 Giessen, Germany}
\address[kl]{Fachbereich Physik, Technische Universit\"{a}t Kaiserslautern, D-67653 Kaiserslautern, Germany}
\begin{abstract}
The progress and the chronology in understanding the influence of electron
correlations on the electronic structure of atoms and the dynamics of atomic
processes is reviewed focusing on benchmark rare-gas atoms. The contributions
and the chronological development of Photon-Induced Fluorescence Spectroscopy
(PIFS), measuring {} dispersed-fluorescence emission cross sections
upon excitation by single photons provided by monochromatized synchrotron
radiation is described. Selected experimental results obtained by
complementary techniques are also discussed for comparison. The basic suites
of computer programs used for the investigation of the many-electron effects
in atoms and the obtained results are analyzed. Special attention is paid to
the Configuration Interaction Pauli-Fock approximation with Core Polarization
(CIPFCP) method used to interpret the PIFS data.
\end{abstract}
\begin{keyword}
{Photon-induced fluorescence spectroscopy} (PIFS)\sep Satellite production\sep Photoionization of atoms\sep Alignment and orientation of ions\sep Configuration interaction Pauli-Fock approximation with core polarization (CIPFCP)\sep Interference\sep Many-electron correlations\sep Intershell interaction\sep Interchannel interaction
\end{keyword}
\end{frontmatter}

\newpage%

\tableofcontents



\section{Introduction}

\label{sec:Intro}

In many-electron atoms, several electrons are often essentially involved in
the atomic processes which, therefore, are too complex to be described in the
single-electron picture. These many-electron correlations were recognized to
influence strongly the atomic processes about 50 years ago. Taking into
account the correlations in the calculations changes {sometimes} the results by up to two orders of magnitude and leads to a qualitatively new
understanding of {processes in many-electron atoms}.

This review focuses on the development in the understanding of the
consequences of many-electron correlations {for} the electronic structure of atoms and {for} complex atomic processes {achieved by parallel developments of advanced experimental techniques and theoretical modeling}. We concentrate on the dynamics of the processes, i.e. their dependence on the energy of the exciting photon, since
their kinematics was considered already in several reviews by, e.g.,
\cite{greene82}, \cite{schmidt92}, \cite{kabachnik07}, and is elaborated in
great detail in the book of \cite{balashov00}.

We present and discuss the major theoretical approximations and packages of
computer {codes} mostly used to describe atomic processes {} taking into account many-electron correlations. In some detail, we describe the
Configuration Interaction Pauli-Fock approximation with Core Polarization
(CIPFCP) which was {developed} by authors of the present review.

The {} development of Photon-Induced
Fluorescence Spectroscopy (PIFS) as one of the most important experimental
methods for the investigation of {electron correlative} processes will be described. The PIFS results will be {documented and} discussed {} in comparison with {} complementary electron spectroscopic {data}. PIFS began to develop in the 60ies and was
strongly advanced by the use of synchrotron radiation. PIFS, using {}
photons for excitation, introduces a relatively small perturbation only into
the {atomic system}, and the recorded fluorescence photons are not
so sensitive to the surrounding experimental conditions as, e.g., electrons in
other methods.

The ability to {state-selectively} measure {photoionization} cross sections at low photoelectron energies makes PIFS an unprecedented method for the study of {near-}threshold phenomena. This becomes especially interesting at the present time, when the creation and
investigation of sub-micron objects requires the development of methods
accurately recording cross sections of processes in the threshold region. {Whereas PIFS has certain advantages at energies close to thresholds}, a variety of corrections are necessary {at exciting-photon energies exceeding threshold energies substantially}, e.g., for radiative cascades{}. Thus, the methods of photoelectron spectroscopy and PIFS provide complementary
information. We also review cases where additional information about the
dynamics of atomic processes was obtained by other methods, such as the
dual-laser plasma technique {(see section \ref{sec:Iso})} and the two-colour method {(see section \ref{sec:PWA})}.

In the present review, we limited {the discussion to} rare-gas atoms as benchmark systems. This decision also bypasses the influence of multiplet
effects on the investigated processes as much as possible. At present, the
Density Functional Theory (DFT) in the local density modification is often
used to interpret the processes occurring in molecules and solids under the
influence of electromagnetic radiation. In this case, ready-made suites of
programs are used sometimes without taking into account many-electron effects.
However, it is strongly encouraged to consider the many-electron effects found
in atomic processes also in molecules or in solids. In this regard, a review
of how our understanding of the dynamics of the atomic processes has evolved
appears especially worthwhile and relevant today.

\section{Photon-Induced Fluorescence Spectroscopy (PIFS)}

\label{sec:PIFSexp}

The main characteristics of PIFS should be mentioned at the beginning: (i)
using photons for excitation, PIFS causes a little perturbation only of the
investigated system, (ii) {PIFS records dispersed-fluorescence intensities as functions of the energy of the exciting photons}, thereby determining precisely the excited state and its decay processes. Below we review chronologically the development of the
PIFS method with a description of its basic details.

\subsection{From electron and proton impact to synchrotron radiation
excitation}

Fluorescence spectroscopy of atoms using a variety of excitation modes is
known to allow the observation of a wealth of spectral lines beyond those
observed in {photoabsorption}, enabling state-selective investigations of processes and energies of levels which {may not necessarily be} connected to the ground state by an electric dipole transition. The energies of these levels can be compared with
calculated ones and used for a first test of the accuracy of quantum
mechanical calculations. A more stringent test of the approximations applied
in these calculations is based on the comparison of the measured optical
transition probabilities with the corresponding theoretical quantities which
are essentially determined by the electric dipole matrix element with
\emph{both} the eigenfunctions of the initial and the final states. The
spontaneous emission probability (Einstein $A_{i,j}$ coefficient) measured on
absolute scale is required for a quantitative test. However, even if the
intensities of the emission lines are measured absolutely, the $A_{i,j}$
coefficients cannot be extracted because of the lack of quantitative
information on the population density of the excited level where the emission
starts. In classical emission spectroscopy where various designs of gas
discharges are used for excitation, the excitation conditions are not well
defined since electrons of a large range of energy are involved in the impact excitation.

More than half a century ago when dedicated synchrotron radiation sources
($3^{\mathrm{rd}}$ generation sources), which provide spectrally continuous
radiation of known intensity distribution, were not available yet, other
experimental methods were devised to tackle the problem of quantitatively
defining the excitation conditions.

\begin{figure}[ptb]
\begin{center}
\includegraphics[width=0.7\textwidth]{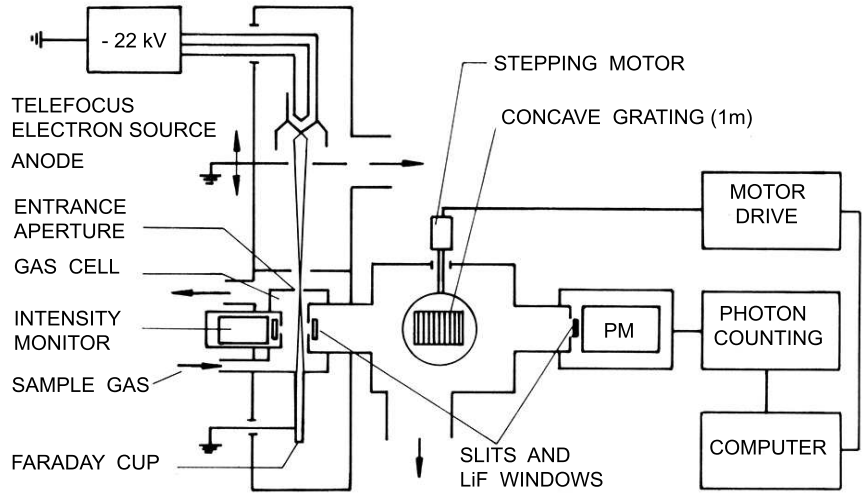}
\end{center}
\caption{Experimental setup for keV-electron impact excited fluorescence
spectrometry in the VUV spectral region (adapted from \citep{schmoranzer80})}%
\label{fig:HS_fig1_HS}%
\end{figure}

The impact of monoenergetic fast electrons or fast protons can provide defined
excitation conditions similar to a light source with a flat continuous
frequency distribution as \cite{fermi24} recognized by Fourier transformation
of the electric field of the moving charge at the location of the atom. The
atoms absorb the proper frequencies out of this continuum so that the electron
bombardment results in populating {predominantly} those excited atomic levels that are coupled
to the ground state by optically allowed transitions. The population density
of the initial levels of the emission is then determined by the optical
absorption probabilities (Einstein $B_{i,j}$ coefficients). A full quantum
mechanical treatment of the inelastic electron scattering from atoms by
\cite{bethe30} (for a modern formulation see also \citep{inokuti71,inokuti78})
within the $1^{\mathrm{st}}$ Born approximation reveals the close connection
between the doubly differential (i.e. with respect to scattering angle and
momentum transfer) electron scattering cross section, proportional to the
so-called generalized oscillator strength, and the optical oscillator strength
for small momentum transfer. Bethe's approximation is very good for
keV-electrons (e.g. for energies of 15 keV or higher) and very small
scattering angles where the scattering is peaked. The concept of keV-electron
excited fluorescence spectroscopy (sometimes called ``poor man's synchrotron''
excitation, see Fig.~\ref{fig:HS_fig1_HS}) was first applied by one of us
\citep{reich64,reich65,schmoranzer73,schmoranzer75} to the hydrogen molecule
and later to rare gas excimers \citep{schmoranzer80,barzen87}.

\begin{figure}[ptb]
\begin{center}
\includegraphics[width=0.7\textwidth]{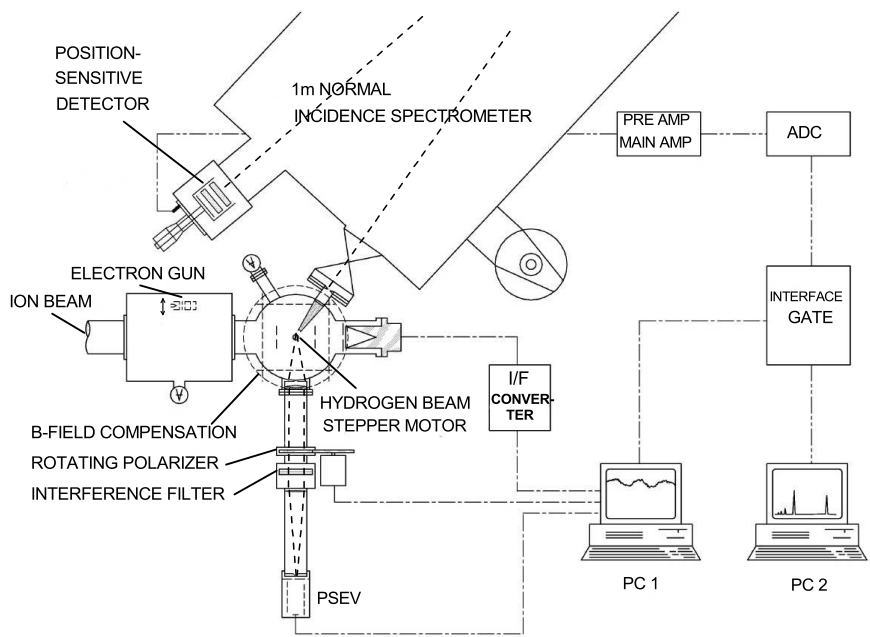}
\end{center}
\caption{Experimental setup for measuring the electron and proton impact,
respectively, induced polarization fraction of the Balmer $\alpha$ transition
(adapted from \citep{werner96})}%
\label{fig:HS_fig2_2PC}%
\end{figure}

For electron and proton impact excitation of atomic hydrogen
(Fig.~\ref{fig:HS_fig2_2PC}), it was demonstrated that 1~MeV protons produce,
e.g., the same linear polarization of the Balmer $\alpha$ line as electrons of
the same velocity \citep{werner96}. In this comparative study the electron
beam replaced the ion beam, keeping all other parameters unchanged. This
technique was also applied in the synchrotron radiation photon impact studies
for calibration purposes.

\begin{figure}[ptb]
\begin{center}
\includegraphics[width=0.9\textwidth]{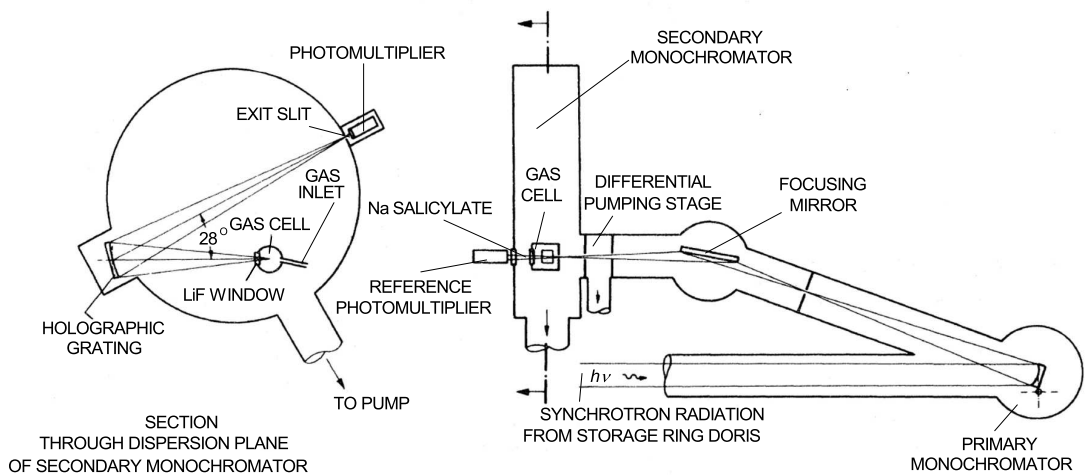}
\end{center}
\caption{Experimental setup for VUV fluorescence spectroscopy following
selective excitation by synchrotron radiation (adapted from
\citep{schmoranzer78})}%
\label{fig:HS_fig3_PIFS}%
\end{figure}

Selective excitation in the vacuum ultraviolet (VUV) spectral region became
feasible when the brilliance of synchrotron radiation sources had been
sufficiently increased to use a monochromator to reduce the bandwidth of the
exciting radiation. First attempts were made at bending magnets of the
electron storage ring DORIS at HASYLAB, Hamburg, mainly to obtain excitation
spectra for undispersed fluorescence, ionization, or radiating dissociation
products of small molecules \citep{sroka73a,sroka73}. Selectively
photon-excited fluorescence spectroscopy (see Fig.~\ref{fig:HS_fig3_PIFS}) was
started by using a specially designed and home-built high-luminosity secondary
monochromator (asymmetric Pouey mount \citep{pouey78}) to scan the
fluorescence spectrum of molecular hydrogen at a resolution of 1.5~nm across a
slit in front of a VUV photomultiplier \citep{schmoranzer78}. The count rates
in this two-monochromator experiment were very low but the extremely low noise
of the VUV detector allowed for a reasonable signal-to-noise ratio. The
efficiency of the VUV fluorescence spectrometry was greatly increased when the
wavelength-sequential detection was replaced by parallel photon counting
within a spectral range of typically 20~nm \citep{schmoranzer86}. Here the VUV
fluorescence spectrum was imaged onto a photocathode in front of a stack of
microchannel plates and the multiplied photoelectrons were recorded by a
one-dimensionally position-sensitive charge partitioning anode of a
backgammon-like pattern with analog processing electronics. When undulators
became available as radiation sources with much enhanced brilliance, the
{exciting-photon energy} step size could be decreased to a fraction of the bandwidth of the primary monochromators (using toroidal or spherical gratings). By
combining series of measured fluorescence spectra recorded at stepwise varied
{exciting-photon} energy, {the intensities of dispersed fluorescence are obtained as functions of the exciting-photon energy (dispersed fluorescence excitation functions)}. From these
results, one obtains by a section at constant {exciting-photon energy} the fluorescence spectrum excited at this fixed {photon energy} whereas a section at constant fluorescence wavelength represents the {excitation function of the fluorescence for this wavelength}. Note that the fluorescence intensity, as recorded by single-photon counting, is a digital quantity which
can be plotted in grey or colour scale or three-dimensionally as a
surface on top of a narrow mesh of pairs of the {exciting-photon energies} and fluorescence wavelengths (see, e.g., \citep{ukai95,liebel00,schmoranzer01}).

\subsection{Measurement of absolute cross sections}

\begin{figure}[tb]
\begin{minipage}[t]{0.32\textwidth}
\begin{center}
\includegraphics[width=\textwidth]{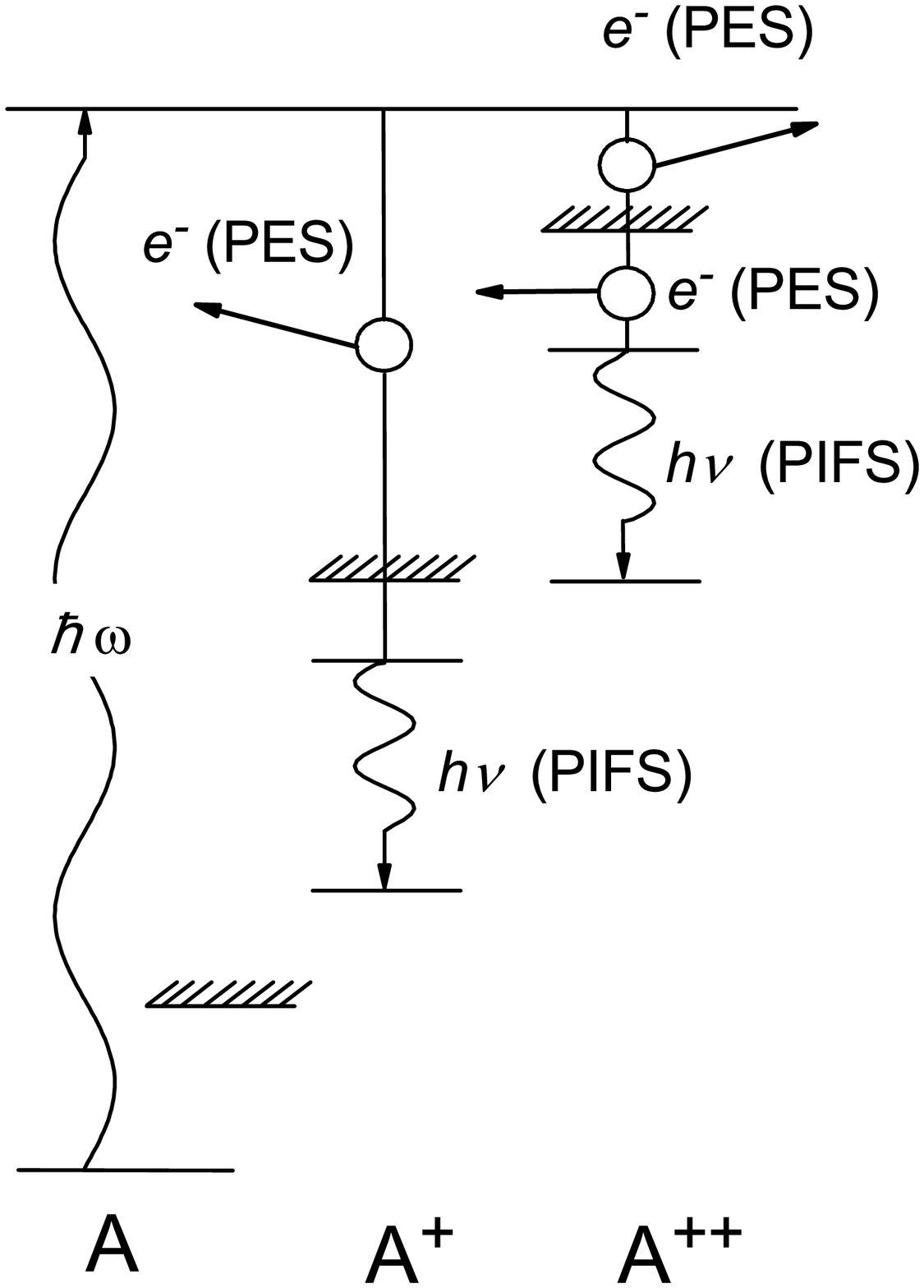}
\end{center}
\caption{Equivalence of PIFS, registering photons $h\nu$, and
photoelectron spectroscopy (PES), registering electrons $e^-$  for single or double ionization by
photons of energy $\hbar \omega$ (adapted from
\citep{schartner90}).}\label{fig:HS_fig4_schem}%
\end{minipage} \hfill\begin{minipage}[t]{0.63\textwidth}
\begin{center}
\includegraphics[width=\textwidth]{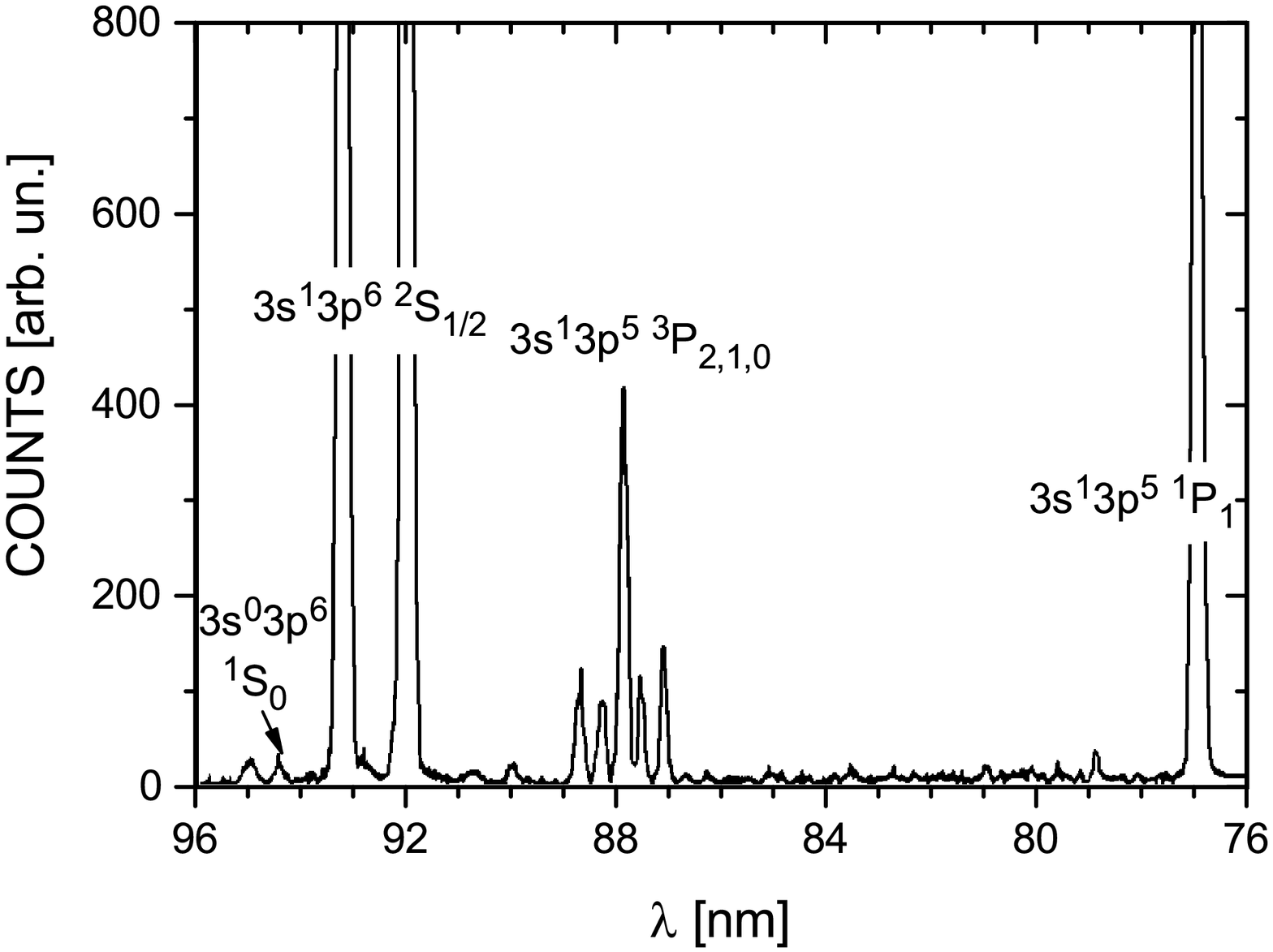}
\end{center}
\caption{VUV fluorescence spectrum of Ar excited by 100~eV photons.
Lines result from the 3s-electron and the 3s3p-electron single and double
photoionization (adapted from
\cite{mobus94}).}\label{fig:HS_fig5_spectr}\end{minipage}
\end{figure}

Photon-induced fluorescence spectroscopy (PIFS), as applied here to rare
gases, is based on a quantitative analysis of the emission probability and
wavelength of the fluorescence photon $h\nu$ emitted by an excited ion created
by photoionization through photons $\hbar\omega$. In contrast to photoelectron
spectroscopy, the photoelectron is not observed (Fig.~\ref{fig:HS_fig4_schem}%
). Double photoionization studies thus need no coincidence methods like
PhotoElectron-PhotoElectron COincidence (PEPECO) spectroscopy, which is an
advantage of PIFS. Fig.~\ref{fig:HS_fig5_spectr} displays the fluorescence
spectrum of Ar resulting from single and double photoionization after
excitation by photons of 100~eV \citep{mobus94}. The doublet at 91.9~nm and
93.2~nm following the $3s$ electron photoionization is clearly resolved. A
second advantage is the fact that the spectral resolution of the exciting
photons and the spectral resolution of the fluorescence spectrometer are
independent of each other. In this way, PIFS is superior for investigations of
ionization threshold phenomena to photoelectron spectrometries where
photoelectrons of strongly varying low energy with strongly varying detection
efficiencies have to be analyzed and interpreted.

When PIFS is used for the investigation of atomic photoionization processes,
emission cross sections of dispersed fluorescence from an excited level of an
ion are measured. For comparison with a theory describing processes to
populate a particular ionic level, however, these emission cross sections have
to be converted into population cross sections or population probabilities.
These population cross sections contain information on all possible population
paths of the excited ionic level and sum up the electron emission processes
into all possible space directions. They are therefore sensitive to quantum
mechanical interference processes and to electron correlative phenomena. For a
quantitative comparison with a theory, we therefore need -- in the best case
-- experimentally determined absolute emission cross sections and a conversion
of emission cross sections into population cross sections caused by the
process of interest, both highly non-trivial aspects of the method.

Here we will describe first different setups recording ionic fluorescence
after photoionization processes initialized by monochromatized synchrotron
radiation, then state-of-the-art methods to determine absolute emission cross
sections, and finally aspects of the conversion of emission cross sections
into population cross sections for a particular population process or a
combination of several ones. In this sense, it is necessary to convert the
recorded fluorescence emission cross section to population cross sections for
the investigated process.

In order to determine emission cross sections of dispersed fluorescence for
state-selective photoionization studies, the fluorescence resolution must be
as high as possible, allowing additionally for a decent photon flux as beam
time allocations at synchrotron radiation or free-electron laser sources are
limited. A good compromise between resolution and flux was achieved by using a
1m-normal-incidence spectrometer (1m-NIM, McPherson~225), with a resolving
power $\lambda/ \Delta\lambda$ of more than 1000.

\begin{figure}[ptb]
\begin{center}
\includegraphics[width=0.65\textwidth]{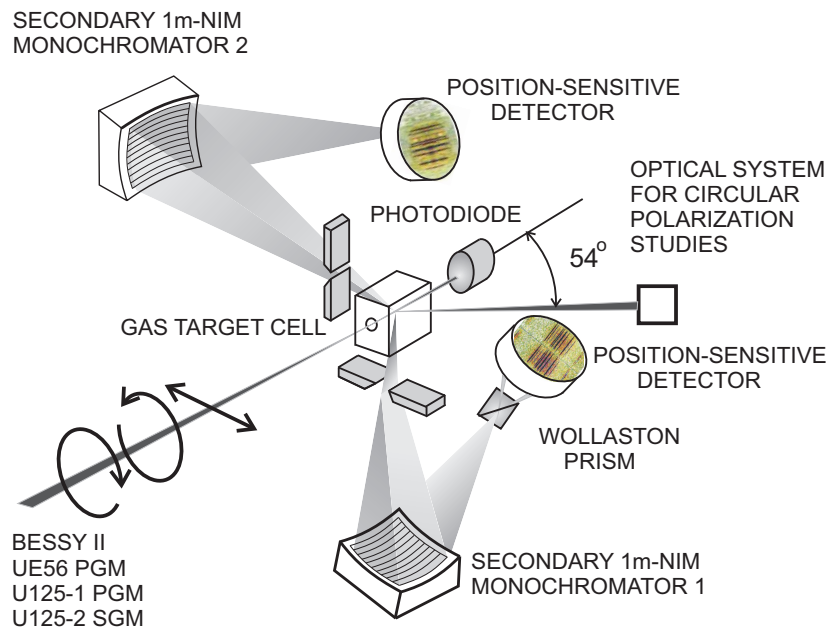}
\end{center}
\caption{Schematic display of the main components of the PIFS setup at BESSY
II. Both 1m-NIMs with their detectors, a Wollaston prism for polarization
measurements, the target cell (with openings for observation of the
interaction area under 90$^{\circ}$, an optical system with photomultiplier
and a photodiode for flux monitoring are displayed. Adapted from
\citep{zimmermann04}.}%
\label{fig:HS_fig6_polset}%
\end{figure}

This type of spectrometer, originally developed for spectrometry in the VUV
spectral range, allows also spectroscopy in the visible range. The
installation of a second 1m-NIM, perpendicularly mounted to the first one,
enabled the simultaneous observation of fluorescence in two different spectral
regions or to alternatively investigate the angular dependence of dispersed
fluorescence emission. Gratings with different blaze wavelengths and coatings
were used for signal optimization. Accordingly, two-dimensionally
position-sensitive detectors, equipped with photocathodes optimized for the
spectral region of interest, with resistive, wedge-and-strip \citep{kraus89},
and recently also delay line anodes for dispersed time resolved measurements
combined with digital processing were used. Position-sensitive recording in
combination with a dispersive element using a setup with a target cell crossed
by the exciting synchrotron radiation beam images the complete source volume
as seen in the light of the different wavelengths. Integration along the lines
of the two-dimensionally recorded spectrum delivers the conventional spectrum.
Fig.~\ref{fig:HS_fig6_polset} shows the final setup.

Due to the multiple observation angles, the target chamber is correspondingly
complex. It contains two slits with separately adjustable slit bars, small
tubes as differential pumping elements for the in- and outgoing
exciting-photon beam, the gas inlet and a semiconductor photodiode
(alternatively an Al Faraday cup) to record the intensity of the transmitted
exciting-photon beam. As the monochromatized radiation of standard synchrotron
radiation or undulator beam lines is not only composed of photons of the
nominal photon energy, but also of photons of multiples of this energy
(so-called higher orders), a component suppressing these unwanted
higher-energy photons can be mounted in front of the target chamber
\citep{schmoranzer01}. It consists of an absorber cell containing a rare gas
at comparatively high pressure separated from the target chamber and from the
undulator beam line by tubes of 2~mm diameter for efficient differential
pumping. All three parts of the absorber stage are separately pumped by turbo
pumps. For 24~eV photon energy, a reduction of the third-order radiation by a
factor of 175 was achieved using He at 30~mbar as absorber gas.

The absorption of electromagnetic radiation by a gas is described by the
well-known Lambert-Beer law%

\begin{equation}
I\left(  \hbar\omega,x\right)  =I_{0}\cdot\mathrm{exp}\left(  -\sigma
_{t}\left(  \hbar\omega\right)  \cdot n\cdot x\right)  \label{eq:Lambert}%
\end{equation}
where $I\left(  x\right)  $ is the intensity of the transmitted radiation as a
function of path length x through the gas, $I_{0}$ is the intensity of the
impinging radiation, $n={p/\left(  kT\right)  }$ is the number volume density
of atoms in the gas with $p,$ $T$ being the actual pressure and temperature of
the gas in the cell during measurements. The total cross section $\sigma_{t}$
describes all processes leading to a weakening of the impinging radiation in
its forward direction. This includes the cross sections for elastic and
inelastic scattering $\sigma_{s}$ and $\sigma_{a}$, i.e. $\sigma_{t}%
=\sigma_{s}+\sigma_{a}$. In the exciting-photon energy range of interest for
this review, scattering processes can be neglected as compared to absorptive
processes. Absorptive processes can be categorized into resonant excitations
with re-emissions of photons, and photoionization processes into non-radiating
and radiating ionic states, i.e. $\sigma_{a}=\sigma_{res}+\sigma_{PInr}%
+\sigma_{PIrad}$. As the energies of the exciting photons used in the
described investigations are usually higher than the photoionization
thresholds of the investigated atoms, we disregard resonant absorption and
re-emission within the neutral atom. Fluorescence spectrometry is sensitive to
the radiation emitted by excited ionic fragments, therefore we also disregard
photoionization processes leading to non-radiating final ionic states. When
dispersed fluorescence is recorded, the method is immediately sensitive to the
final ionic states populated by the investigated process as such states do
emit fluorescence of well-defined wavelengths. Therefore the cross section for
photoionization processes populating excited ionic states sums up the
population contributions $\sigma_{PIrad,n}\left(  \hbar\omega\right)  $ of all
accessible excited ionic states $n$.%

\begin{equation}
\sigma_{PIrad}\left(  \hbar\omega\right)  =\sum_{n}\sigma_{PIrad,n}\left(
\hbar\omega\right)  . \label{eq:sPIrad}%
\end{equation}

The population cross sections of ionic states include all possible pathways of
population at the exciting-photon energy $\hbar\omega$ and integration over
all photoelectron emission directions. For a determination of population cross
sections via measured fluorescence emission cross sections, the branching into
different de-excitation channels $i$ has to be considered such that%

\begin{equation}
\sigma_{PIrad,n}=\sum_{i}\sigma_{PIrad,n,i} \label{eq:sPIradn}%
\end{equation}
where different fluorescent de-excitation channels are characterized by
different emission wavelengths $\lambda_{i}$. In each case, also possible
non-radiative de-excitation processes competing with the radiative processes
have to be discussed. With photon-induced fluorescence spectrometry, it is
possible to determine from measured fluorescence intensities individual
dispersed-fluorescence emission cross sections $\sigma_{PIrad,n,i}$ {as functions of the exciting-photon energy}.

The measured fluorescence intensity $I_{meas} $ on the used position-sensitive
detection systems is connected with the dispersed-fluorescence emission cross
section by%

\begin{equation}
I_{meas}(\lambda_{i};x,y;\vartheta,\varphi;\Pi)=\Phi_{0}\cdot k\cdot
n\cdot\eta_{SD}(\lambda_{i};x,y)\cdot\sigma_{PIrad,n,i}(\lambda_{i})\cdot
f(\lambda_{i};\vartheta,\varphi;\Pi). \label{eq:Imeas}%
\end{equation}

$I_{meas}$ depends on the quantum efficiency of the spectrometer-detector
combination $\eta_{SD}(\lambda_{i};x,y)$ at wavelength $\lambda_{i}$ recorded
at position ($x,y$) of the position-sensitive detector, on the incoming photon
intensity $I_{0}$, on the length of the source volume $l$ and the solid angle
$\Omega$ of the source volume, expressed by the geometrical factor $k$%

\begin{equation}
k=l\cdot\frac{\Omega}{4\pi}, \label{eq:geom_k}%
\end{equation}
and a factor $f$ considering a possible angular dependence of the emitted
fluorescence intensity when observing at angles $\vartheta,\varphi$ and a
possible quantum efficiency change for different fluorescence polarizations
$\Pi$. For each measurement, a possible position dependence of the quantum
efficiency to record fluorescence of wavelength $\lambda_{i}$ has to be tested
with the spectrometer-detector combination. The dependence of quantum
efficiencies $\eta_{SD}(\lambda_{i};x,y)$\ on the spectral feature's position
($x,y$)\ on the detector surface can easily be determined by moving the
feature across the detector surface at otherwise constant experimental
conditions. Therefore, this position dependence is regarded as known in the
following, expressed by $\eta_{SD}(\lambda_{i};x,y)=\eta_{SD}(\lambda_{i})$.
Usually the dependence of the efficiency of fluorescence detection on
observation angles and fluorescence polarization is also small (provided the
solid angle of observation remains unchanged) so that we will for the moment
not discuss the factor $f$ of equation~(\ref{eq:Imeas}). The most important
remaining tasks for absolute dispersed-fluorescence emission cross sections
determination are: \textbf{(i)} to determine the relative quantum efficiency
of the spectrometer-detector combination used in the experiments as a function
of wavelength for the fluorescence wavelength interval of interest,
\textbf{(ii)} to determine the geometrical factor $k$ and the target gas
volume density $n$, \textbf{(iii)} to determine the incoming photon flux, and
\textbf{(iv)} to determine at least for one spectral feature at least at one
exciting-photon energy the absolute cross section. When \textbf{(i)} to
\textbf{(iii)} have been achieved, \textbf{(iv)} allows for the determination
of absolute dispersed-fluorescence emission cross sections for \emph{all}
observed lines excited by photons of \emph{all} exciting-photon energies as
long as the cross section for \emph{one} feature at \emph{one}
exciting-photon energy can be determined.

\begin{description}
\item[(i)] The determination of the relative quantum efficiency of a
spectrometer-detector combination is a severe experimental task, especially
for the spectral range of the VUV. In contrast to the visible spectral range
with the tungsten ribbon lamp as solution, there exists no similar calibration
light source in the VUV. Therefore, a transferable secondary intensity
standard \citep{schartner87} was developed and applied here. Fluorescence is
induced by impact of 2~keV and 3~keV electrons from a transportable electron
source. For these electron energies, absolute fluorescence emission cross
sections $\sigma_{rad,eimpact}\left(  \lambda_{i}\right)  $ for 20 wavelengths
$\lambda_{i}$ between 46 nm and 120 nm have been determined {in cooperation} with the
radiometry lab of the Physikalisch-Technische Bundesanstalt (PTB) using the
synchrotron radiation of BESSY as a primary calibration light source
\citep{jans95,jans97}. The accuracy of the absolute cross sections ranging
between 4.6\% and 8.7\% is superior to the results of earlier similar efforts
\citep{risley89,vanderburgt89}. For a determination of the relative quantum
efficiencies of the spectrometer-detector combination, an electron source has
been integrated in the setup of Fig.~6. Solving equation~(\ref{eq:Imeas}) for
$\eta_{SD}(\lambda_{i})$ and replacing photon-excited cross sections by
electron impact excited ones, relative quantum efficiencies with respect to an
efficiency $\eta_{SD}(\lambda_{ref})$ at an arbitrarily chosen reference
wavelength $\lambda_{ref}$ can be determined by the ratio of
equation~(\ref{eq:eta_ratio}), using the known electron impact excited
fluorescence emission cross sections and the measured intensities at the
corresponding wavelengths:%

\begin{equation}
\frac{\eta_{SD}(\lambda_{i})}{\eta_{SD}(\lambda_{ref})}=\frac{I_{meas}%
(\lambda_{i})}{I_{meas}(\lambda_{ref})}\cdot\frac{\sigma_{rad,eimpact}\left(
\lambda_{ref}\right)  }{\sigma_{rad,eimpact}\left(  \lambda_{i}\right)  }.
\label{eq:eta_ratio}%
\end{equation}

By building this ratio, it is neither necessary to determine the absolute
intensity of the exciting electron beam neither the geometrical factor $k$ of
equation~(\ref{eq:Imeas}) nor the absolute pressure in the target cell, as
long as they remain constant for the intensity measurements of fluorescence at
the different wavelengths.

\item[(ii)] The quantities $k$ and $n$ are not trivial to determine and are
usually a large source of experimental uncertainty. To avoid the absolute
determination of these quantities, the apparatus of Fig. 6 has been equipped
by an electron source which may replace the synchrotron beam under the same
geometrical conditions of source volume observation by the position-sensitive
detector. It is then possible to compare synchrotron radiation excited
fluorescence spectra with the spectra after electron impact excitation as long
as the intensities of the photon and the electron beams are known.

\item[(iii)] The flux of the exciting photons has been determined by
calibrated flux monitors, either by a GaAsP photodiode or by an Al Faraday
cup. The quantum efficiencies of both monitors are different in two respects
and have been chosen according to the needs of the measurement task: (a) the
quantum efficiency of the photodiode is higher by one to two orders of
magnitude as compared to that of the Faraday cup material and (b) the quantum
efficiency of the photodiode is increasing with increasing photon energy,
whereas the quantum efficiency of the Faraday cup is strongly decreasing with
increasing photon energy. The quantum efficiency dependence on the
exciting-photon energy is explained by the way how charges are created by the
photons, in the photodiode due to the inner photo-effect creating
electron-hole pairs, in the Faraday cup due to the external photo-effect,
where at increasing photon energies the photon penetration depth increases but
the short escape length of the emitted electrons essentially stays constant.
Due to these differences the photodiode seems to be the best choice for the
exciting-photon flux measurement. Here another particular aspect of
fluorescence spectrometry after excitation by synchrotron or free electron
laser radiation has to be considered: as the measured fluorescence signal does
not tell the experimentalist whether it is excited by photons of nominal
energy or by higher-order photons, it is important to consider corresponding
effects in the fluorescence as well as in the exciting-photon flux signal.
Therefore the higher quantum efficiency at higher exciting-photon energy
possesses a severe drawback as it amplifies the influence of higher-order
photons in the signal of the exciting-photon flux monitor, whereas the
decreasing quantum efficiency of the Faraday cup amplifies the signal of the
photons of nominal energy. Therefore the photodiode is the best choice when
the exciting-photon flux is small and when higher-order radiation is not
present or is efficiently suppressed by other means, whereas the Faraday cup
is advantageous at higher exciting-photon intensities and present higher-order radiation.

\item[(iv)] With \textbf{(i)} to \textbf{(iii)}, it is possible to determine
the exciting-photon flux and the relative quantum efficiencies of the
spectrometer-detector combination. If it is now possible to determine for one
spectral line at one exciting-photon energy the absolute dispersed
fluorescence emission cross section, emission cross sections for all other
fluorescence lines in the spectral range of the spectrometer-detector
combination can be determined. One trivial way to do this is to use existing
literature data of known cross sections. In this case, cross checks of
different literature sources are possible as well as consistency tests of the
calibration procedure. The second way is to compare the known electron impact
induced dispersed-fluorescence emission cross sections with the photon induced
cross sections by replacing the exciting-photon beam by the exciting electron
beam keeping the geometrical conditions for fluorescence detection unchanged.
For this procedure it is important that the widths and depths of the source
volumes during electron and photon excitation are the same, therefore it is
important to take care of the divergences of the exciting beams.
\end{description}

The {preparational step for a comparison between} theoretical models and fluorescence data is to convert the determined absolute dispersed-fluorescence emission
cross sections to population cross sections for a particular process or a
combination of processes. For this it is necessary to quantify the probability
of the excited state to decay into different energetically lower-lying states
(branching ratios), to quantify a possible angular distribution of the emitted
fluorescence, and to discuss the probability of the excitation of high-lying
levels, which by themselves will decay into the emitting levels (cascades).

\subsection{Measurement of alignment and orientation}

\begin{figure}[tb]
\begin{minipage}[b]{0.47\textwidth}
\begin{center}
\includegraphics[width=\textwidth]{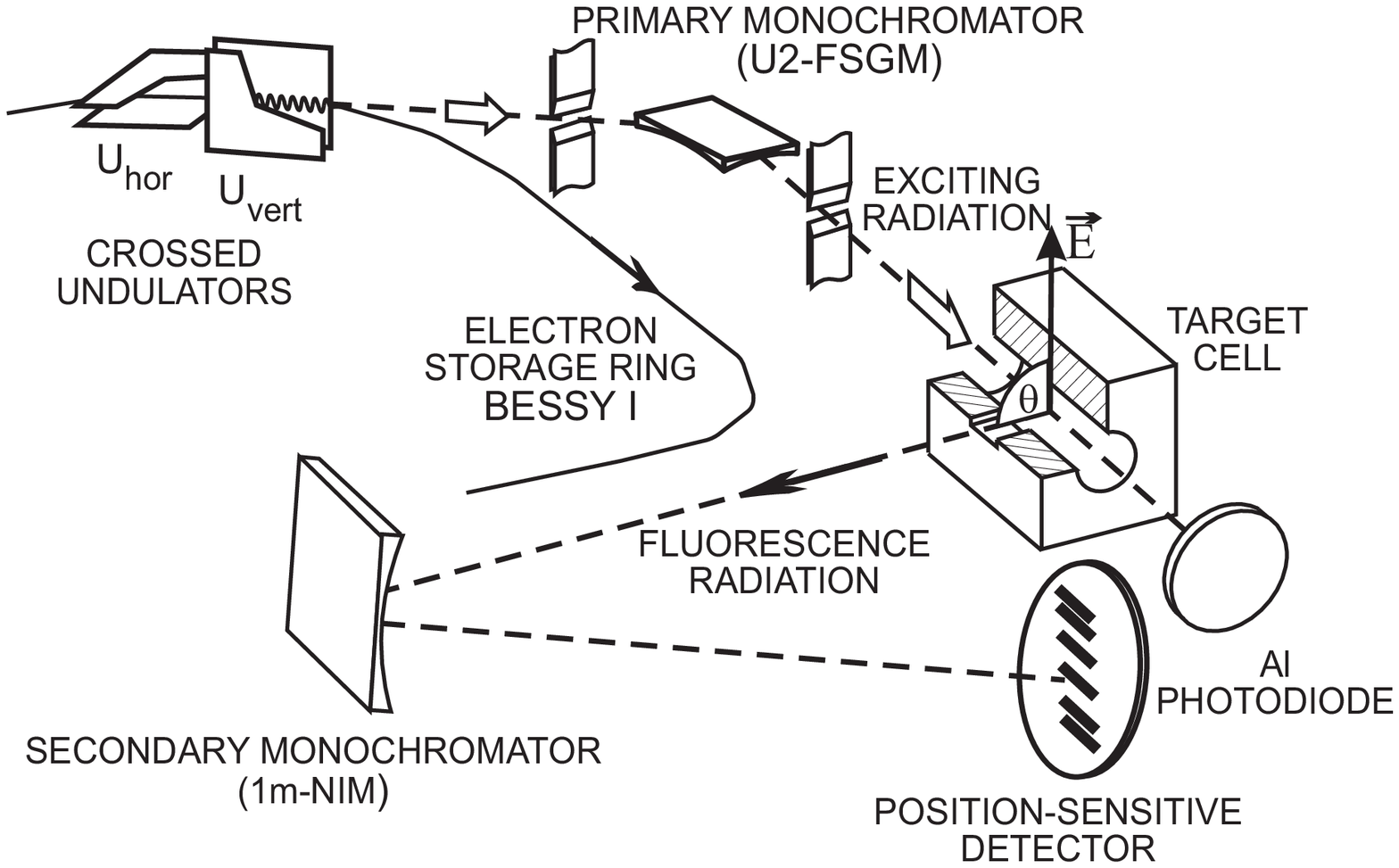}
\caption{Experimental setup for studying the angular distribution of the
fluorescence radiation in PIFS. The polarization shown corresponds to the
operation of the vertical undulator (adapted from
\citep{schmoranzer97a}).}\label{fig:HS_fig7_cdrbw}\end{center}
\end{minipage}
\hfill\begin{minipage}[b]{0.47\textwidth}
\begin{center}
\includegraphics[width=\textwidth]{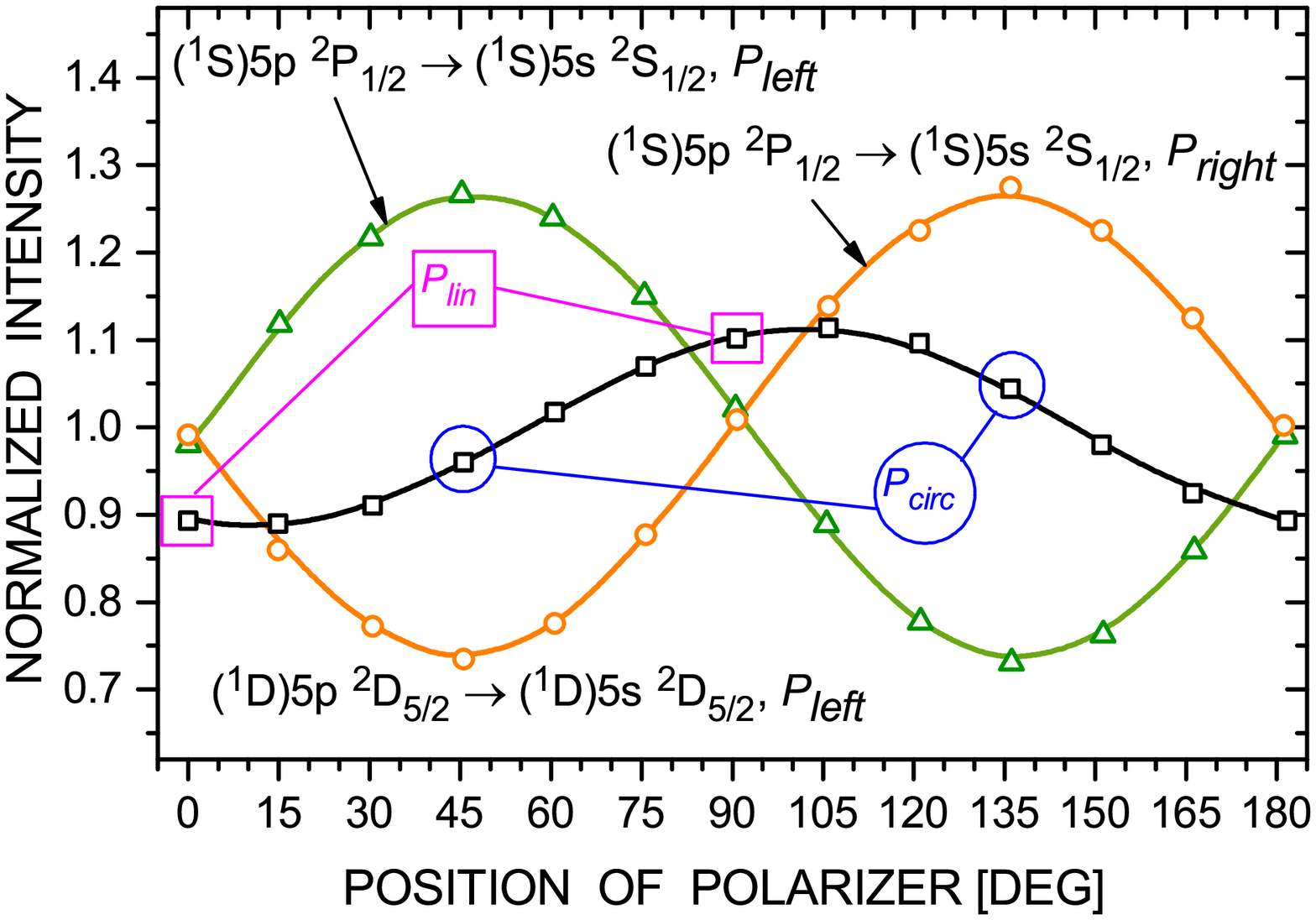}
\caption{Fluorescence intensities as function of the polarizer angle. The
linear and circular polarization fractions were evaluated from the
rectangle-rounded and ring-rounded data points (see text) (adapted from
\citep{schartner05}).}\label{fig:HS_fig8_polang}\end{center}
\end{minipage}
\end{figure}

The PIFS technique was also applied to investigate the angular distribution of
the emitted radiation after excitation by linearly or circularly polarized
undulator radiation. It results from the non-statistical population of the
magnetic sublevels and allows via the alignment parameter $A_{20}$ and the
orientation parameter $O_{10}$ access to a partial wave analysis (PWA) of the
emitted photoelectron. We used different methods to investigate the angular
distribution of the fluorescence. By using one or the other of two crossed
undulators the electric field vector of the linearly polarized synchrotron
radiation can be chosen either parallel or perpendicular to the direction of
the detected fluorescence radiation. This technique is also suited for
fluorescence in the spectral range of the VUV (see Fig.~\ref{fig:HS_fig7_cdrbw}).

If only one polarization {direction} of the exciting photons is available, the insertion
of a Wollaston prism opens the access to a polarization analysis, as indicated
in Fig.~\ref{fig:HS_fig6_polset}. This method needs photons in the visible
spectral range. The two spectra with the electric field vector either parallel
or perpendicular to the electric field vector of the undulator radiation are
simultaneously recorded (see Fig.~\ref{fig:HS_fig6_polset}). In case the
undulator radiation is circularly polarized, both the alignment parameter
$A_{20}$ and the orientation parameter $O_{10}$ can be determined. The
fluorescence radiation is observed here under 54$^{\circ}$ with respect to the
direction of the incoming beam by an optical system containing a quarter-wave
plate, an interference filter, and a rotating polarizer foil
\citep{schartner05}. In this way, a partial wave analysis of the emitted
photoelectron becomes possible. We applied this technique in studies of
resonant Raman Auger transitions of Kr. Fig.~\ref{fig:HS_fig8_polang} shows an
example of the fluorescence intensity recorded by the photomultiplier of the
optical system for two transitions in Kr$^{+}$ with $J=1/2$ and $J=5/2$ as
function of the polarizer angle, induced by circularly polarized photons of
different helicity. Circular and linear polarization fractions {for the $J=5/2$ state} are derived
from the signals at 45$^{\circ}$ and 135$^{\circ}$ and at 0$^{\circ}$ and
90$^{\circ}$, respectively.

\section{Theoretical methods for interpretation of experimental results}

\subsection{Calculation of atomic orbitals}

In order to take into account many-electron correlations in the interpretation
of the observed experimental features different theoretical techniques have
been applied. Most of them are based on the use of the atomic orbitals (AOs)
of the central field approximation, having as basic assumption that the
movement of each electron takes place in the spherical field of the nucleus
and the average field of the other electrons \citep{bethe57,slater60,cowan81}.
This assumption could lead to either Hartree-Fock (HF) or Dirac-Fock (DF)
equations. Although the HF \citep{fock30} and DF \citep{swirles35} equations
have been derived shortly after the creation of quantum mechanics, their
intensive application started after the availability of fast computers. We
present here some of the numerous papers containing detailed descriptions of
the physical and numerical methods used in computing atomic processes.

In order to accelerate the computing, many authors used the simplified
versions of the HF approach, treating the relativistic effects in the Breit
approximation or \textquotedblleft localizing\textquotedblright\ non-local
Coulomb interaction using the idea of \cite{slater51} resulting in the
Hartree-Slater (HS) approach. Atomic orbitals and potentials computed in the
HS approximation by \cite{herman63} for atoms with $2\leq Z\leq103$ have been
widely used in numerous applications.

Numerical procedures used for solving HF equations with non-local electron
potentials have been described by \cite{froese63}. {These}
numerical methods have been {applied} in numerous {investigations using} the HF and DF
methods, including the widely applied multi-configuration HF approach (MCHF)
created by Froese-Fischer \citep{fischer70,fischer72}.

\cite{cowan67} introduced statistical approximations for exchange and
correlation, including {in addition} relativistic effects in the manner described by \cite{herman63}. {Subsequently}, \cite{cowan76} developed an approximation where the non-local part of the electron-electron interaction was taken into account in the HF approach without the statistical approximation, the most valuable
relativistic correction being included directly in the HF equation within the
Pauli approximation \citep{bethe57}.

\cite{grant70} has reviewed the state of relativistic calculations of atomic
structures at the end of the sixties. A little later, \cite{desclaux71}
published the description of two independent computer {codes implementing} the relativistic DF method. This numerical procedure has been applied
later to create the relativistic DF multiconfiguration {code}
\citep{desclaux75} which may be considered as the relativistic equivalent of
the HF code of \cite{fischer70}. The numerical techniques utilized by
\cite{desclaux75} have later been applied by \cite{grant80} to create an
atomic multiconfigurational DF package. Further development of the computer
code \citep{grant80} resulted in several suites of programs for the
relativistic atomic calculations on the basis of the DF AOs, e.g., GRASP92 of
\cite{parpia96}, allowing to compute various atomic properties.

\subsection{Suites of computer programs for the calculation of atomic
structures}

\label{sec:suites_PC}

Having a set of AOs at hand, one needs to apply any kind of many-electron
theories created since the earlier fifties (see, e.g., the work
\citep{lowdin55,lowdin55a,lowdin55b} where the density and transition matrices
techniques are introduced).

\cite{kelly66,kelly66a} used the original HF code and created a many-body
theory of atomic properties generalized for open-shell atoms on the basis
of the Feynman diagram technique.

\cite{amusia69,amusia70,amusia71,amusia72a,amusia72} have created a suite of
programs computing many spectral features on the basis of the original HF
program and the random-phase approximation (RPA) \citep{altick64}. Their
approximation, called random-phase approximation with exchange (RPAE), and the
application of the RPAE are reviewed in \citep{amusia74,amusia75} and
summarized in a book by \cite{amusia90a}.

\cite{burke71a} have presented a theory of electron scattering by complex
atoms based on using the R-matrix and described the interconnection of this
technique with the K-matrix and S-matrix methods. \cite{allison72} and
\cite{burke75} applied the R-matrix theory to investigate atomic
polarizabilities and photoionization, respectively.

A many-body theory of atomic transitions based on the transition matrices
technique of \cite{lowdin55} has been created by \cite{chang76a,chang76}. In
these papers, detailed interconnections between the created theory and RPA,
time dependent HF (TDHF), and many-body perturbation theory (MBPT) have been established.

\cite{lundqvist74} and \cite{wendin76} have presented computing techniques of
photoelectron spectra using the whole spectral weight function built by
applying the diagram technique and Green functions. Later on, this technique
has been widely applied to interpret a variety of X-ray photoelectron
spectra (XPES) \citep{ohno80,ohno80a,yarzhemsky92,ohno00,ohno00a,ohno01}.

The suite of computer programs including calculations of AOs in the DF
approximation and relativistic random-phase approximation (RRPA) has been
described by \cite{johnson79} and \cite{johnson80b}. This program package is
the relativistic generalization of the RPA method using the DF AOs. The
time-dependent local-density version of this suite (RTDLDA) has been presented
by \cite{parpia84a,parpia84}.

\cite{dyall82a,dyall82b} have described the realization of the
configuration-interaction (CI) HF approach {accounting for} the
relaxation of AOs {within the} theory of non-orthogonal orbitals where AOs of the initial and final states are computed in different cores (likewise,
e.g., in the work of \cite{sachenko65,aberg67,sukhorukov79}). The AOs used in
computing atomic spectra have been obtained by applying the HF program of
\cite{mayers68}.

In the last quarter of the 20th century, several {suites for} the calculation of polarization phenomena {as, e.g., angular distribution of fluorescence or photoelectrons, alignment, orientation etc.,} have been created. Among
these suites we mention three. {(i)} \cite{tulkki92a}
used the multichannel multiconfiguration Dirac-Fock method (MMCDF) in a
combination of configuration-interaction and K-matrix \citep{starace82}
techniques for the calculation of continuum wave functions. Wave functions of
the ion core were computed using the MCDF method of \cite{grant80}.

{(ii)} In order to compute alignment and orientation of ions under irradiation by polarized synchrotron radiation, \cite{vanderhart98,vanderhart99,vanderhart02}
combined the MCHF program of \cite{fischer72} with the R-matrix technique
described in the review of \cite{aymar96}.

{(iii)} During the last decades, \cite{fritzsche01,fritzsche12} created a suite of programs for relativistic calculations of many atomic properties, RATIP. This
package allows users to calculate about twenty different atomic quantities
(such as: parameters of the Auger decay, Einstein $A_{i,j}$ and $B_{i,j}$
coefficients, photoionization cross sections, parameters of angular
distribution and spin polarization of photoelectrons, alignment and
orientation of photoions, energy levels, etc.) using AOs generated by the
GRASP92 program \citep{parpia96}.

\subsection{Configuration-Interaction Pauli--Fock approximation with Core
Polarization (CIPFCP)}

\label{sec:CIPFCP}

The Configuration Interaction Pauli-Fock approximation with Core Polarization
(CIPFCP) is based on using Pauli-Fock AOs like in \citep{cowan76}.
Configuration interaction theory to build wave functions of the ionic core and
of doubly excited states and K-matrix theory to describe interchannel and
intrachannel interactions were implemented {to account for}
many-electron correlations. By parts the CIPFCP is described in
\citep{schmoranzer93,sukhorukov94,lagutin96,kau97,petrov99,petrov03,demekhin05,sukhorukov10,ehresmann10,sukhorukov12}.
Below we show the main points of this approach.

\subsubsection{Atomic orbitals}

\label{sec:theory_AO}

In central field approximation (CFA), atomic orbitals (AOs) are presented in
hydrogen-like form $\phi_{n\ell sm_{\ell}m_{s}}(r,\theta,\varphi)=\frac{1}%
{r}P_{n\ell}(r)Y_{\ell m_{\ell}}(\theta,\varphi)\chi_{m_{s}}(s)$, but with a
nonhydrogenic radial part $P_{n\ell}(r)$ which is the solution of the equation%

\begin{equation}
\left(  {-}\frac{\mathrm{d}^{2}}{\mathrm{d}r^{2}}{+}\frac{{\ell(\ell+1)}%
}{r^{2}}{+}V_{n\ell}^{\mathrm{CFA}}(r)\right)  P_{n\ell}(r)=\varepsilon
_{n\ell}P_{n\ell}(r). \label{eq:PCF}%
\end{equation}

Here and below, atomic units are used except for the energies, for which we
adopt Rydberg units ($1~\mathrm{Ry}=13.6057~\mathrm{eV}$); $\ell$\ is the
orbital angular momentum quantum number; $\varepsilon_{n\ell}$ is a
variational parameter corresponding to the single-electron energy. The HF
central field potential $V_{n\ell}^{\mathrm{CFA}}(r)$ consists of several parts:%

\begin{equation}
V_{n\ell}^{\mathrm{CFA}}(r)=V_{n\ell}(r)-X_{n\ell}(r). \label{eq:VCF}%
\end{equation}

The potential $V_{n\ell}(r)$ is represented by the {Coulomb potential of the nucleus} $-2Z/r$ and the local\ part of the electron--electron interaction; the term $X_{n\ell }(r)$ describes the exchange part of the electron--electron interaction.

Relativistic corrections were included in equation (\ref{eq:PCF}) using the
Breit--Pauli operator \citep{bethe57}. The major relativistic terms which
influence the distribution of the core electron density are one-electron
mass--velocity $H_{n\ell}^{\mathrm{m}}(r)$ and Darwin $H_{n\ell}^{\mathrm{D}%
}(r)$ corrections. The expressions for these corrections are obtained by
excluding the `small' component from the system of Dirac--Fock equations. The
action of the $H_{n\ell}^{\mathrm{m}}(r)$ and $H_{n\ell}^{\mathrm{D}}(r)$ on
the $P_{n\ell}(r)$ is the following \citep{cowan81}:%

\begin{equation}
H_{n\ell}^{\mathrm{m}}(r)P_{n\ell}(r)=-\frac{{\alpha^{2}}}{4}\left(
\varepsilon_{n\ell}-V_{n\ell}(r)\right)  ^{2}P_{n\ell}(r), \label{eq:Hm}%
\end{equation}

\begin{equation}
H_{n\ell}^{\mathrm{D}}(r)P_{n\ell}(r)=-\delta_{\ell,0}\frac{\alpha^{2}}%
{4}\left[  1+\frac{\alpha^{2}}{4}\left(  \varepsilon_{n\ell}-V_{n\ell
}(r)\right)  \right]  ^{-1}r\frac{\mathrm{d}V_{n\ell}(r)}{\mathrm{d}r}%
\frac{\mathrm{d}\left[  P_{n\ell}(r)/r\right]  }{\mathrm{d}r}, \label{eq:HD}%
\end{equation}

In these equations, ${\alpha}=1/137.036$ is the fine-structure constant. The
terms $H_{n\ell}^{\mathrm{m}}$ and $H_{n\ell}^{\mathrm{D}}$ have spherical
symmetry and therefore do not change the usual nonrelativistic configuration.
The spin--orbit operator $H_{n\ell j}^{\mathrm{SO}}(r)$, where $j$ is the
total angular momentum quantum number of the electron ($\vec{j}=\vec{\ell
}+\vec{s}$), reads:%

\begin{equation}
H_{n\ell j}^{\mathrm{SO}}(r)=\frac{j(j+1)-\ell(\ell+1)-s(s+1)}{2}\cdot
\frac{{\alpha^{2}}}{2}\left[  1+\frac{\alpha^{2}}{4}(\varepsilon_{n\ell
}-V_{n\ell}(r))\right]  ^{-1}\frac{1}{r}\frac{\mathrm{d}V_{n\ell}%
(r)}{\mathrm{d}r}. \label{eq:HSO}%
\end{equation}

Inserting (\ref{eq:Hm},\ref{eq:HD},\ref{eq:HSO}) into (\ref{eq:PCF}) results
in the `Pauli--Fock' radial functions $P_{n\ell j}(r)$. A detailed derivation of
equations (\ref{eq:Hm},\ref{eq:HD},\ref{eq:HSO}) can be found in \citep{selvaraj84,lagutin98}.

The PF equations (\ref{eq:PCF}) are solved {by} numerical methods
described in \citep{amusia74,amusia75}. Rewritten for continuum AOs
$P_{\varepsilon l}(r)$, equation (\ref{eq:PCF}) is solved in the frozen core
approximation with the following normalization condition:%

\begin{equation}
P_{\varepsilon\ell}(r)\overset{r\rightarrow\infty}{\longrightarrow}\sqrt
{\frac{2}{\pi k}}\mathrm{sin}(kr-\frac{\ell\pi}{2}+\frac{Z_{as}}{k}%
\mathrm{ln}(2kr)+\delta_{\ell}). \label{eq:asympt}%
\end{equation}

Here $k$ is the wave number of the continuum electron in a.u.; $Z_{as}\ $is
the asymptotic charge of the ion;$\ \delta_{\ell}$ represents the sum:%

\begin{equation}
\delta_{\ell}=arg~\Gamma\left(  \ell+1-\imath\frac{Z_{as}}{k}\right)
+\varphi_{\ell} \label{eq:phase}%
\end{equation}
where $\varphi_{\ell}$\ is the short-range phase shift.

The phase shift $\delta_{\ell}$ is computed via a nonrelativistic procedure
with relativistically corrected wave vector $k$ and effective charge $Z_{eff}$
as \citep{aaberg82}:%

\begin{equation}
k^{2}=\varepsilon(1+\frac{\alpha^{2}}{4}\varepsilon),\qquad Z_{eff}%
=Z_{as}(1+\frac{\alpha^{2}}{2}\varepsilon). \label{eq:Zas}%
\end{equation}

In computing the AOs for atoms with $Z\gtrsim30$, the influence\ of terms
$H_{n\ell}^{\mathrm{m}}$ (\ref{eq:Hm}) and $H_{n\ell}^{\mathrm{D}}$
(\ref{eq:HD}) is found to be considerable \citep{kau97,petrov99}. It is also
important to take into account the finite size of the nucleus which is considered
as a homogeneously charged sphere with radius $R_{\mathrm{n}}=2.2677\times
10^{-5}\,A^{1/3}\,a_{0}$, where $A$ is the atomic nucleon number \citep{desclaux75}.

The influence of many-electron correlations on $P_{n\ell}(r)$ {is considered by including} the core polarization potential $V_{\ell}^{\mathrm{CP}}(r)$ in equation (\ref{eq:PCF}). This potential has been
derived in \citep{petrov99} by applying the variational principle for the
total energy of the atom obtained using the second-order correlational
corrections as described in, e.g., \citep{sukhorukov94,lagutin96}. The
\emph{ab initio} core polarization potential derived in \cite{petrov99} has
the asymptotic form connected with the dipole polarizability $\alpha
_{\mathrm{d}}$ of the ionic core as $V^{\mathrm{CP}}(r)=-\alpha_{\mathrm{d}%
}/2r^{4}$ like in, e.g.,
\citep{weisheit72,norcross73,aymar78,laughlin78,aymar84}, but a constant value
of about 1~Ry in the inner core region, in contrast to the above references
where the cut-off radius is used at small distances. We note that in some
calculations of atomic structures the complete set of AOs was computed without
the core polarization potential. In this case the CP is omitted and
the approximation is called CIPF.

\subsubsection{Wave functions of the ionic core.}

\label{sec:theory_functions}

In this section, we outline some major points of the wave function
calculations following \cite{lagutin96}. As an example
illustrating the inclusion of many-electron correlations in the calculations,
we use the $3s$ photoionization of Ar described in the lowest order of
perturbation theory by the transition:%

\begin{equation}
3s^{2}3p^{6}\dashrightarrow3s^{1}3p^{6}\varepsilon p. \label{eq:Ar-lowest}%
\end{equation}

The wave function of the final state is known to be strongly influenced by the
interaction between the $3s^{1}3p^{6}$ and $3s^{2}3p^{4}n\ell$ configurations \citep{minnhagen63}.

The multi-configuration wave functions corresponding to $3p^{4}n\ell~(E_{c}J)$
states are represented by:%

\begin{equation}
\left\vert E_{c}J\right\rangle =\sum_{\alpha LS}\left\langle \alpha
LSJ\left\vert E_{c}J\right.  \right\rangle \left\vert \alpha LSJ\right\rangle
\label{eq:core_WF}%
\end{equation}
where the set of single-configuration wave functions $\left\vert \alpha
LSJ\right\rangle $\ ($\alpha$ denotes all internal quantum numbers including
electron configuration) consists of the $\left\vert 3s^{1}3p^{6~2}%
S_{1/2}\right\rangle $\ main level and the $\left\vert 3s^{2}3p^{4}%
n\ell~^{2S+1}L_{J}\right\rangle $ $\left(  \ell=s,d\right)  $ $\left(
J=\frac{1}{2};\frac{3}{2};\frac{5}{2};\frac{7}{2};\frac{9}{2}\right)
$\ satellites (only these states are accessible from the Ar $^{1}S_{0}$ ground
state according to the dipole selection rule). The coefficients $\left\langle
\alpha LSJ\left\vert E_{c}J\right.  \right\rangle $ are the solution of the
ordinary secular equation:%

\begin{equation}
\sum_{\alpha^{\prime}L^{\prime}S^{\prime}}\left\langle \alpha^{\prime
}L^{\prime}S^{\prime}J\left\vert E_{c}J\right.  \right\rangle \left(
\left\langle \alpha^{\prime}L^{\prime}S^{\prime}J\left\vert \mathbf{H}%
\right\vert \alpha LSJ\right\rangle -E\ \delta\left(  \alpha^{\prime}%
L^{\prime}S^{\prime},\alpha LSJ\right)  \right)  =0 \label{eq:secular}%
\end{equation}
with matrix elements computed using AOs as described in section
(\ref{sec:theory_AO}).

In order to take into account the residual part of the Coulomb interaction,
the matrix elements entering equation (\ref{eq:secular}) are {refined by including} the influence of high-lying excited configurations.
Usually all single- and double-excitations of the ionic core are taken into
account using the second order of perturbation theory (PT).

The correction for the center of gravity of the electron configuration $K$ has
the meaning of a correlational energy:%

\begin{equation}
E_{C}(K)=\frac{1}{g(K)}\sum\limits_{m\alpha LSJ}\frac{(2J+1)\left\langle
m\right\vert \mathbf{H}^{ee}\left\vert \alpha LSJ\right\rangle ^{2}}{E(\alpha
LSJ)-E(m)} \label{eq:theor_Vii}%
\end{equation}
where $\alpha$ is the set of internal quantum numbers including the electron
configuration $K$, $g(K)$ is the statistical weight of the configuration $K$,
and the sum over $m$ includes summation over discrete and integration over
continuum states and their quantum numbers. Summation over $LS$ belonging to
the fixed configuration $K$ can be performed in closed form if one assumes
$E(\alpha LSJ)\approx E(K)$ and uses the \textquotedblleft transition
array\textquotedblright\ technique \citep{bauche-arnoult79,bauche-arnoult82,bauche-arnoult85,karazija91}.

Nondiagonal matrix elements of the Coulomb operator $\mathbf{H}^{ee}$ can be
corrected by solving the K-matrix equation :%

\begin{equation}
\left\langle \alpha^{\prime}LSJ\left\vert \mathbf{H}_{eff}^{ee}\right\vert
\alpha LSJ\right\rangle =\left\langle \alpha^{\prime}LSJ\left\vert
\mathbf{H}^{ee}\right\vert \alpha LSJ\right\rangle +\sum\limits_{m}%
\frac{\left\langle \alpha^{\prime}LSJ\left\vert \mathbf{H}^{ee}\right\vert
m\right\rangle \ \left\langle m\left\vert \mathbf{H}_{eff}^{ee}\right\vert
\alpha LSJ\right\rangle }{E(\alpha LSJ)-E(m)}. \label{eq:theor_Vik}%
\end{equation}

\cite{lagutin98} used a simplification of this equation introducing the factor
$\chi\left(  \alpha,\alpha^{\prime}\right)  $ which scales the directly
computed matrix element $\left\langle \alpha^{\prime}LSJ\left\vert
\mathbf{H}^{ee}\right\vert \alpha LSJ\right\rangle $ to its effective value
$\left\langle \alpha^{\prime}LSJ\left\vert \mathbf{H}_{eff}^{ee}\right\vert
\alpha LSJ\right\rangle $ and assumed that $\chi\left(  \alpha,\alpha^{\prime
}\right)  $, which enters equation (\ref{eq:theor_Vik}), depends on the
quantum numbers $\alpha,\alpha^{\prime}$\ only. With this assumption, equation
(\ref{eq:theor_Vik}) becomes:%

\begin{equation}
\frac{\left\langle \alpha^{\prime}LSJ\left\vert \mathbf{H}^{ee}\right\vert
\alpha LSJ\right\rangle }{\chi\left(  \alpha,\alpha^{\prime}\right)
}=\left\langle \alpha^{\prime}LSJ\left\vert \mathbf{H}^{ee}\right\vert \alpha
LSJ\right\rangle +\frac{\Delta\left\langle \alpha^{\prime}LSJ\left\vert
\mathbf{H}^{ee}\right\vert \alpha LSJ\right\rangle }{\chi\left(  \alpha
,\alpha^{\prime}\right)  }. \label{eq:theor_eqx}%
\end{equation}
This equation has a simple solution for $\chi\left(  \alpha,\alpha^{\prime
}\right)  $:%

\begin{equation}
\chi\left(  \alpha,\alpha^{\prime}\right)  =1-\frac{\Delta\left\langle
\alpha^{\prime}LSJ\left\vert \mathbf{H}^{ee}\right\vert \alpha
LSJ\right\rangle }{\left\langle \alpha^{\prime}LSJ\left\vert \mathbf{H}%
^{ee}\right\vert \alpha LSJ\right\rangle } \label{eq:theor_x}%
\end{equation}
where the correction is computed in 2nd order of PT:%

\begin{equation}
\Delta\left\langle \alpha^{\prime}LSJ\left\vert \mathbf{H}^{ee}\right\vert
\alpha LSJ\right\rangle =\sum\limits_{m}\frac{\left\langle \alpha^{\prime
}LSJ\left\vert \mathbf{H}^{ee}\right\vert m\right\rangle \ \left\langle
m\left\vert \mathbf{H}^{ee}\right\vert \alpha LSJ\right\rangle }{E(\alpha
LSJ)-E(m)}. \label{eq:theor_Delta}%
\end{equation}

The reduction factor can be computed directly or using the effective operator
methods \citep{judd67,lindgren86}. This technique allows one to represent the
main part of the correction to the matrix element $\left\langle \alpha
^{\prime}LSJ\left\vert \mathbf{H}^{ee}\right\vert \alpha LSJ\right\rangle $ as
corrections $\Delta F^{k}$ and $\Delta G^{k}$ to the Slater integrals $F^{k}$
and $G^{k}$ determining this matrix element. Fairly good agreement between
direct and simplified calculations of the reduction factors has been found for
the case investigated by \cite{petrov03}.

\subsubsection{Photoionization cross sections.}

\label{sec:theory_cross}

{The consideration of} many-electron correlations complicates the scheme (\ref{eq:Ar-lowest}) of the {Ar} $3s$ photoionization:%

\begin{equation}%
\begin{array}
[c]{ccc}%
\begin{array}
[c]{cc}%
\left(  \mathbf{0}\right)  & 3s^{2}3p^{6}\\
& ground~state
\end{array}
& \dashrightarrow & \left\{
\begin{array}
[c]{cc}%
3s^{2}3p^{5}\varepsilon\left(  s/d\right)  & \left(  3\mathbf{p}\right) \\
3s^{1}3p^{6}\left(  n/\varepsilon\right)  p & \left(  3\mathbf{s}\right)
\end{array}
\right. \\
\updownarrow &  & \updownarrow\\
\left.
\begin{array}
[c]{cc}%
\left(  \mathbf{a}\right)  & 3s^{2}3p^{4}\left\{  s/d\right\}  ~\left(
n/\varepsilon\right)  (s/d)\\
\left(  \mathbf{b}\right)  & 3s^{1}3p^{5}n(s/d)~\left(  n^{\prime}%
/\varepsilon\right)  (p/f)\\
\left(  \mathbf{c}\right)  & 3s^{2}3p^{4}\left\{  p/f\right\}  ~\left(
n^{\prime}/\varepsilon\right)  (p/f)
\end{array}
\right\}  &  & \left\{
\begin{array}
[c]{cc}%
3s^{2}3p^{4}n(s/d)~\varepsilon(p/f) & \left(  \mathbf{sat}\right) \\
3s^{2}3p^{4}n(s/d)~n^{\prime}(p/f) & \left(  \mathbf{2ex}\right)
\end{array}
\right. \\
\mathbf{ISCI} &  & \mathbf{FISCI}%
\end{array}
\label{eq:Ar-scheme}%
\end{equation}
where the horizontal dashed arrow denotes the electric-dipole interaction and
the double arrows denote the Coulomb interaction; $\left\{  \ell\right\}  $
means a complete set of intermediate AOs, over which summation and integration
are carried out. Electric-dipole interaction between the states in the
\emph{lower} part of the diagram marked by single brace is neglected. The total
and intermediate momenta of all states are omitted in
scheme~(\ref{eq:Ar-scheme}) to simplify the notation. Usually, correlations
including a continuum are taken into account by the K-matrix technique. If the
correlation entering scheme (\ref{eq:Ar-scheme}) contains divergent
continuum-continuum integrals, the technique of correlational functions has
been used along the lines of \cite{sukhorukov94}, following the procedure
described in \citep{chang75,laughlin78,aymar78}.

Scheme (\ref{eq:Ar-scheme}) describes the influence of the most significant
correlations on the $3s\dashrightarrow\varepsilon p$ transition. For instance,
the pathway $\left\langle \mathbf{0\leftrightarrow a}\dashrightarrow
3\mathbf{p}\right\rangle $ describes the intrashell correlation influencing the
(intermediate) $3p$ photoionization; the pathway $\left\langle \mathbf{0}%
\dashrightarrow3\mathbf{p\leftrightarrow3s}\right\rangle $ describes the
intershell correlation qualitatively changing the near-threshold $3s$
photoionization having also an abundant resonance structure due to the
$\left\langle \mathbf{0}\dashrightarrow3\mathbf{s\leftrightarrow
2ex}\right\rangle $ pathway, etc.

The photoionization cross section for the $\left\vert E_{c}J\right\rangle $
state (\ref{eq:core_WF}) is:%

\begin{equation}
\sigma_{E_{c}J}(\omega) =\sum_{\ell,j}\sigma_{E_{c}J}^{\varepsilon\ell
j}(\omega)=\frac{4}{3}\pi^{2}\alpha a_{0}^{2}\omega^{\pm1}\sum_{\ell
,j}\left\vert D(E_{c}J\varepsilon\ell j)\right\vert ^{2} \label{eq:PICS}%
\end{equation}

\begin{equation}
D(E_{c}J\varepsilon\ell j) =\left\langle \overline{E_{c}J\varepsilon\ell
j}\left\Vert \mathbf{D}\right\Vert 0\right\rangle \label{eq:ampl_D}%
\end{equation}
where\ the signs ($+$) and ($-$) correspond to the length and velocity forms
of the transition dipole operator $\mathbf{D}$, respectively; $\omega$
determined by $E_{0}+\omega=E_{c}+\varepsilon$\ stands for the exciting-photon
energy in atomic units, $\alpha=1/137.036$ is the fine-structure constant, and
the square of the Bohr radius $a_{0}^{2}=28.0028$~Mb converts the atomic units
to cross sections in $\mathrm{Mb}=10^{-22}$~m$^{2}$.

A line over the final state wave function $\left\vert \overline{E_{c}%
J\varepsilon\ell j}\right\rangle $ entering equation~(\ref{eq:ampl_D}) denotes
that the wave function is modified by interaction with both all resonances via
the pathway $\left\langle \mathbf{0}\dashrightarrow3\mathbf{s\leftrightarrow
2ex}\right\rangle $ in scheme~(\ref{eq:Ar-scheme}) and other continua. This
wave function is computed applying the K-matrix technique \citep{starace82}
and the theory of interacting resonances in the complex calculus form
\citep{sorensen94} as:%

\begin{equation}
\left\vert \overline{E_{c}J\varepsilon\ell j}\right\rangle =\left\vert
E_{c}J\varepsilon\ell j\right\rangle +\sum_{i}\frac{\left\langle \overline
{i}\left\vert \mathbf{H}^{ee}\right\vert E_{c}J\varepsilon\ell j\right\rangle
}{E-E^{(i)}}\left[  \left\vert \overline{i}\right\rangle +\sum_{\beta}\int
dE^{\prime}\frac{\left\langle \beta E^{\prime}\left\vert \mathbf{H}%
^{ee}\right\vert \overline{i}\right\rangle }{E-E^{\prime}-\imath\delta
}\left\vert \beta E^{\prime}\right\rangle \right]  \label{eq:wf_cont}%
\end{equation}
where the summation over all resonances $\left\vert \overline{i}\right\rangle
$\ and continua $\left\vert \beta E\right\rangle =\left\vert E_{c}%
J\varepsilon\ell j\right\rangle ~\left(  \ell=p,f\right)  $; $\left\vert
4p_{J}^{5}\varepsilon\ell^{\prime}j^{\prime}\right\rangle $ $\left(
\ell^{\prime}=s,d\right)  \ $is performed. The total energy $E$ entering
equation~ (\ref{eq:wf_cont}) is connected with the photoelectron energy
$\varepsilon$ and the threshold energy $E_{c}$ via the usual relation:
$E=E_{c}+\varepsilon$.

After building the basis set of doubly excited states ($\mathbf{2ex}$)
$\left\vert m\right\rangle $, the functions of the resonances modified via the
interaction between continua and between resonances through continua were
computed in the form:%

\begin{equation}
\left\vert \overline{i}\right\rangle =\sum_{m}b_{m}^{\left(  i\right)
}\left\vert m\right\rangle . \label{eq:wf_disc}%
\end{equation}

Complex energies of resonances $E^{(i)}$ and complex coefficients
$b_{m}^{\left(  i\right)  }$ were obtained as the solution of the secular
equation with a complex symmetric (and therefore non-Hermitian) matrix:%

\begin{equation}
\sum_{m}\left[  \left(  E_{m}-E^{\left(  i\right)  }\right)  \delta
_{mm^{\prime}}+\left\langle m\left\vert \mathbf{H}^{ee}\right\vert m^{\prime
}\right\rangle +\sum_{\beta}\int dE^{\prime}\frac{\left\langle m\left\vert
\mathbf{H}^{ee}\right\vert \beta E^{\prime}\right\rangle \left\langle \beta
E^{\prime}\left\vert \mathbf{H}^{ee}\right\vert m^{\prime}\right\rangle
}{E-E^{\prime}+\imath\delta}\right]  b_{m}^{\left(  i\right)  }=0
\label{eq:secular_cmp}%
\end{equation}

The complex energy of each resonance contains its position $E_{i}$ and width
(FWHM) $\Gamma_{i}$ as%

\begin{equation}
E_{i}=\mathrm{Re}E^{(i)},\qquad\Gamma_{i}=-2\mathrm{Im}E^{(i)}.
\label{eq:energ_width}%
\end{equation}
The function~(\ref{eq:wf_disc}) {enables} to compute the complex transition amplitude $D^{(i)}$ and Fano parameters \citep{fano61,fano65}\ for the
resonance $\left\vert \overline{i}\right\rangle $ as:%

\begin{equation}
D^{(i)}=\left\langle \overline{i}\left\Vert \mathbf{D}\right\Vert
0\right\rangle ,\qquad q_{i}=-\frac{\mathrm{Re}D^{(i)}}{\mathrm{Im}D^{(i)}%
},\qquad\sigma_{0}\rho_{i}^{2}=\frac{\left(  \mathrm{Im}D^{(i)}\right)  ^{2}%
}{\pi\Gamma_{i}/2} \label{eq:Fano_par}%
\end{equation}
where the so-called background cross section $\sigma_{0}$ is computed via
equation~(\ref{eq:PICS}) using the `non-resonant' continuum wave functions
$\left\vert E_{c}J\varepsilon\ell j\right\rangle $. The parameters $q_{i}$,
$\rho_{i}^{2}$, and $\sigma_{0}$\ enter the well known formula
\citep{fano61,shore68} and can be used for the parameterization of computed
line shapes as:%

\begin{equation}
\sigma_{E_{c}J}\left(  \omega\right)  =\sum_{i}\sigma_{0}\rho_{i}^{2}\left[
\frac{\left(  q_{i}+\epsilon_{i}\right)  ^{2}}{1+\epsilon_{i}^{2}}-1\right]
+\sigma_{0} \label{eq:cross_shore}%
\end{equation}
where $\epsilon_{i}$ is the reduced energy connected with energy and width of
the resonance $E_{i}$\ and $\Gamma_{i}$, respectively, by $\epsilon
_{i}=2\left(  E-E_{i}\right)  /\Gamma_{i}$.

\section{Photoionization Cross Sections (PICSs) of the outer shells and their
interpretation}

In this chapter we present a short review of the state of the experimental and
theoretical investigations preceding the development of the photon-induced
fluorescence spectroscopy (PIFS) emphasizing the areas where this technique
contributes substantially.

\subsection{Earlier measurements of the absolute PICSs of rare-gas atoms}

Since the first measurement of the absolute photoabsorption cross sections of
some alkali atoms by \cite{mohler29}, hundreds of papers on this subject have
been published, among which exist very detailed reviews (see, e.g.,
\citep{samson66,marr76,samson76,schmidt92,henke93,samson02}). In the present
section, we concentrate on the papers mainly devoted to some ideas improving
our understanding of the photoeffect in the outer shells of rare gases,
particularly Ne, Ar, Kr, and Xe.

\begin{figure}[ptb]
\begin{minipage}[t]{0.48\textwidth}
\begin{center}
\includegraphics[width=\textwidth]{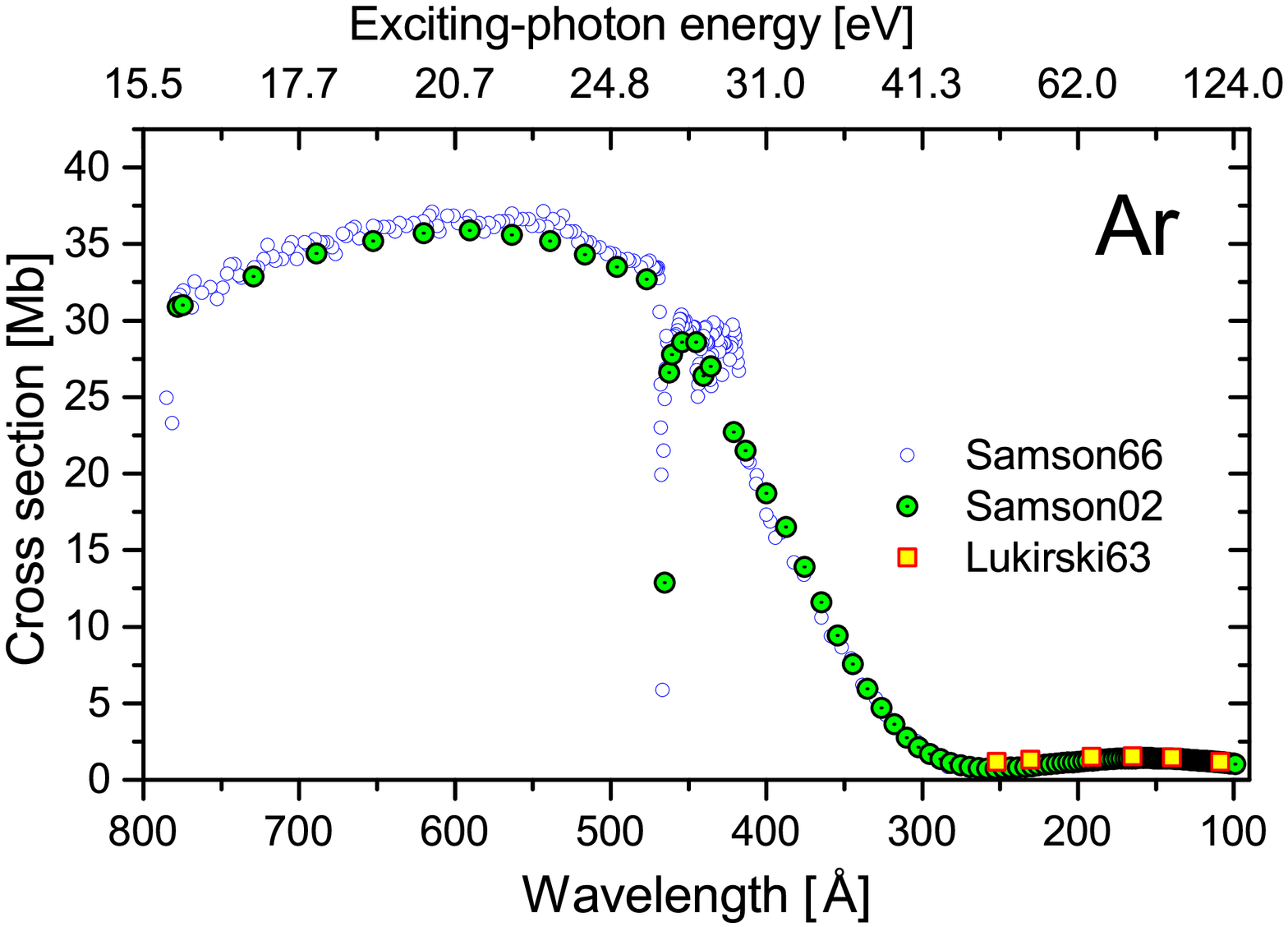}
\caption{Photoionization cross section of Ar. Experimental data are from
\citep{lukirskii63,samson66,samson02}.}\label{fig:ABS_Samson66}\end{center}
\end{minipage} \hfill\begin{minipage}[t]{0.48\textwidth}
\begin{center}
\includegraphics[width=\textwidth]{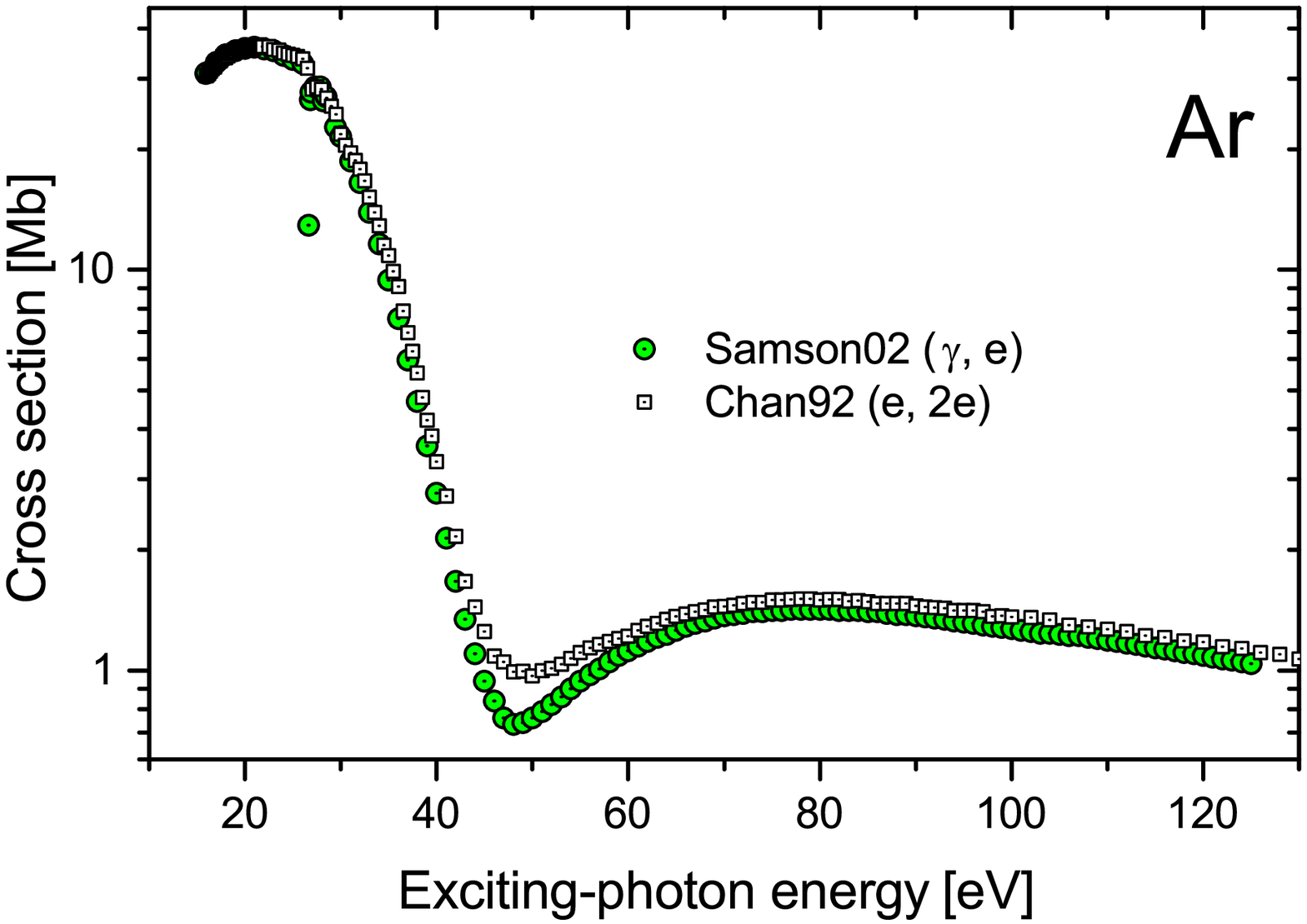}
\caption{Comparison between photoionization cross sections measured in (e, 2e) \citep{chan92a} and ($\gamma,~e$) \citep{samson02} experiments. Adapted from
\citep{samson02}}\label{fig:ABS_Samson02}\end{center}
\end{minipage}
\end{figure}

The total PICSs of Ar, Kr, and Xe have been measured in the near-threshold region
by both \cite{samson63,samson64} who used a high-voltage spark discharge in Ar
and in the soft X-ray region and by \cite{lukirskii63}. The most important
data for rare gases (\emph{Rg}) and some other gases and metal vapors were
collected in the review of \cite{samson66}. For Ar, the PICS are depicted in
Fig.~\ref{fig:ABS_Samson66}. In this figure one can recognize the structure
connected with the $3s^{1}np$ resonances at about 27~eV and the Cooper minimum
connected with the sign-reversal behaviour of the transition amplitude
\citep{cooper62} at about 50 eV. Later these experimental features became a
subject of investigation in numerous papers. The \emph{Rg} PICS measurements
have been revisited several times (see, e.g., the reviews
\citep{marr76,samson89}). At present, the state-of-the-art PICS are published
by \cite{samson02} for all \emph{Rg} with a quoted accuracy of 1-3\%. In
Fig.~\ref{fig:ABS_Samson02} the Ar PICS of \cite{samson02} are compared with the
PICS of \cite{chan92a} obtained in an (e, 2e) experiment. One can see good
overall agreement between cross sections measured by photon and electron
impact. However, in the region of the Cooper minimum these methods yield
substantially different results.

\begin{figure}[ptb]
\begin{minipage}[t]{0.50\textwidth}
\begin{center}
\includegraphics[width=\textwidth]{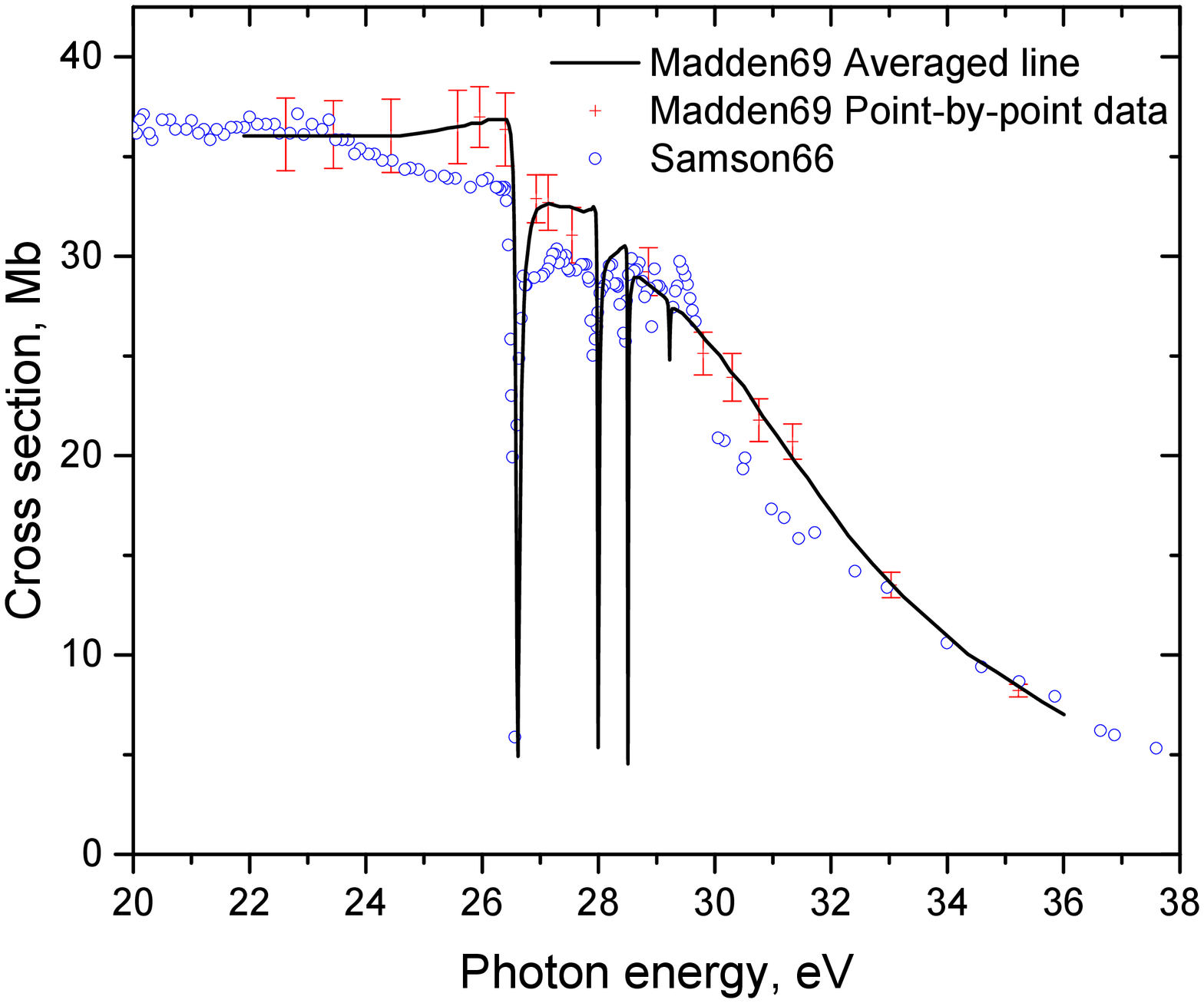}
\caption{Photoionization cross section of Ar influenced by $3s^{1}np$
resonances. Resonances connected with
doubly excited states at energies above 29 eV are not shown. Open circles are data of
\cite{samson66}. Adapted from
\citep{madden69}}\label{fig:Madden69_3p}\end{center}
\end{minipage} \hfill\begin{minipage}[t]{0.47\textwidth}
\begin{center}
\includegraphics[width=\textwidth]{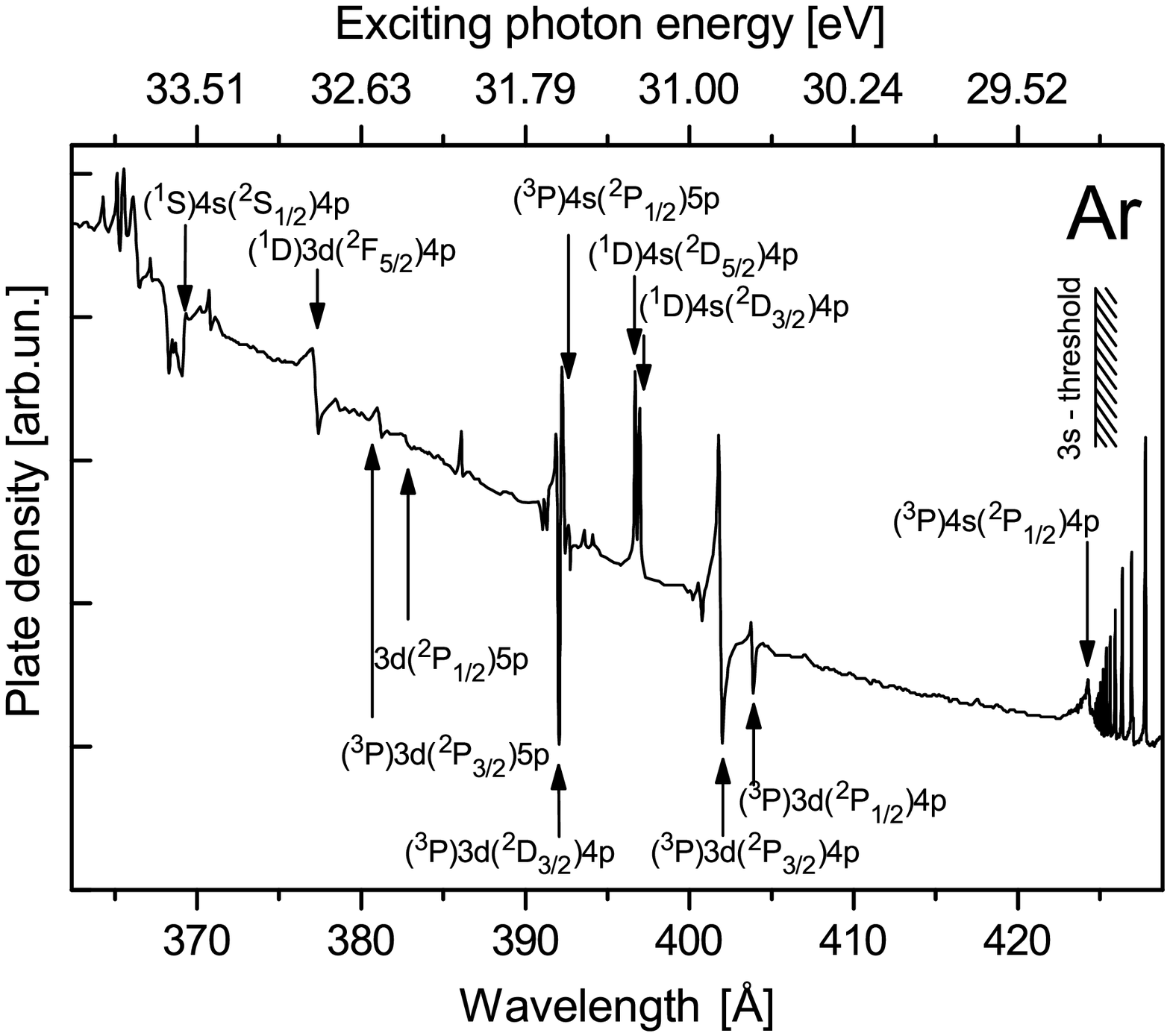}
\caption{Resonance structure in the 3p photoionization of Ar above the $3s$ threshold connected with doubly excited states. Figure with a densitometer trace adapted from
\citep{madden69}.}\label{fig:Madden69_2ex}\end{center}
\end{minipage}
\end{figure}

A rich resonance structure has been observed in He \citep{madden63}, Ne
\citep{codling67}, Ar \citep{madden69}, and Kr and Xe
\citep{codling71,codling72} PICS utilizing synchrotron radiation, which later
became the routine excitation source. As an example of earlier observations,
we show the Ar PICS in the vicinity of the $3s^{1}np$ resonances
(Fig.~\ref{fig:Madden69_3p}) and a densitometer trace just above the Ar $3s$
threshold (Fig.~\ref{fig:Madden69_2ex}) from \citep{madden69}. Many of the
resonances observed in Fig.~\ref{fig:Madden69_2ex} have been attributed to the
doubly excited states $3p^{4}n(s/d)n^{\prime}\ell$ of which the energy
positions are known from optical spectra of \cite{minnhagen63}.

\begin{figure}[tb]
\begin{center}
\includegraphics[width=0.7\textwidth]{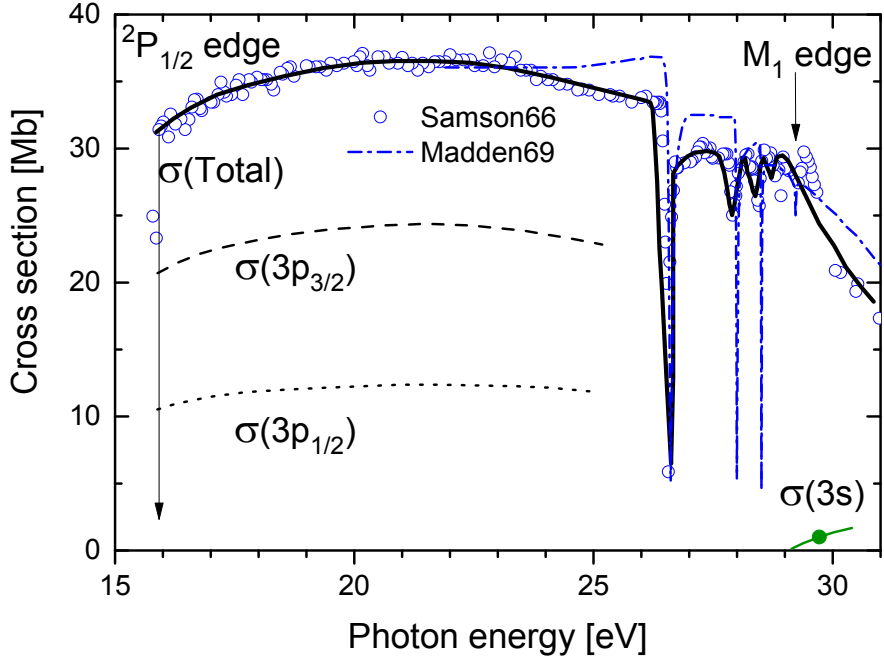}
\end{center}
\caption{Total and partial photoionization cross sections $\sigma_{3p_{3/2}}$
and $\sigma_{3p_{1/2}}$ of Ar. Filled circle in the right lower corner
represents $\sigma_{3s}$. Adapted from \citep{samson68}. Total
photoionization cross sections from \citep{samson66,madden69} are also shown.
Vertical arrows mark the $3p_{1/2}$ and $3s$ thresholds.}%
\label{fig:Samson68_3p}%
\end{figure}

\cite{samson68} have applied photoelectron spectroscopy, a relatively new
method at that time, invented by \cite{vilesov61} and Turner
\citep{al-joboury63} to measure the partial photoionization cross sections
(PICS). In the right lower corner of Fig.~\ref{fig:Samson68_3p}, adapted from
\cite{samson68}, one can recognize one point which was associated with the
$3s$ partial PICS $\sigma_{3s}$. In this paper, the authors pointed out that
the threshold value of $\sigma_{3s}$ is of the order of 1~Mb. However, they
attributed the large value to the experimental inaccuracy because the
calculation of \cite{manson68} existing at this time predicted $\sigma_{3s}$
close to zero. Only four years later the strong influence of many-electron
correlations on $\sigma_{3s}$ was discovered theoretically by \cite{amusia72}
predicting an energy dependence of the cross section that differs
qualitatively from the calculation of \cite{manson68}.

\subsection{Earlier calculations of the PICSs of rare-gas atoms}

\label{sec:Intershell}

Among the initiators of the \emph{ab initio} calculations of photoionization
were \cite{bates41}, who have computed the photoionization of Ca and Ca$^{+}$
in HF approximation, and \cite{cooper62}, who used the core AOs known from
literature and the originally computed model continuum AOs to calculate
photoionization cross sections of He, Ne, Ar, and Kr, and some other systems.
In the latter work, the atomic subshells were classified using different
spectral distributions of oscillator strengths and it was stressed that
changing the sign in some transition amplitudes results in cross sections with
a pronounced minimum later called Cooper minimum. \cite{altick64} were the
first who clearly showed the importance of taking into account many-electron
correlations in the calculation of the photoionization cross sections applying
the RPA approximation for the description of the photoeffect in the outer shells
of Be, Mg, Ca, and Sr. \cite{manson68} have performed a large-scale
calculation of partial photoionization cross sections utilizing the
one-electron local density model with a Herman-Skillman (HS) \citep{herman63}
central potential. The most important papers existing at that time have been
reviewed by \cite{fano68}.

\begin{figure}[ptb]
\begin{minipage}[t]{0.48\textwidth}
\begin{center}
\includegraphics[width=\textwidth]{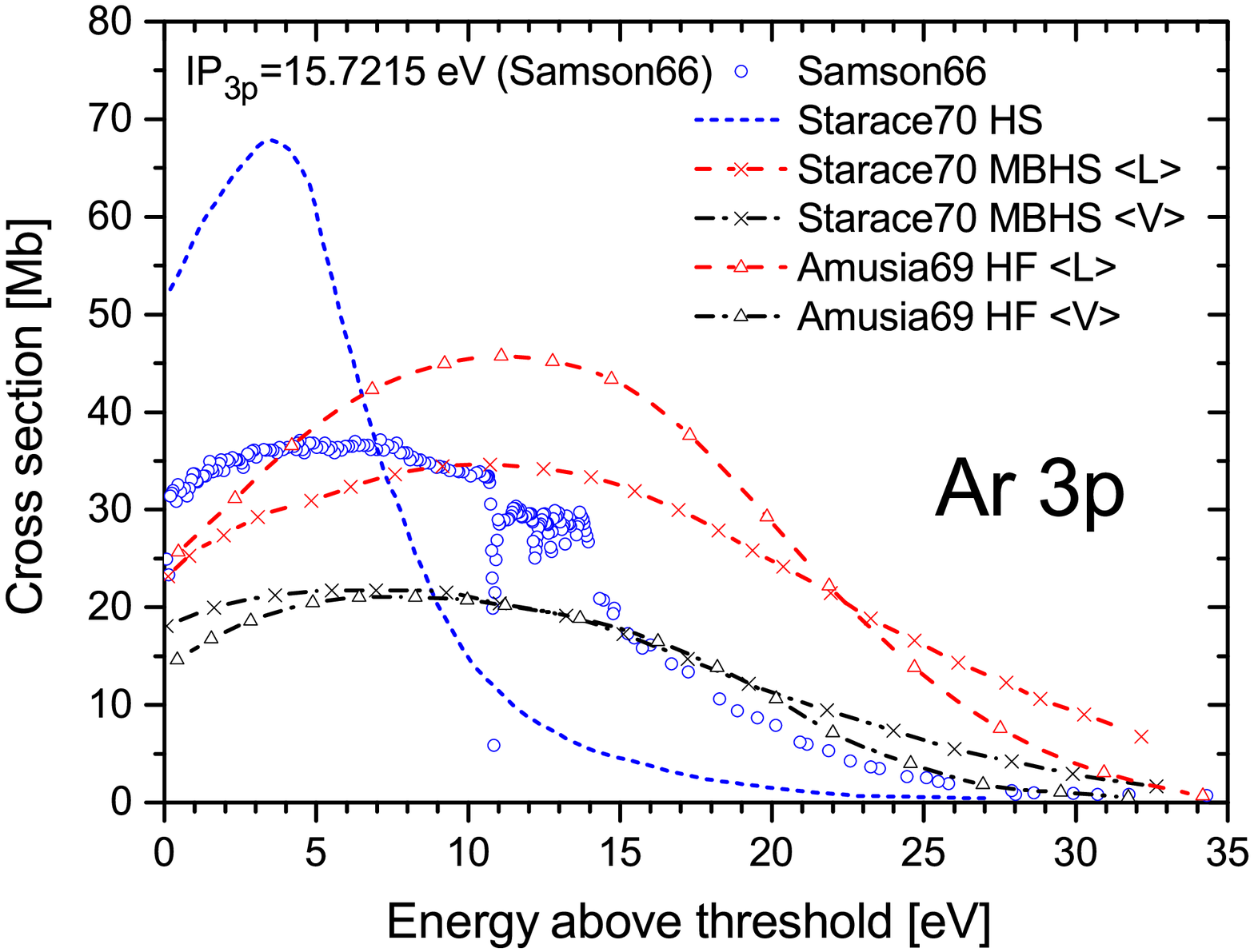}
\caption{Transformation of $\sigma^{HS}(3p)$  computed with HS potential of \cite{herman63} to model cross sections $\sigma_{L,V}^{mod}(3p)$ via taking into account perturbation (\ref{eq:V_HF-HS}). HF cross sections $\sigma_{L,V}^{HF}(3p)$ of \cite{amusia69} and experimental data of \cite{samson66} are shown for comparison. Adapted from
\citep{starace70}}\label{fig:3pStarace70}\end{center}
\end{minipage} \hfill\begin{minipage}[t]{0.48\textwidth}
\begin{center}
\includegraphics[width=\textwidth]{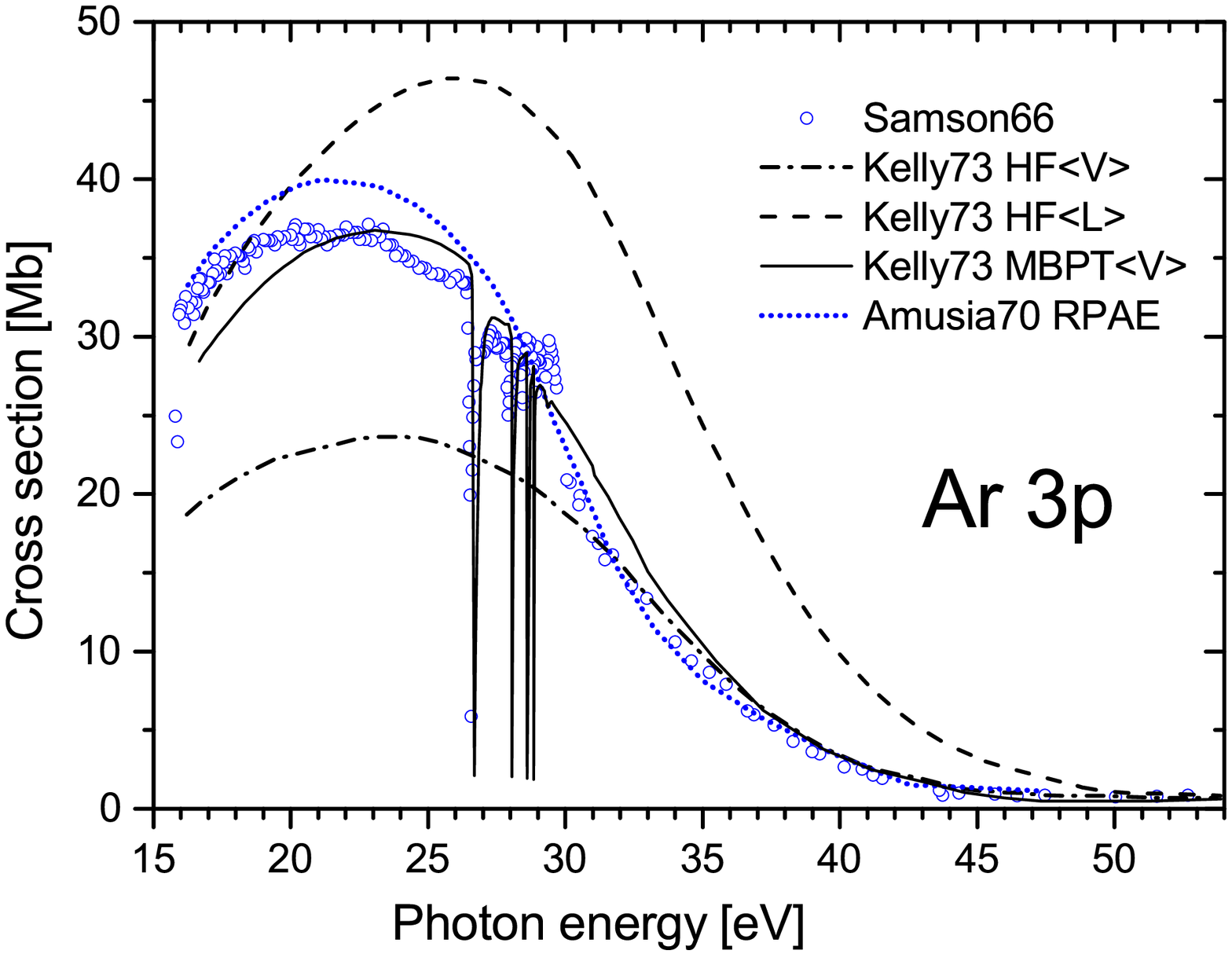}
\caption{Comparison between cross sections $\sigma_{L,V}^{HF}(3p)$ computed in HF approach and with taking into account intrashell correlations \citep{amusia70,kelly73}. Calculation of \cite{kelly73} includes also $3s^13p^6np$ resonances. Experimental data of \cite{samson66} are shown for
comparison.}\label{fig:3pKelly73}\end{center}
\end{minipage}
\end{figure}

The disadvantage of using the HS calculation for the description of
photoionization was revealed by \cite{amusia69} who illustrated that the
nucleus in the HS potential is `too opened' which results in too sharp and
large PICS of the valence shells in $Rg$. In Ar, the PICS $\sigma^{HS}(3p)$ is
shifted towards the photoionization threshold and has $\sim70$~Mb at its
maximum instead of $\sim35$~Mb observed in experiment (see
Fig.~\ref{fig:3pStarace70}). Such a difference between HS theory
($\sigma_{max}^{HS}(5p)=153$~Mb) and experiment ($\sigma_{max}^{exp}%
(5p)\sim30$~Mb) is even more impressive in Xe \citep{starace70}.

The spectral shape of the $\sigma^{HF}(3p)$ computed in the HF approach is
much closer to the observed $\sigma^{exp}(3p)$. However, it is different for
the length $\sigma_{L}^{HF}(3p)$ and velocity $\sigma_{V}^{HF}(3p)$\ gauges.
\cite{starace70} used HS AOs as a basis set, the Hamiltonian%

\begin{equation}
V=-\sum_{i}\left(  Z/r_{i}+V_{HS}(r_{i})\right)  +\sum_{i>j}1/r_{ij}
\label{eq:V_HF-HS}%
\end{equation}
as a perturbation, and applied the K-matrix technique to calculate $\sigma
_{L}(3p)$ and $\sigma_{V}(3p)$. Using perturbation (\ref{eq:V_HF-HS}) {includes}
the single-electron excitations of the Brillouin type ($\Delta\ell=0$, e.g.
$3p\varepsilon\ell-3p\varepsilon^{\prime}\ell$) and results in cross
sections $\sigma_{L}^{mod}(3p)$ and $\sigma_{V}^{mod}(3p)$ which are very
close to $\sigma_{L}^{HF}(3p)$ and $\sigma_{V}^{HF}(3p)$, respectively
(see Fig.~\ref{fig:3pStarace70}). Although the HS AOs {yield}
similar results {as calculations with the} HF AOs, using the HF AOs is {advantageous} because it reduces the {calculatory work} of
the cumbersome K-matrix technique. Therefore, starting from the seventies,
theoreticians working in atomic physics prefer to use the HF (or DF) AOs as a
basis set or even as a final step. The large-scale calculation of
\cite{kennedy72} performed in the HF approach has become the next `benchmark' in
the calculations of atomic cross sections replacing the previous ones of
\cite{manson68}. The paper of \cite{kennedy72} contains also the calculation
of the parameters of angular distribution of photoelectrons.

Intensive calculations of the influence of many-electron correlations on
the atomic photoionization have been started after the {seminal} work of \cite{amusia70} who took into account intrashell correlations {when} computing $\sigma_{3p}$ of Ar
(the largest contribution in the Ar $\sigma_{3p}$ case stems from the
interference between the $3p^{6}\dashrightarrow3p^{5}\varepsilon d$ and
$3p^{6}\rightarrow3p^{4}\varepsilon^{\prime}d\ \varepsilon d\dashrightarrow
3p^{5}\varepsilon d$ channels, see also scheme~(\ref{eq:Ar-scheme})).
\cite{amusia70} have extended the RPA approach of \cite{altick64} by using the
HF AOs computed with exchange which is called the RPAE approach. In the paper
of \cite{amusia70}, it has been shown that intrashell correlations result in
practical coincidence between $\sigma_{L}^{RPAE}(3p)$ and $\sigma_{V}%
^{RPAE}(3p)$ of Ar and in good agreement between theory and experiment (see
Fig.~\ref{fig:3pKelly73}). Later on, the RPAE approach has been applied to the
description of the valence shell photoionization of Kr and Xe
\citep{amusia71}. The next step in understanding the valence shell
photoionization has been made by \cite{kelly73} who included in the
calculation of $\sigma_{3p}$ of Ar, in addition to intrashell correlations,
the $3s^{1}3p^{6}np$ resonances explaining their `window' type observed in
experiment \citep{samson63,samson66,madden69}.

\begin{figure}[ptb]
\begin{minipage}[t]{0.48\textwidth}
\begin{center}
\includegraphics[width=\textwidth]{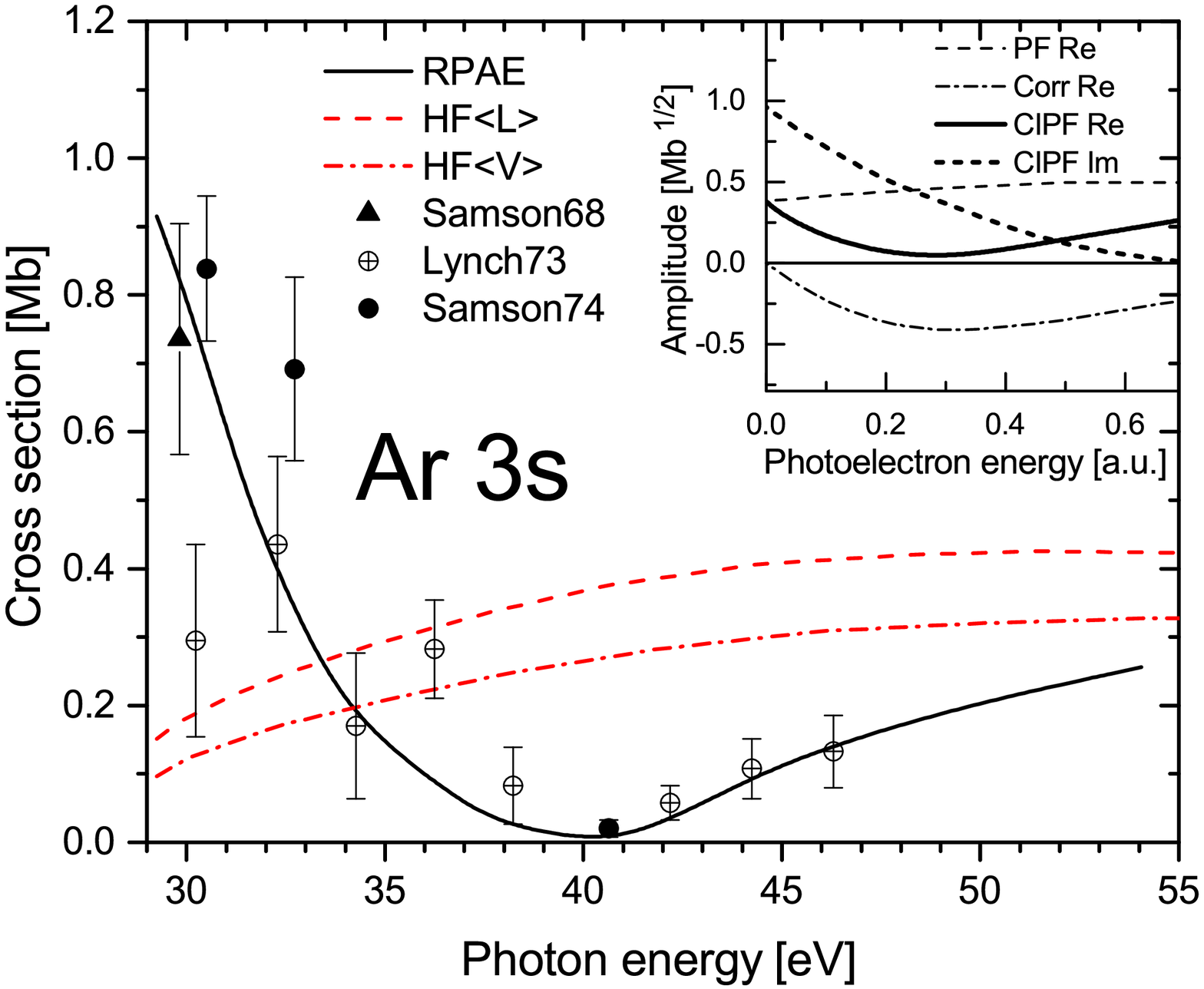}
\caption{Comparison between cross sections $\sigma_{3s}$ of Ar computed in HF and RPAE approaches by \cite{amusia74} with experimental data \citep{samson68,lynch73,samson74}. Insert shows the partial terms of the transition amplitude computed in CIPF approach by
\cite{lagutin99}.}\label{fig:RPAE_3s_Ar}\end{center}
\end{minipage} \hfill\begin{minipage}[t]{0.48\textwidth}
\begin{center}
\includegraphics[width=\textwidth]{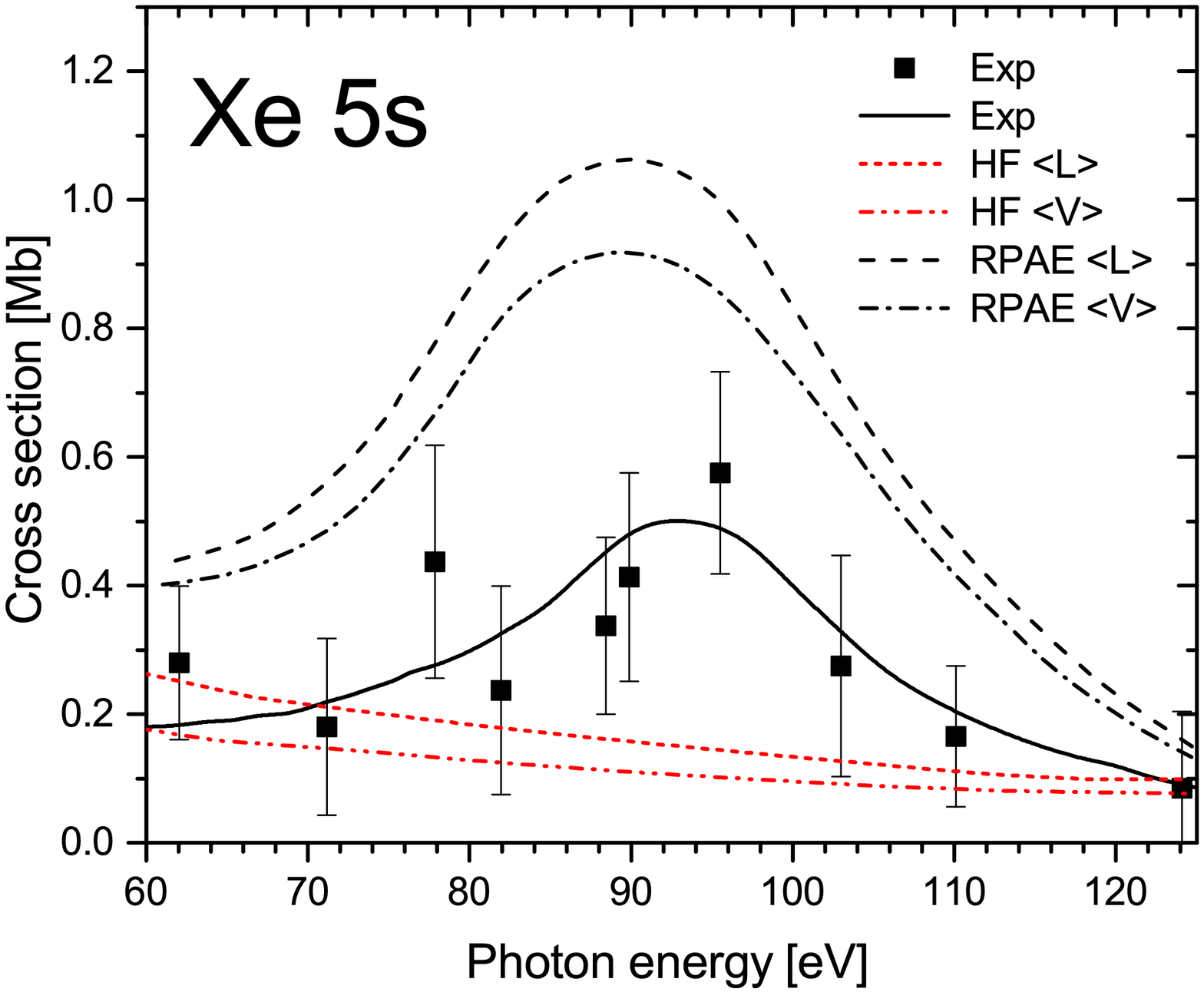}
\caption{Comparison between experimental \citep{west76a} and theoretical
\citep{amusia74} $\sigma_{5s}$ photoionization cross sections. Calculations are performed in HF and RPAE approaches,
respectively.}\label{fig:RPAE_5s_Xe}\end{center}
\end{minipage}
\end{figure}

In a pioneering work, \cite{amusia72} found that the large admixture of the
intershell transition $3s^{2}3p^{6}\dashrightarrow3s^{2}3p^{5}\varepsilon
d\rightarrow3s^{1}3p^{6}\varepsilon p$ to the main $3s^{2}3p^{6}%
\dashrightarrow3s^{1}3p^{6}\varepsilon p$\ photoionization channel totally
changes the HF $\sigma_{3s}$\ cross sections in Ar (see
Fig.~\ref{fig:RPAE_3s_Ar}). The theoretical prediction of \cite{amusia72} has
stimulated numerous experimental efforts. At first, \cite{lynch73} measured
$\sigma_{3s}$\ of Ar and confirmed the prediction, also recalling the earlier
measurement of \cite{samson68} which got an explanation after four years (!)
(see solid triangle in Fig.~\ref{fig:RPAE_3s_Ar}). Secondly, \cite{samson74}
have measured photoionization cross sections of the subvalence shells
$\sigma_{ns}$ of Ar, Kr and Xe {using} the photoelectron spectrometry and revealing a similar behaviour of $\sigma_{ns}$ for all heavy \emph{Rg} (for
Ar, typical experimental results are shown in Fig.~\ref{fig:RPAE_3s_Ar}).

The minimum in $\sigma_{3s}$ looks similar to that in $\sigma_{3p}$ (see
Figs.~\ref{fig:ABS_Samson66}, \ref{fig:ABS_Samson02}) and, therefore, is
usually called `Cooper minimum' in literature. However, we emphasize that the
nature of these two minima is different: the minimum in $\sigma_{3p}$
essentially stems from the `sign-reversal' behaviour of the transition
amplitude $3p\dashrightarrow\varepsilon d$, whereas the minimum in
$\sigma_{3s}$ stems from the interference between the real parts of the direct
(PF Re) and correlational (Corr Re) amplitudes. Both these amplitudes are not
`sign-reversal' as well as the result of their interference (CIPF Re). One can
recognize this from the insert of Fig.~\ref{fig:RPAE_3s_Ar} where the partial
terms of the $3s\dashrightarrow\varepsilon p$ transition amplitude from
\citep{lagutin99} are depicted. An additional rise of the $\sigma_{3s}$
photoionization cross section in the threshold region stems from the imaginary
part of the transition amplitude also shown in the insert of
Fig.~\ref{fig:RPAE_3s_Ar}. We note that in the case of Kr, similar complex
interference results in `sign-reversal' behaviour of the real part of the
$4s\dashrightarrow\varepsilon p$ amplitude. In Ne, however, the $\sigma_{2s}$
photoionization cross section does not exhibit a minimum \citep{amusia72,amusia74}.

Since the revealing of the complex behaviour of $\sigma_{ns}$, theoretical and
experimental investigations of the \emph{Rg} photoionization have been
split in three directions: (i) total and partial photoionization cross
sections; (ii) angular distribution of photoelectrons; and (iii) main line and
satellites resonance structure.

\subsection{Total and partial PICSs}

\label{sec:PICS}

\begin{figure}[ptb]
\begin{minipage}[t]{0.49\textwidth}
\begin{center}
\includegraphics[width=\textwidth]{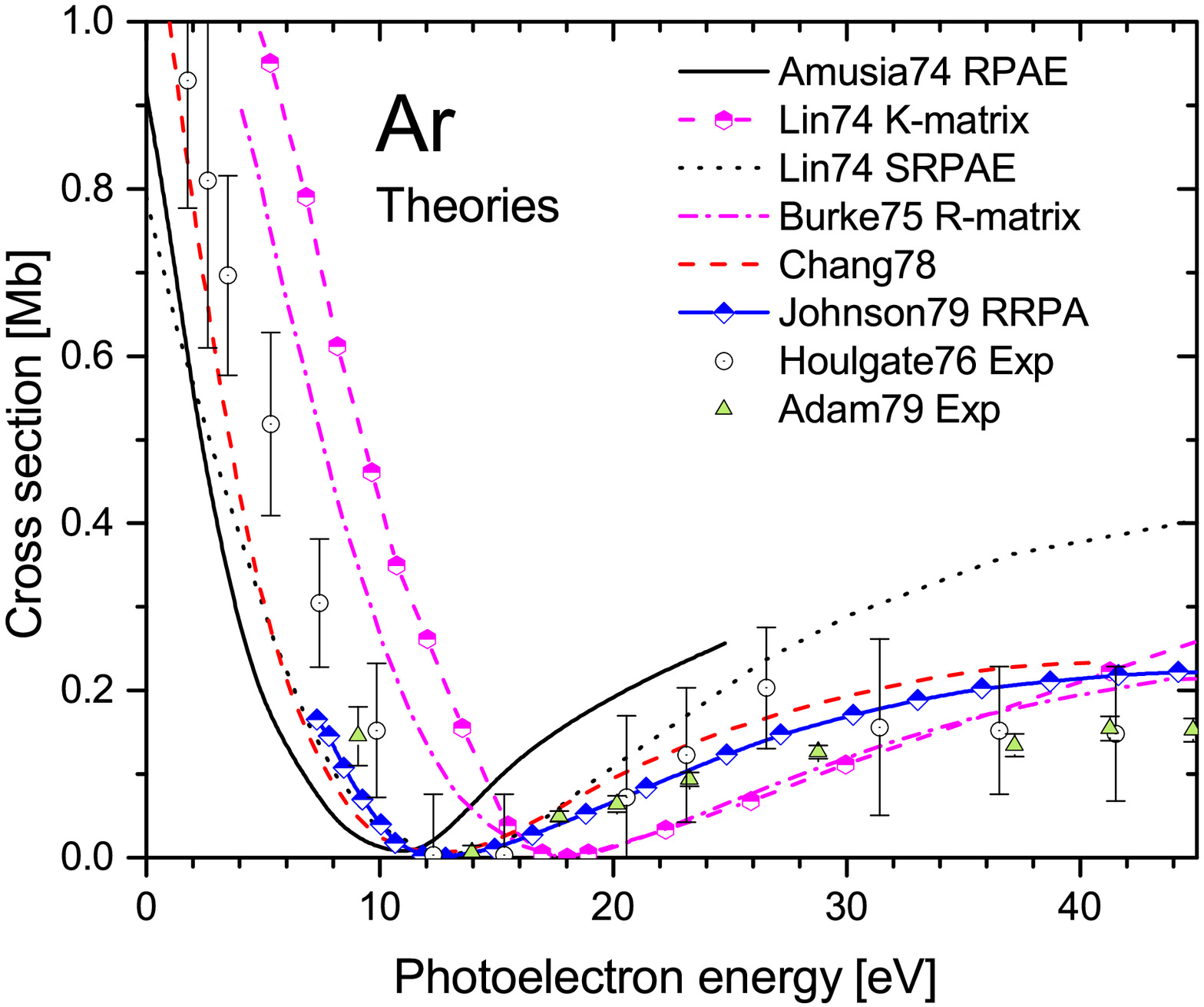}
\caption{$3s$ photoionization cross sections of Ar computed in different many-electron approaches \citep{amusia72,lin74,burke75,chang78}. Experimental data of \cite{houlgate76,adam79} are shown for
comparison.}\label{fig:Ar3sPICS_Theor}\end{center}
\end{minipage} \hfill\begin{minipage}[t]{0.49\textwidth}
\begin{center}
\includegraphics[width=\textwidth]{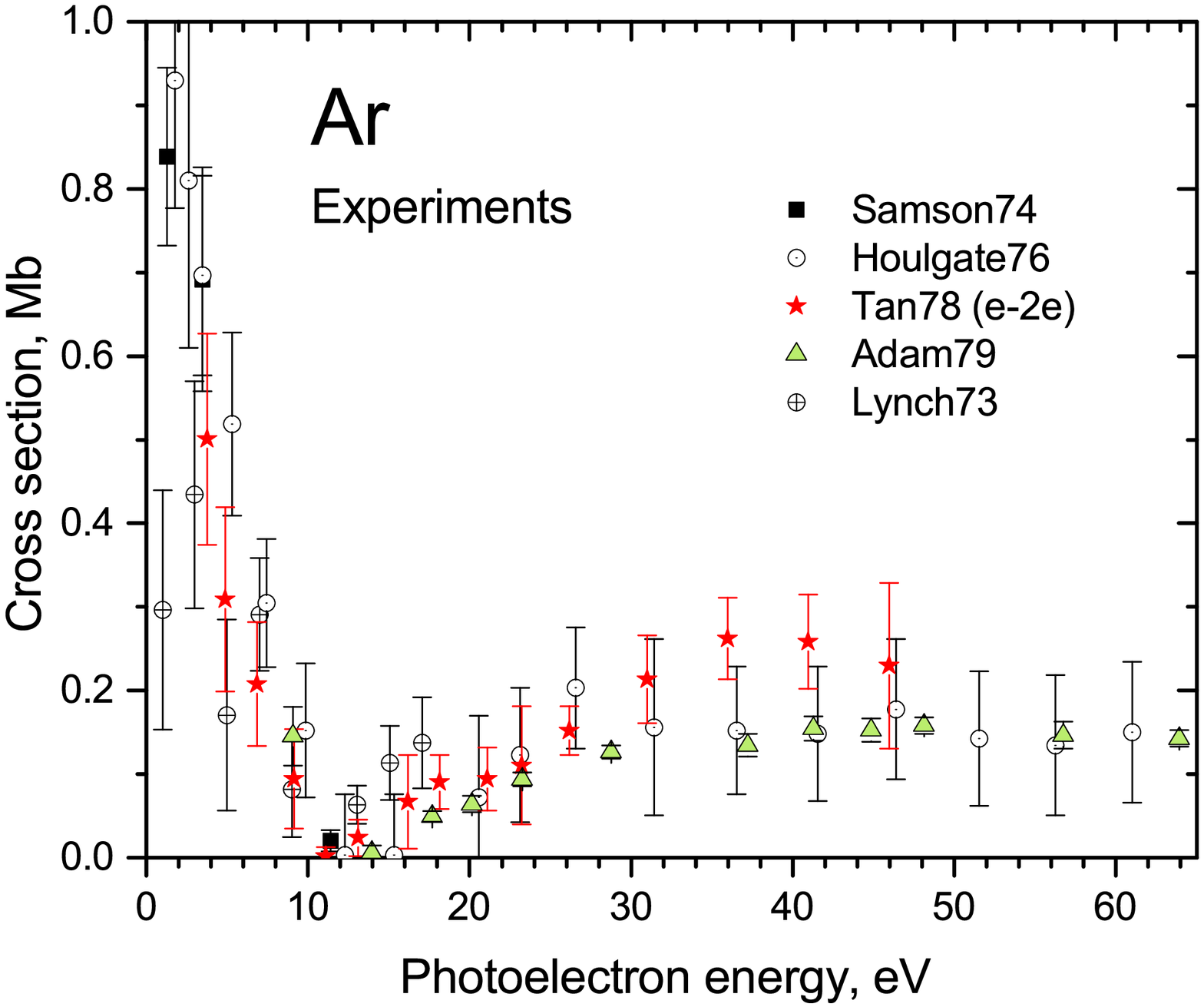}
\caption{$3s$ photoionization cross sections of Ar measured by different authors
\citep{samson74,houlgate76,tan78,adam79,lynch73}.}\label{fig:Ar3sPICS_Exp}%
\end{center}
\end{minipage}
\end{figure}

The strong impact of intershell correlations on the photoionization of `weak'
shells revealed by \cite{amusia72} has been summarized in
\citep{amusia74,amusia75}. Since then, numerous theoretical and experimental
papers devoted both to the creation of new computational methods and to
measuring the absolute cross sections of the `weak shells' have appeared.
\cite{lin74} applied the K-matrix technique {to this problem} and developed the Simplified RPAE technique (SRPAE) which is similar to (\cite{amusia72});
\cite{burke75} created the R-matrix technique; \cite{chang76a,chang76} applied
the transition matrices technique. A relativistic version of the random phase
approach (RRPA) created by \cite{johnson79} allowed them to get adequate
results not only for Ne and Ar but also for the heavier \emph{Rg} Kr and Xe. Their
calculation performed for the Xe $\sigma_{5s}$ confirmed results of
\cite{amusia74} and \cite{amusia75} concerning strong intershell correlations
between $5s\dashrightarrow\varepsilon p$\ and $4d\dashrightarrow\varepsilon
f$\ transitions, which {explained} the maximum in the Xe $\sigma_{5s}(\omega)$ at $\omega\sim93$~eV connected with the giant resonance in the
$4d\dashrightarrow\varepsilon f$\ channel. Measurements performed by
\cite{west76a} confirmed the prediction of \cite{amusia74} qualitatively,
resulting, however, in absolute cross sections almost twice less than the
computed ones (see Fig.~\ref{fig:RPAE_5s_Xe}). In
Fig.~\ref{fig:Ar3sPICS_Theor}, we compare cross sections computed {by different authors} for Ar because in this case relativistic effects are assumed to be
small. One can see that all calculations lead to qualitatively similar
results. However, sometimes substantial quantitative differences remain,
especially for R-matrix and K-matrix results.

Existing measurements of the {\emph{Rg} subvalence $ns$ photoionization cross sections $\sigma_{ns}(\omega)$}
\citep{samson68,lynch73,houlgate74,samson74} have been extended by
\cite{houlgate76,tan78,adam79} for Ar and by \cite{gustafsson77} for Xe. Some
years later \cite{aksela87} {remeasured} the $\sigma_{4s}$
for Kr. Extended measurements have been concentrated on the Cooper minimum
range. {The results showed} smooth curves above the Cooper minimum and exhibited a substantial spread below it {(Fig.~\ref{fig:Ar3sPICS_Exp})}. The XPS data of \cite{houlgate76} and (e, 2e) data of \cite{tan78} are consistent with each other within the error bars, however being about twice larger than the XPS cross sections of \cite{adam79}.

\begin{figure}[tb]
\begin{center}
\includegraphics[width=0.7\textwidth]{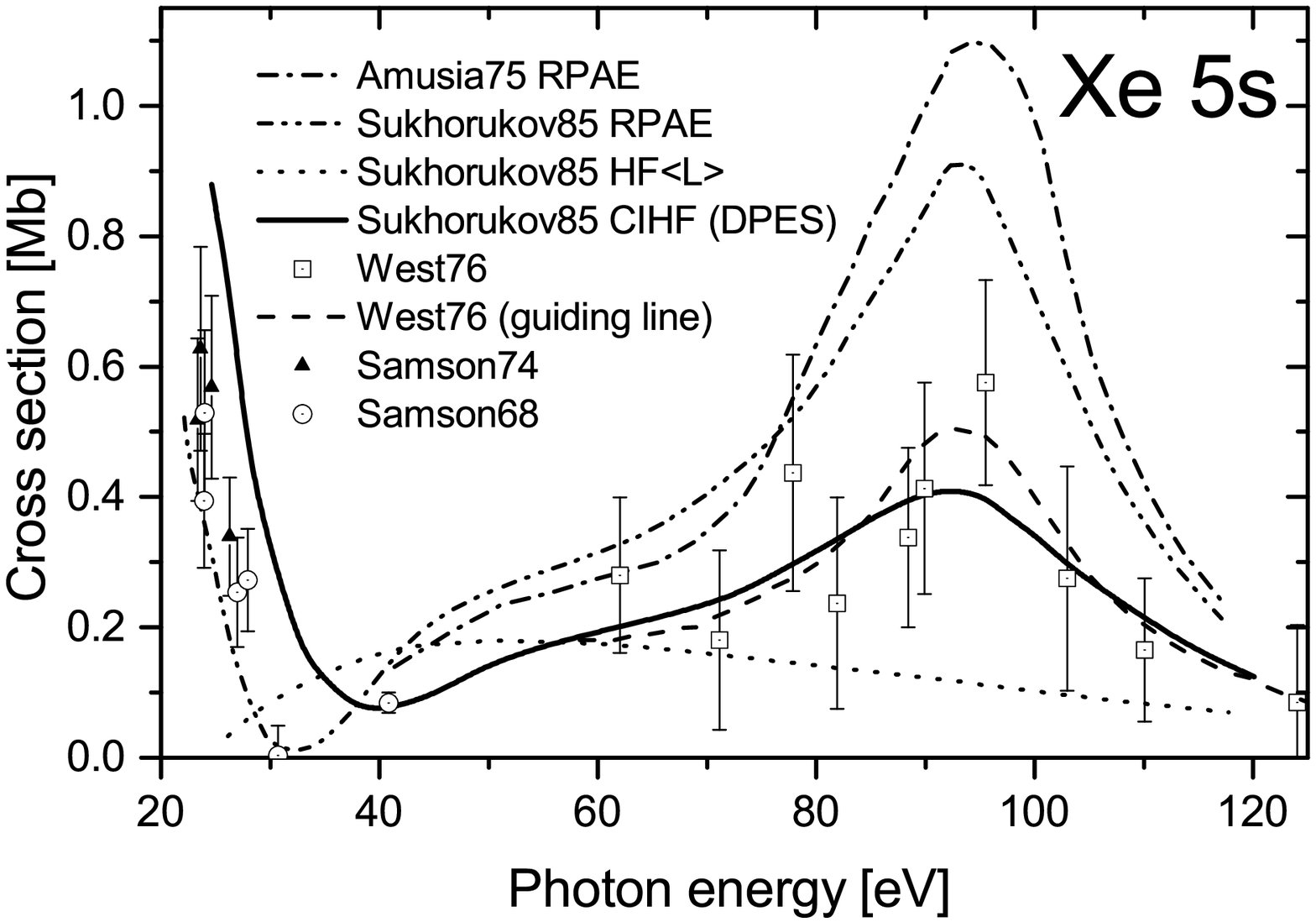}
\end{center}
\caption{Comparison between the {$5s$ photoionization cross sections of Xe $\sigma_{5s}$} computed in different approximations \citep{amusia75,sukhorukov85a} and measured by different
authors \citep{samson68,samson74,west76a}. The figure illustrates the
influence of DPES on the $\sigma_{5s}$. Adapted from \citep{sukhorukov85a}.}%
\label{fig:RPAE_5s_Xe_CIHF}%
\end{figure}

The calculations of \cite{amusia72,lin74,amusia75,burke75,chang78,johnson79a}
qualitatively agreed with the existing measurements but did not explain (i)
the observed spread of the experimental data near threshold and (ii) why the
computed cross sections above the Cooper minimum were larger than the measured ones.

The reason of the latter disagreement has been investigated by
\cite{sukhorukov85a} who used the configuration-interaction HF approximation
(CIHF) and computed the photoionization cross section $\sigma_{5s}(\omega)$
for Xe. They took into account that the main $5s$ level of Xe has a complex
structure due to the large influence of correlation described by the
$5p5p-5snd$ excitations, sometimes called dipole polarization of electron
shells (DPES, see section~\ref{sec:CorrSat} for details). As a result, the
wave function of the main $5s$ level $\left\vert 5s~^{2}S\right\rangle $ becomes%

\begin{equation}
\left\vert 5s\right\rangle =0.680\left\vert 5s^{1}5p^{6}~^{2}S\right\rangle
+0.666\left\vert 5s^{2}5p^{4}5d~^{2}S\right\rangle +... \label{eq:5sWF-Xe}%
\end{equation}

Intershell correlation (\ref{eq:5sXe-abs}) in the Xe case is described by the
interference between the main (\ref{eq:5sXe-abs}a) and the intershell
(\ref{eq:5sXe-abs}b)\ photoionization channels%

\begin{equation}%
\begin{array}
[c]{clc}%
5s^{2}5p^{6}\dashrightarrow5s\ \varepsilon p & 5s-direct & (a)\\
5s^{2}5p^{6}\dashrightarrow5s^{2}5p^{5}\varepsilon^{\prime}d\rightarrow
5s\ \varepsilon p & 5p-intershell & (b)\\
5s^{2}5p^{6}\rightarrow5s^{2}5p^{4}5d\ \varepsilon^{\prime}d\dashrightarrow
5s\ \varepsilon p & 5p-ISCI & (c)
\end{array}
\label{eq:5sXe-abs}%
\end{equation}

As a result, the direct channel (\ref{eq:5sXe-abs}a) is reduced by the
spectroscopic factor of $0.462=0.680^{2}$ because the electric dipole
transition $5s^{2}5p^{6}\dashrightarrow5s^{2}5p^{4}5d$ is forbidden. The
intershell channel (\ref{eq:5sXe-abs}b) remains practically unchanged because
both Coulomb transitions $5s^{2}5p^{5}\varepsilon^{\prime}d\rightarrow
5s^{1}5p^{6}\varepsilon p$ and $5s^{2}5p^{5}\varepsilon^{\prime}%
d\rightarrow5s^{2}5p^{4}5d\ \varepsilon p$ are allowed. The `$5p-ISCI$'
channel (\ref{eq:5sXe-abs}c) also contributes to the transition amplitude near
the $5s$ threshold appreciably due to the large admixture of the $5s^{2}%
5p^{4}5d$ configuration to the wave function $\left\vert 5s\right\rangle $
(\ref{eq:5sWF-Xe}). Therefore, the near-threshold cross section determined
mainly by the intershell channel is changed only a little, whereas the
far-threshold cross section is reduced by about a factor of two. Thus,
simultaneously taking into account the complex structure of the main $5s$
level due to DPES and intershell correlations provided a good overall
agreement between computed and measured photoionization cross section
$\sigma_{5s}(\omega)$ in a wide spectral range (see
Fig.~\ref{fig:RPAE_5s_Xe_CIHF}). Including DPES in the calculation
substantially increased the value of $\sigma_{5s}(\omega)$ at the Cooper
minimum, slightly changing also its position. We remind that the origin of the
maximum near $\omega=93$~eV is connected with the trace of the
$4d\dashrightarrow\varepsilon f$\ giant resonance due to intershell
correlation as obtained by \cite{amusia74,amusia75}.

\subsection{Angular distribution of the $ns\dashrightarrow\varepsilon p$
photoelectrons}

The investigation of the angular distribution of photoelectrons near the
Cooper minima in the {subvalence} $ns\dashrightarrow\varepsilon p$ photoionization of
\emph{Rg}, especially in the $5s\dashrightarrow\varepsilon p$ photoionization
of Xe, received much interest. The reason is that the nonrelativistic value of
the angular distribution parameter $\beta_{5s}(\omega)$ should equal 2,
whereas in relativistic approximation a deviation from 2 can be expected
\citep{walker74}. {Therefore the details of the photoelectron angular distribution close to the Cooper minima are a measure of the relativistic effects in the photoelectron emission.} The expression for the $\beta_{5s}(\omega)$ can be
transformed to%

\begin{equation}
\beta_{5s}(\omega)=2-\frac{\left\vert D_{3/2}(\omega)-\sqrt{2}D_{1/2}%
(\omega)\right\vert ^{2}}{\sigma_{5s}(\omega)} \label{eq:5sBET}%
\end{equation}

\begin{equation}
\sigma_{5s}(\omega)=\left\vert D_{1/2}(\omega)\right\vert ^{2}+\left\vert
D_{3/2}(\omega)\right\vert ^{2} \label{eq:5sSIG}%
\end{equation}
where in nonrelativistic approximation $D_{3/2}(\omega)=\sqrt{2}D_{1/2}%
(\omega)$ and $\beta_{5s}(\omega)=2$ (see, e.g., \citep{johnson78,lagutin96}).
When the spin-orbit interaction of the $\varepsilon p$ electron is taken into
account, the $\varepsilon p_{1/2}$ and $\varepsilon p_{3/2}$ AOs slightly
differ from each other \citep{seaton51} and the numerator in equation
(\ref{eq:5sBET}) does not equal zero. Thus, $\beta_{5s}(\omega)$ may
substantially deviate from 2 in the vicinity of the Cooper minimum.
\cite{walker74} had computed the Xe $\beta_{5s}(\omega)$ using the
Dirac-Slater (DS) approximation without taking into account many-electron correlations.

\begin{figure}[tb]
\begin{minipage}[t]{0.48\textwidth}
\begin{center}
\includegraphics[width=\textwidth]{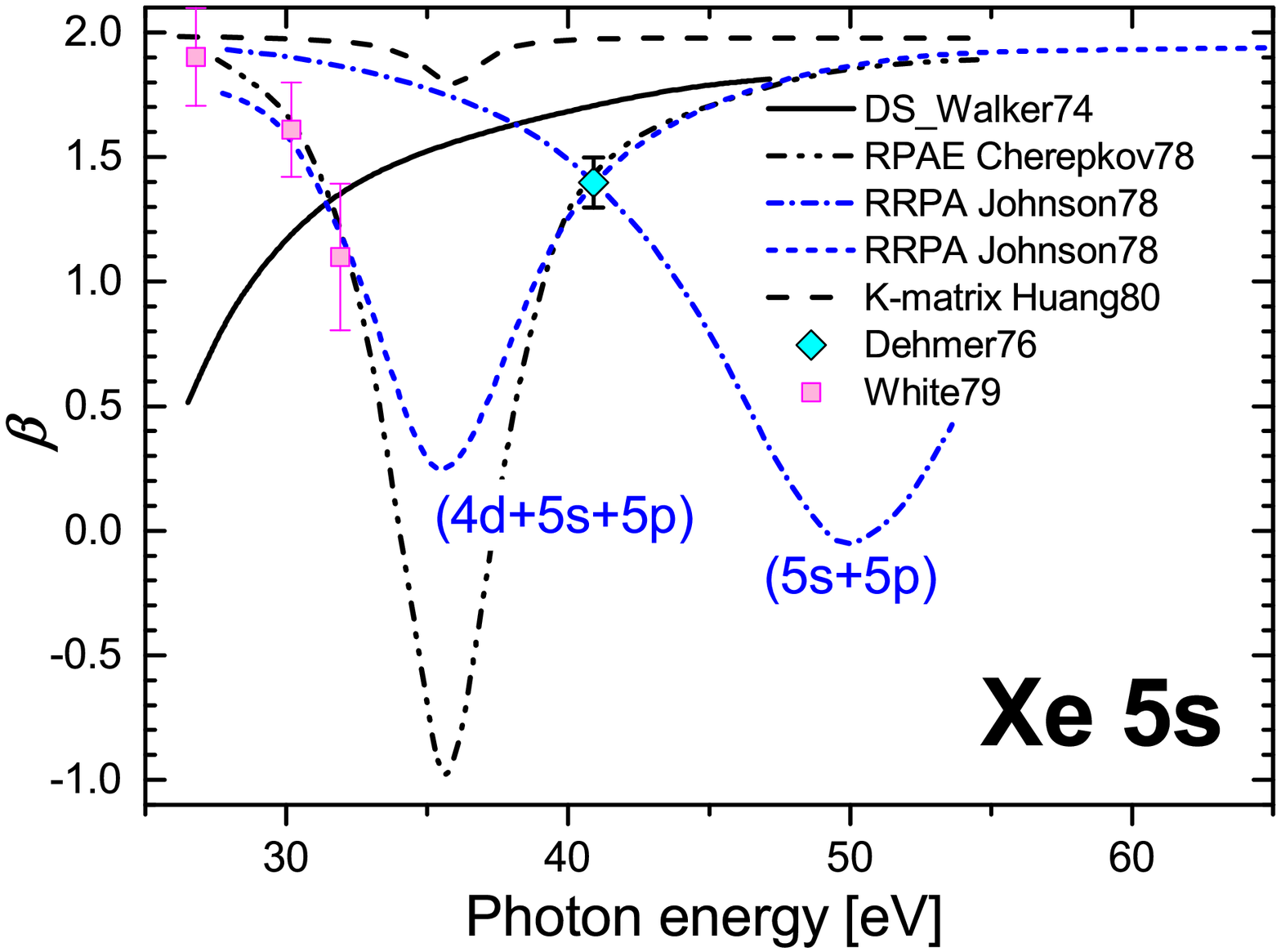}
\caption{Earlier measurements and calculations of the Xe $\beta_{5s}$. Calculations: Dirac-Slater (DS) from \citep{walker74};
parameterized RPAE from \citep{cherepkov78}; RRPA from \citep{johnson78};
K-matrix from \citep{huang80a}. Experiments are from \citep{dehmer76,white79}.
Note that \emph{all} earlier calculations cross the experimental point of
\cite{dehmer76} which later appeared to be wrong (!) (see
Fig.~\ref{fig:5sBEGparp}). Adapted from
\citep{johnson78}.}\label{fig:5sBEGjohn}\end{center}
\end{minipage} \hfill\begin{minipage}[t]{0.47\textwidth}
\begin{center}
\includegraphics[width=\textwidth]{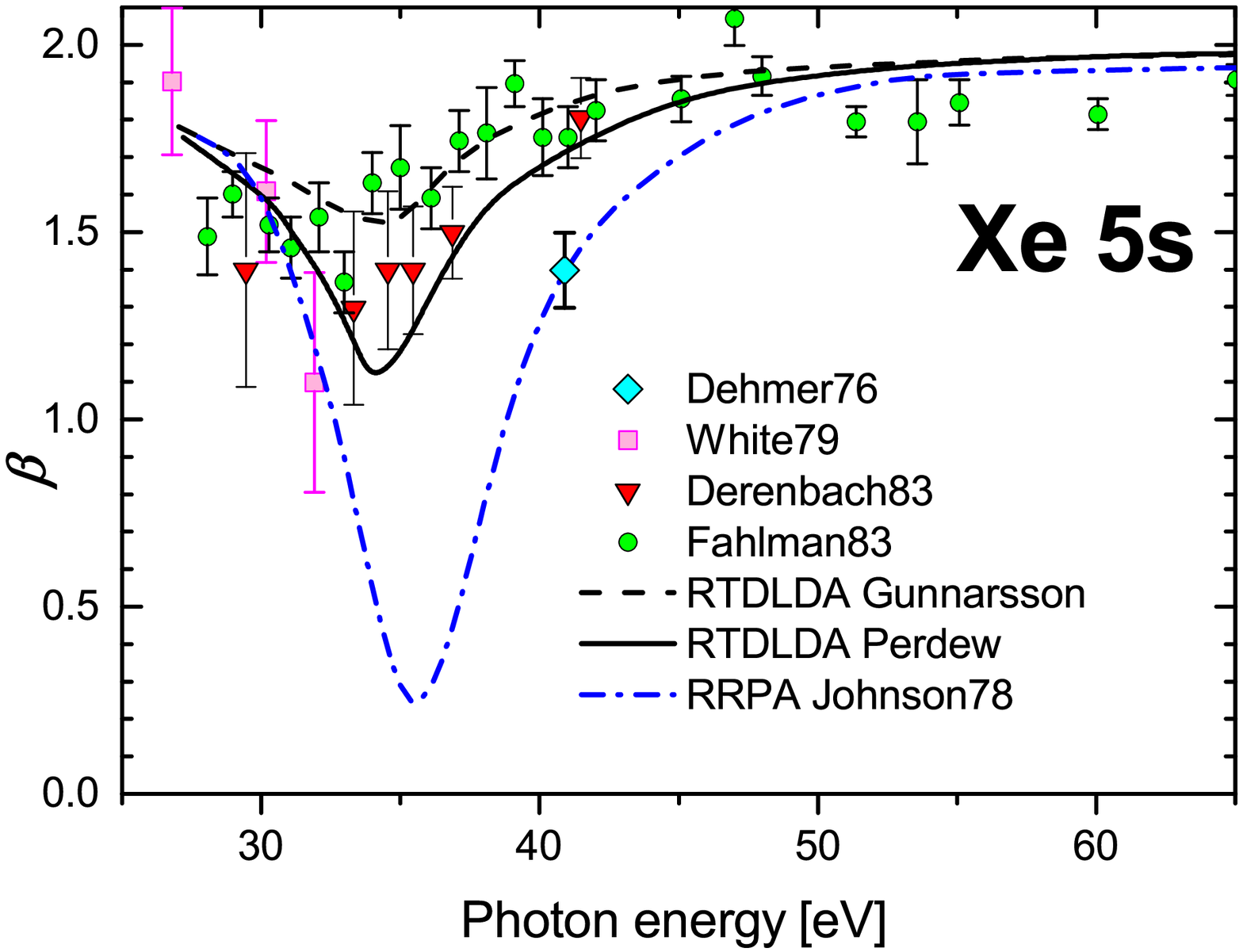}
\caption{Comparison between later measurements of the Xe $\beta_{5s}$
\citep{fahlman83,derenbach83} and calculations performed with the
Gunnarsson-Lundquist and Perdew-Zunger local density potentials. Adapted from
\citep{parpia84}.}\label{fig:5sBEGparp}\end{center}
\end{minipage}
\end{figure}

\cite{dehmer76} have measured $\beta_{5s}(\omega)$ at one point supporting
the prediction of \cite{walker74}. However, the calculation by
\cite{cherepkov78} with taking into account the spin-orbit interaction of the
$\varepsilon p$ electron and intershell correlations showed that at the energy
of the Cooper minimum $\beta_{5s}(\omega)$ should also {display a}
minimum with negative derivative at threshold. This result has been confirmed
by RRPA calculations of \cite{johnson78} and by measurements of \cite{white79}
just above the $5s$ threshold (see Fig.~\ref{fig:5sBEGjohn},
\ref{fig:5sBEGparp}).

In Fig.~\ref{fig:5sBEGjohn}, $\beta_{5s}(\omega)$ computed by \cite{huang80a} {in K-matrix technique with a HF basis including $\epsilon p$ spin-orbit interaction as a perturbation is shown.
This calculation resulted in a shallower minimum} in $\beta_{5s}(\omega)$ than that predicted by \cite{johnson78}. \cite{huang80a} {attributed} this difference to the neglection of the Coulomb interaction between $5s^{1}\varepsilon p_{1/2}$ and
$5s^{1}\varepsilon p_{3/2}$ channels. {This detail illustrates nicely that} the investigation of the angular
distribution of the $5s$ electrons turned out to be a sensitive test of theories.

\begin{figure}[tb]
\begin{minipage}[t]{0.48\textwidth}
\begin{center}
\includegraphics[width=\textwidth]{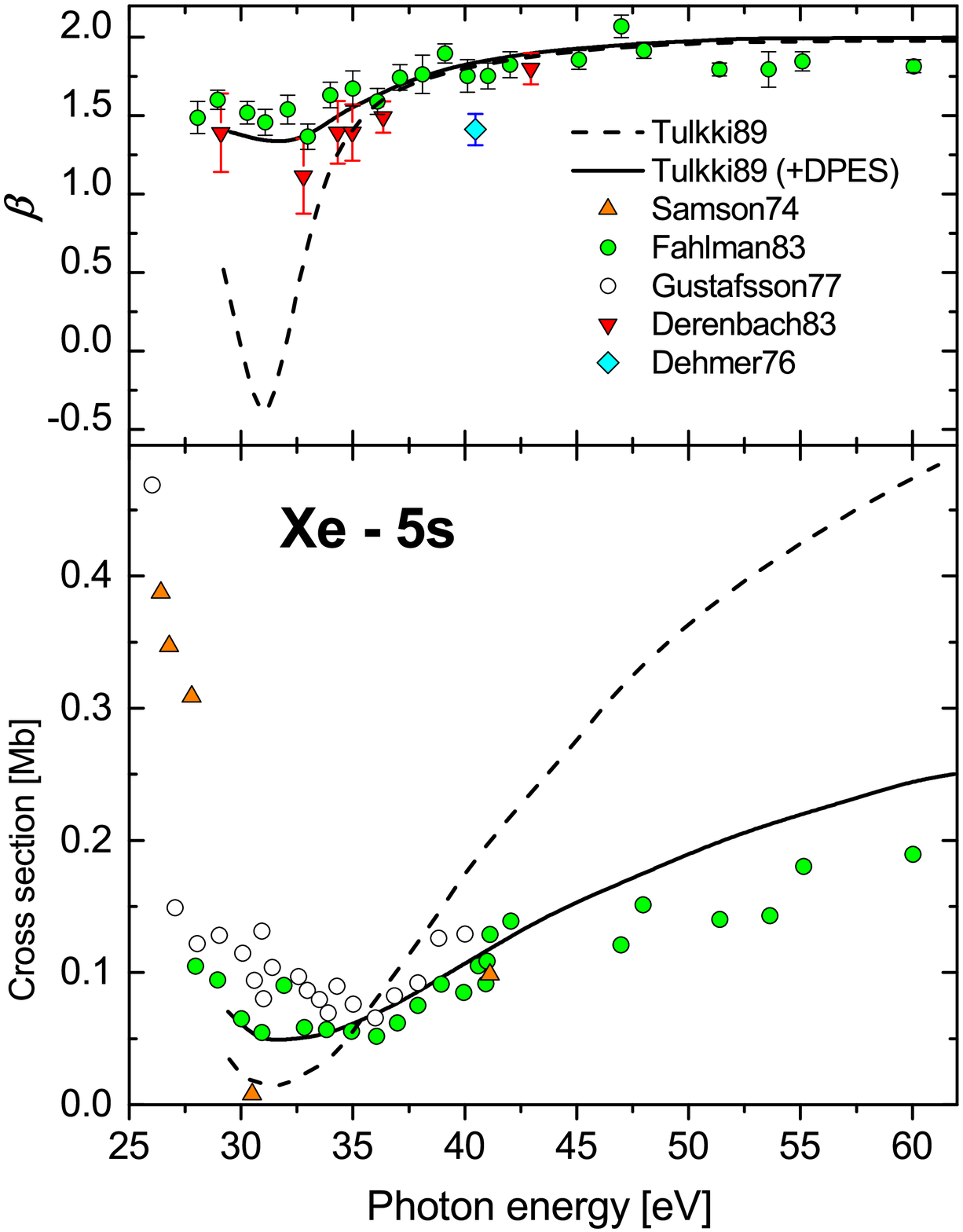}
\caption{Comparison between measured
\citep{samson74,dehmer76,gustafsson77,derenbach83,fahlman83} $\sigma_{5s}$ and $\beta_{5s}$ of Xe and computed by \cite{tulkki89} with (solid line) and without (dashed line) DPES. Adapted from
\citep{tulkki89}.}\label{fig:5sINVtulk}\end{center}
\end{minipage} \hfill\begin{minipage}[t]{0.48\textwidth}
\begin{center}
\includegraphics[width=\textwidth]{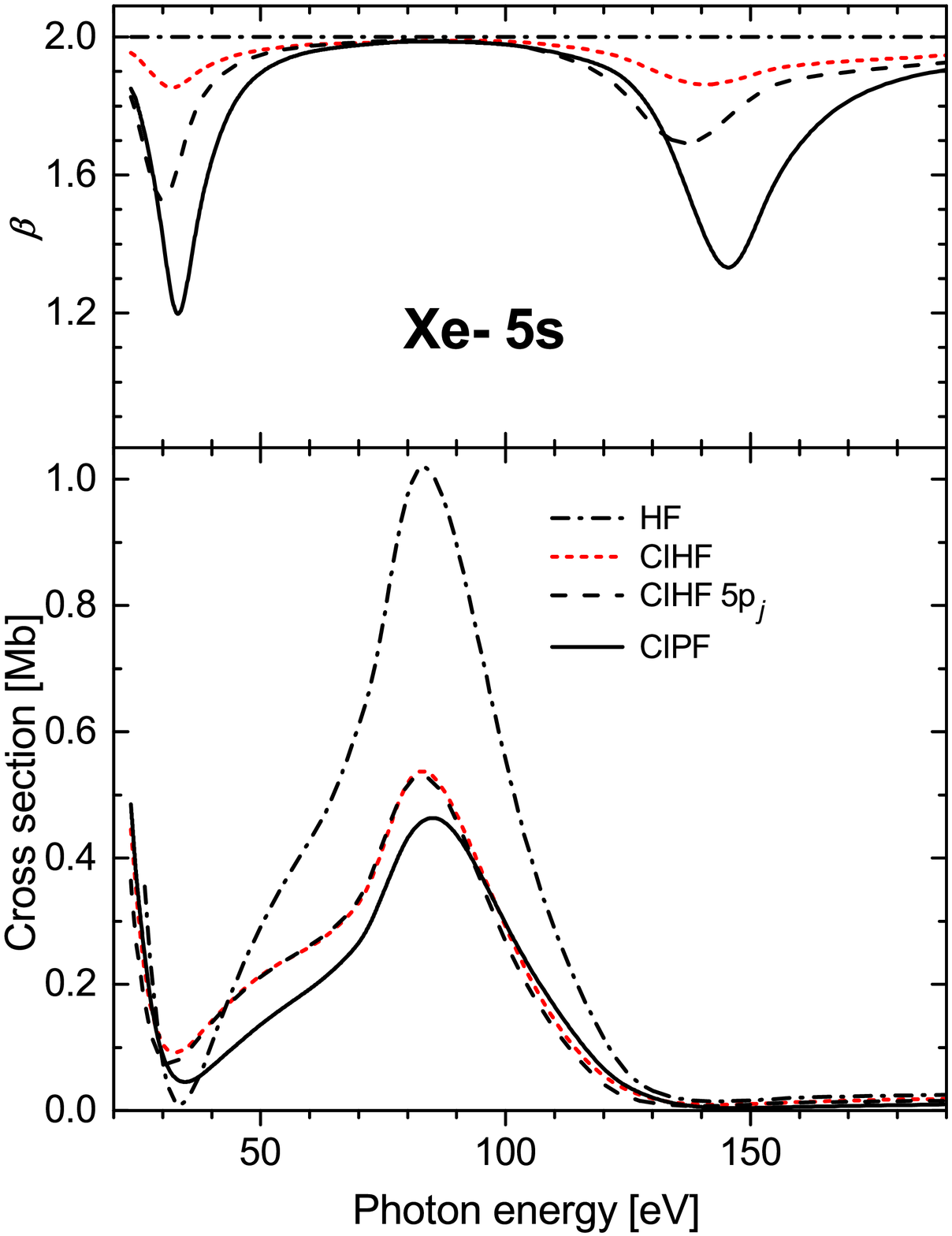}
\caption{$\sigma_{5s}$ and $\beta_{5s}$ of Xe and computed in different approximations.
HF -- nonrelativistic RPAE
calculation with HF AOs; CIHF = HF+DPES+$\varepsilon p_j$; (CIHF $5p_j$) = CIHF+$5p_j$ ; CIPF--all AOs are computed in PF approach. Adapted from
\citep{lagutin98}.}\label{fig:5sINVour}\end{center}
\end{minipage}
\end{figure}

\cite{johnson79a} presented an extended investigation of a sequential inclusion
of interchannel interactions on the $\beta_{ns}(\omega)$ for all $Rg$. In
Fig.~\ref{fig:5sBEGjohn}, one can see that in Xe the additional
$4d\dashrightarrow\varepsilon f$ channel (calculation $4d+5s+5p$%
)\ substantially shifts the minimum in $\beta_{5s}(\omega)$ towards the $5s$
threshold in comparison with the ($5s+5p$) calculation including the
interchannel interaction between $5s\dashrightarrow\varepsilon p$ and
$5s\dashrightarrow\varepsilon\ell$ only. However, both calculations of
\cite{johnson78,johnson79a} and \cite{cherepkov78} agreed with existing
experimental points of \cite{dehmer76} and \cite{white79}, so that a decisive experiment
was required.

Measurements of \cite{fahlman83} and \cite{derenbach83} revealed that the
minimum in the $\beta_{5s}(\omega)$ is {shallower}, indeed (see
Fig.~\ref{fig:5sBEGparp}). Calculations of \cite{parpia84} performed within
the relativistic time-dependent local density approximation RTDLDA and
depicted in Fig.~\ref{fig:5sBEGparp} illustrated that the choice of the
`exchange-correlational' potential results in substantial changes of the
$\beta_{5s}(\omega)$ curve. Their outcome raises the question about the
predictive ability of the LDA approximation if it is used as a final instance.

A detailed investigation of different approximations in computing the
photoionization cross section $\sigma_{5s}(\omega)$\ and angular distribution
of the $5s\dashrightarrow\varepsilon p$\ photoelectrons has been performed by
\cite{tulkki89}. In this paper, \citeauthor{tulkki89} used the relativistic
multichannel multiconfiguration Dirac-Fock (MMCDF) method and took into
account intershell $5s$, $5p$, and $4d$ correlation, relaxation of electron
shells, dipole polarization of the $5s$ main level (DPES) and interchannel
interaction, sequentially. {All together bring experiment and theoretical description close to each other, where particularly the DPES turns out to be important.} In Fig.~\ref{fig:5sINVtulk}, we show the influence
of the DPES correlation on $\sigma_{5s}$ (cf. Fig.~\ref{fig:RPAE_5s_Xe_CIHF})
and on $\beta_{5s}$ of Xe as illustrated by \cite{tulkki89}.

\begin{figure}[tb]
\begin{minipage}[t]{0.48\textwidth}
\begin{center}
\includegraphics[width=\textwidth]{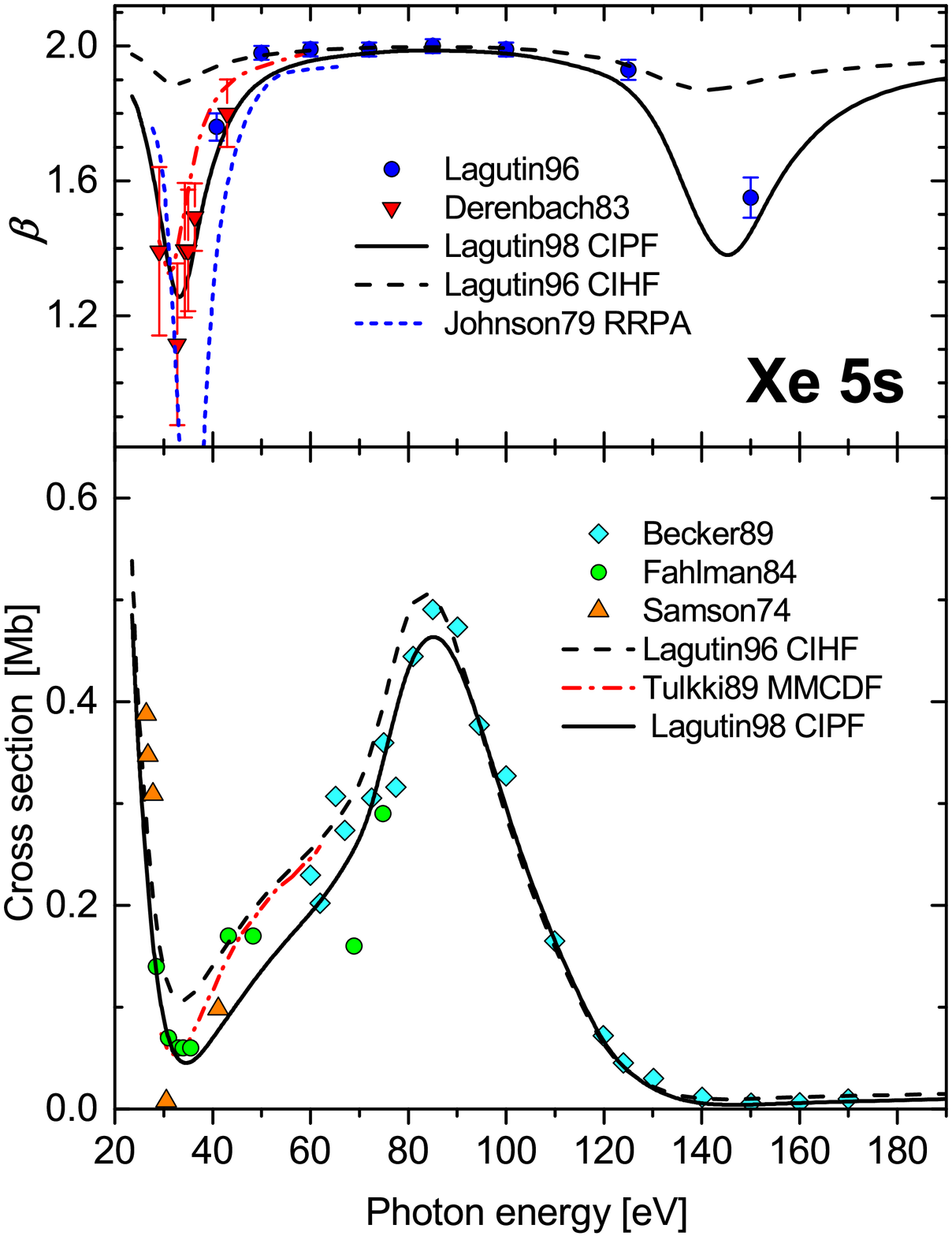}
\caption{Comparison between theory and experiment for $\sigma_{5s}$ and $\beta_{5s}$ of Xe
Experimental data from
\citep{samson74,derenbach83,fahlman84,becker89,lagutin96}; calculations from
\citep{tulkki89,lagutin96,lagutin98}.}\label{fig:5sRESxe}\end{center}
\end{minipage} \hfill\begin{minipage}[t]{0.48\textwidth}
\begin{center}
\includegraphics[width=\textwidth]{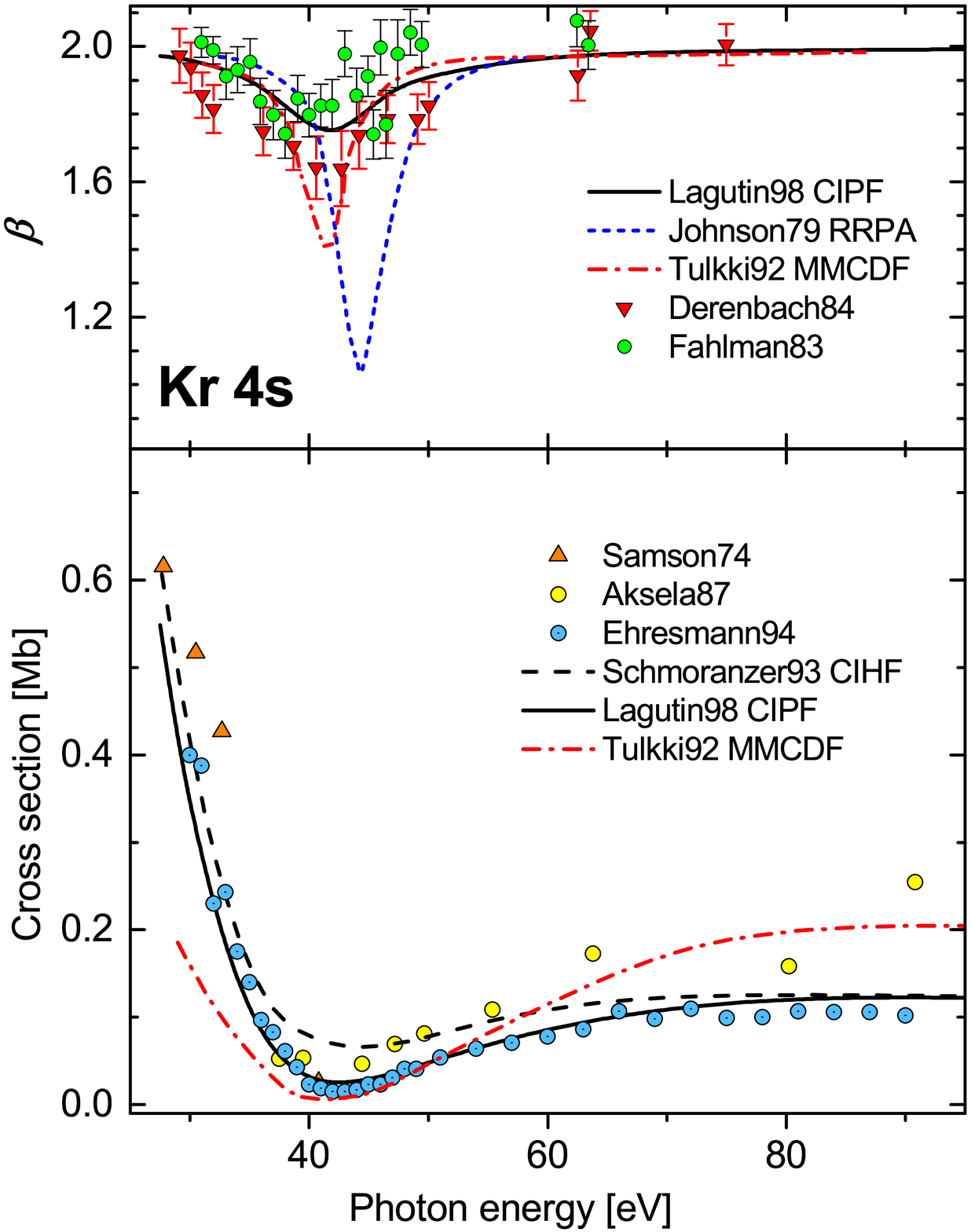}
\caption{Comparison between theory and experiment for $\sigma_{4s}$ and $\beta_{4s}$ of Kr. Adapted from \citep{lagutin98}.
Experimental data from
\citep{samson74,fahlman83a,derenbach84,aksela87,ehresmann94a}; calculations
from
\citep{johnson79a,tulkki92,schmoranzer93,lagutin98}.}\label{fig:5sRESkr}%
\end{center}
\end{minipage}
\end{figure}

\cite{lagutin98} have investigated the Xe $\beta_{5s}(\omega)$ and the Kr
$\beta_{4s}(\omega)$ including different interactions step by step {(Fig.~\ref{fig:5sINVour})}.
The Hartree-Fock approach with intrashell and intershell correlations, denoted
for brevity as HF, results in $\beta_{5s}(\omega)=2$ in the whole energy
region although the cross section $\sigma_{5s}(\omega)$ exhibits two Cooper
minima (at $\omega=33$ and 141~eV). Taking into account DPES and spin-orbit
interaction of the continuum $\varepsilon p_{j}$ AO results in the appearance
of two minima in $\beta_{5s}(\omega)$ corresponding to the minima in the
$\sigma_{5s}(\omega)$ (CIHF in Fig.~\ref{fig:5sINVour}) being, however, too shallow. Those minima are deepened by including the spin-orbit interaction of the $5p$ electron
(CIHF+$5p_{j}$) and even more deepened if all AOs are {considered} in the relativistic configuration-interaction with PF AOs (CIPF) approach (see
section~\ref{sec:theory_AO}). In comparison with the calculation of
\cite{tulkki89}, the CIPF approach of \cite{lagutin98} takes into account
high-order PT corrections by computing the screening of the Coulomb
interaction (see section~\ref{sec:Coulomb_screening}). {This work demonstrated} that the PF approximation {adequately describes}
relativistic effects for atoms with $Z\leq54$.

In Fig.~\ref{fig:5sRESxe}, some selected calculations and experiments for
$\sigma_{5s}(\omega)$ and $\beta_{5s}(\omega)$ in Xe are compared. Good
overall agreement between measured and computed quantities can be stated. {Therefore,} the major mechanisms of the subvalence-shell
photoionization {can be identified}: (i) intrashell and intershell correlations; (ii) dipole polarization of electron shells (DPES); (iii) dependence of the
$5p_{i}$ and $\varepsilon p_{j}$ AOs on the spin-orbit interaction. The latter
effect appeared to be more pronounced in the $\beta_{5s}(\omega)$ than in the
$\sigma_{5s}(\omega)$.

The $4s$ photoionization of Kr has been studied less than the $5s$ one of Xe. The angular distribution of the $4s\dashrightarrow\varepsilon p$ photoelectrons has been predicted by
\cite{johnson79a} in RRPA approach with the same disadvantage as for Xe: the neglection of the interchannel interaction between $4s^{1}\varepsilon p_{1/2}$\ and $4s^{1}\varepsilon p_{3/2}$ channels and the neglection of DPES. In the RTDLDA calculation of \cite{parpia84}, the DPES was
not taken into account, too. Consequently, the calculation of
\cite{johnson79a} predicted too small cross sections $\sigma_{4s}(\omega)$ {for energies close to}
the Cooper minimum resulting in too deep a minimum in the $\beta_{4s}(\omega)$
(see Fig.~\ref{fig:5sRESkr}), according to equation~(\ref{eq:5sBET})
containing $\sigma_{4s}(\omega)$ in the denominator. The Kr $\beta_{4s}%
(\omega)$ measured by \cite{fahlman83} and \cite{derenbach84} lie above the
prediction of \cite{johnson79a}. The MMCDF calculation of \cite{tulkki92} in
the case of $\beta_{4s}(\omega)$ is in fairly good overall agreement with the
measurements of \cite{fahlman83} and \cite{derenbach84}. In the near-threshold
region, the $\sigma_{4s}(\omega)$ of \cite{tulkki92} is substantially lower
than the cross sections measured by \cite{samson74} and by \cite{aksela87},
although in the region above the Cooper minimum one can see fairly good
agreement between the data of \cite{tulkki92} and \cite{aksela87}. The
$\sigma_{4s}(\omega)$ calculation performed by \citep{schmoranzer93} in CIHF
approximation agrees with the experimental data of \cite{samson74} better than
that of \cite{tulkki92}, but is a little too low. At high energies, too, this
calculation resulted in data which are lower than the experimental data of
\cite{aksela87}.

The substantial spreading of the experimental and theoretical cross sections
$\sigma_{4s}(\omega)$ asked for new precise experiments. Such an experiment
has been performed applying the PIFS technique by \cite{ehresmann94a} with
small energy steps. The results of this measurement are depicted in the lower
panel of Fig.~\ref{fig:5sRESkr}. This measurement has stimulated new
calculations \citep{lagutin98} where all correlations discussed above for Xe
were taken into account using relativistic PF AOs. Fig.~\ref{fig:5sRESkr}
demonstrates good overall agreement between the new theory \citep{lagutin98}
and experiments for both $\sigma_{4s}(\omega)$ \citep{ehresmann94a} and for
$\beta_{4s}(\omega)$ \citep{fahlman83,derenbach84}. $\beta_{4s}(\omega)$ of Kr
in comparison with $\beta_{5s}(\omega)$ of Xe exhibits a more shallow minimum
(cf. Fig.~\ref{fig:5sRESkr} and Fig.~\ref{fig:5sRESxe}) as could be expected
in view of its relativistic nature.

\section{Correlation satellites}

\label{sec:CorrSat}

Studying the fluorescence of Ar~II in a broad wavelength interval (12500--2000
\AA ), \cite{minnhagen63} revealed a strong mixing between the $3s^{1}3p^{6}$
and $3s^{2}3p^{4}ns/nd$ configurations. This mixing became later the subject
of thorough investigations in numerous papers because it has a strong impact
on the structure of the X-ray and photoelectron spectra connected with the
transitions of the subvalence $ns$ electrons, on the photoionization cross
sections of the $ns$ shells, the angular distribution of the photoelectrons,
the lifetimes of the $ns$ vacancies etc. The effects connected with this
configuration mixing look so impressive that they received different `names'
from different authors. As examples, we list the following names: `semi-Auger
transitions' \citep{cooper70}; `dipolar fluctuations' and `strong dynamical
effects' \citep{wendin76}; `dipole relaxation process' \citep{verkhovtseva80};
`conjugate shake-up' \citep{dyall82a,dyall82b};
`symmetric-exchange-of-symmetry (SEOS) correlations' \citep{beck82};
`super-Coster-Kronig fluctuations' \citep{chen85a}; `dynamic dipolar
relaxation' \citep{yarzhemsky92}; `dynamic dipole polarization of electron
shells (DPES)' \citep{sukhorukov91}; `particle-hole interaction' effect
\citep{ohno00,ohno00a,ohno01}. In the present paper, we use the term DPES
because it reflects the change of the $3p$ shell orbital {angular} momentum (its polarization) by the $3s$\ inner vacancy and the fact that the main
contribution to the matrix element of the configuration interaction stems from
the dipole part of the Coulomb operator.

\subsection{$L_{2,3}$ fluorescence}

Studying the correlation satellites in X-ray processes of the rare-gas atoms
has been started by X-ray fluorescence spectroscopy. \cite{cooper70} measured
the $L_{2,3}$ X-ray spectrum of Ar ($3s-2p$ transition) and interpreted the
observed satellites by the excitations following the paper of
\cite{minnhagen63} where broad-range (12500--2000 \AA ) fluorescence had
been documented. \cite{cooper70} estimated {a shake-off probability of 10\% and conjectured}, on this basis, {that}
shake satellites {weakly contribute} to the $L_{2,3}$ fluorescence of Ar.

\begin{figure}[ptb]
\begin{center}
\includegraphics[width=0.65\textwidth]{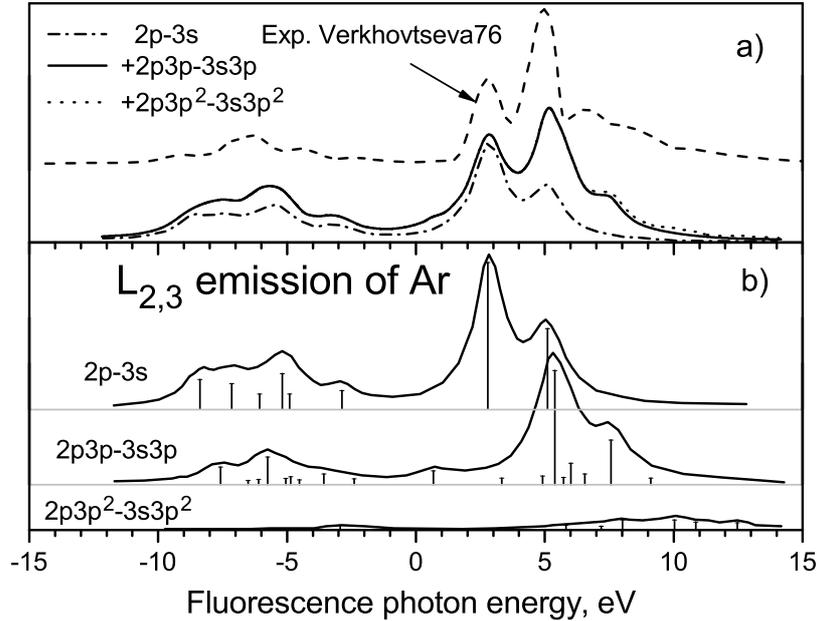}
\end{center}
\caption{a) Comparison between the experimental $L_{2,3}$ X-ray spectrum of Ar
\citep{verkhovtseva76} and spectra computed by sequential addition of
fluorescence with different numbers of vacancies in the $3p$ shell. b) Partial
$L_{2,3}$ emission. Adapted from \citep{sukhorukov85}.}%
\label{fig:Sukhorukov85_ArL23}%
\end{figure}

However, \cite{chen74} found that the radiationless lifetimes for the Ar~III
$2p^{5}3p^{5}$ terms $^{3}S,~^{1}P,~^{3}D$ are by two orders of magnitude
larger than for the $^{1}S,~^{3}P,~^{1}D$ terms. As a result the fluorescence
yield in $^{3}S,~^{1}P,~^{3}D$ terms is anomalously increased and the shake
satellites are reinforced by an order of magnitude, making their intensity
comparable to the intensity of the main $2p-3s$ transitions. The $L_{2,3}$
spectrum of Ar measured at higher resolution by \cite{werme72a} exhibited
additional spectral lines and even more spectral lines have been observed by
\cite{verkhovtseva76}. In particular, the `$2p_{1/2}-3s$' component appeared
to be more intense than the `$2p_{3/2}-3s$' component. This fact was explained
when the shake satellites where taken into account by \cite{karazija79} and by
\cite{sukhorukov85} (see Fig.~\ref{fig:Sukhorukov85_ArL23}). One can see that
the intensity of the `satellite' transitions $2p3p\dashrightarrow3s3p$
representing 77\% of the `main' transition $2p-3s$ contributes strongly to the
position of the `$2p_{1/2}-3s$' component making it `more intense' than the
`$2p_{3/2}-3s$' component. In case the exciting-photon energy will be small
enough to excite an additional $3p$ electron, one can expect drastic changes
of the $L_{2,3}$ fluorescence. To our knowledge, such an experiment has not
been performed so far.

\subsection{Photoelectron spectra (PES) of the rare-gas subvalence $ns$ shells}

\begin{figure}[tb]
\begin{center}
\includegraphics[width=0.7\textwidth]{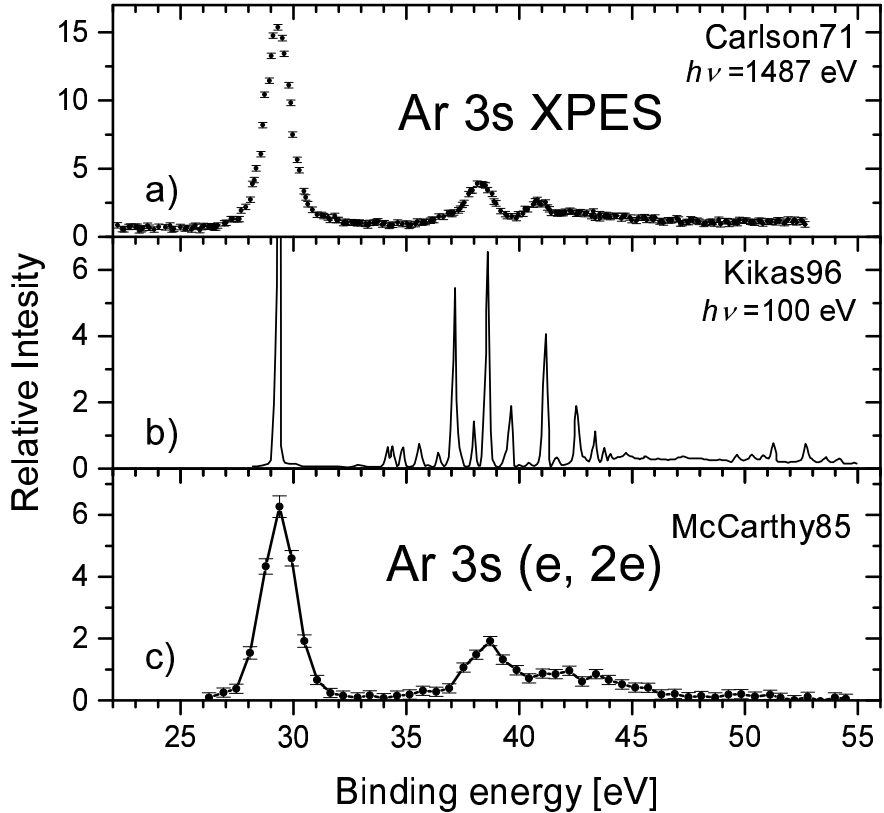}
\end{center}
\caption{$3s$ XPES of Ar excited with Al $K_{\alpha_{1,2}}$ radiation
($\hbar\omega=1487$ eV) at a resolution of about 1 eV adapted from
\citep{carlson71} and with synchrotron radiation $\hbar\omega=100$ eV at a
resolution of 0.137 eV adapted from \citep{kikas96}. ($e,~2e$) spectrum
adapted from \citep{mccarthy85}.}%
\label{fig:Ar3sPES-e2e}%
\end{figure}

The influence of the dipole electron fluctuation on the X-ray photoelectron
spectra (XPES) of the core atomic levels has not been considered as a strong
mechanism in the early seventies. Probably, this was the reason why
\cite{carlson67} and \cite{carlson71} who measured the 2s and 3s XPES of Ne and Ar,
respectively, interpreted the observed correlation satellites as the monopole
shake satellites (see, e.g., \cite{sachenko65,aberg67}). In particular, the
satellites observed at 38.4 eV and 40.6 eV in the $3s$ XPES of Ar (see
Fig.\ref{fig:Ar3sPES-e2e}a) have been interpreted as $3p^{4}(LS)4p$ states
only, whereas further investigations of these satellites revealed their
complex nature including also the $3p^{2}-3sn\ell$ dipole excitations. The
complex structure of the correlation satellites is illustrated in
Fig.\ref{fig:Ar3sPES-e2e}b where a state-of-the-art $3s$ XPES of Ar measured
by \cite{kikas96} is shown.

During the last quarter of the 20th century, measurements of the $ns$ XPES of
\emph{Rg} have been revisited several times. \cite{gelius74} used
monochromatized Al $K_{\alpha}$ radiation to measure the $3d,4p,4d,5s,5p$ XPES
of Xe. For the $5s$ XPES, a resolution of 0.51 eV has been achieved. This
spectrum has been interpreted by \cite{wendin77} who took into account both
monopole excitations and DPES via the theory described in \citep{wendin76}.
The {subvalence} $ns$ PES of Ne, Ar, Kr, and Xe have been measured by \cite{kikas96} by using
synchrotron radiation of $\hbar\omega$ from 60 to 170 eV and a resolution of
about 150 meV for overview spectra and about 70 meV for detailed
spectra in the region of correlation satellites (the overview of the $3s$ XPES
of Ar adapted from this paper is depicted in Fig.\ref{fig:Ar3sPES-e2e}b).
Later on, \cite{alitalo01} measured the $4s$ PES of Kr and the $5s$ PES of Xe
at even higher resolution of about 15 meV. A still higher resolution of about
2 meV has been {achieved} for Kr and Xe studied by high-resolution threshold photoelectron spectroscopy \citep{yoshii05}.

Investigations performed by \cite{kikas96,alitalo01,yoshii07} revealed
hundreds of new correlation satellites. {These satellites have been interpreted using numerous data bases (e.g., \cite{minnhagen63,moore71}) for the energy levels.} The first \emph{ab initio} calculations of correlation satellites (see, e.g.,
\cite{demekhin75,wendin76,wendin77,dyall79,dyall82a,dyall82b}) lacked
sufficient accuracy to interpret the high-resolution experiments because lots
of discrete states \citep{smid81} or even continuum states \citep{smid83}
should be included in the calculation. Against other work, the paper of
\cite{hansen87} stands out: solving the secular equation these authors
optimized the Slater and Trees \citep{racah52,trees52,rajnak63} parameters and
obtained not only the eigenenergies but also the eigenfunctions needed for the
interpretation of the Xe II correlation satellites.

\subsection{PES near the Cooper minimum}

\label{sec:3sPES_Cooper}

\begin{figure}[ptb]
\begin{center}
\includegraphics[width=0.95\textwidth]{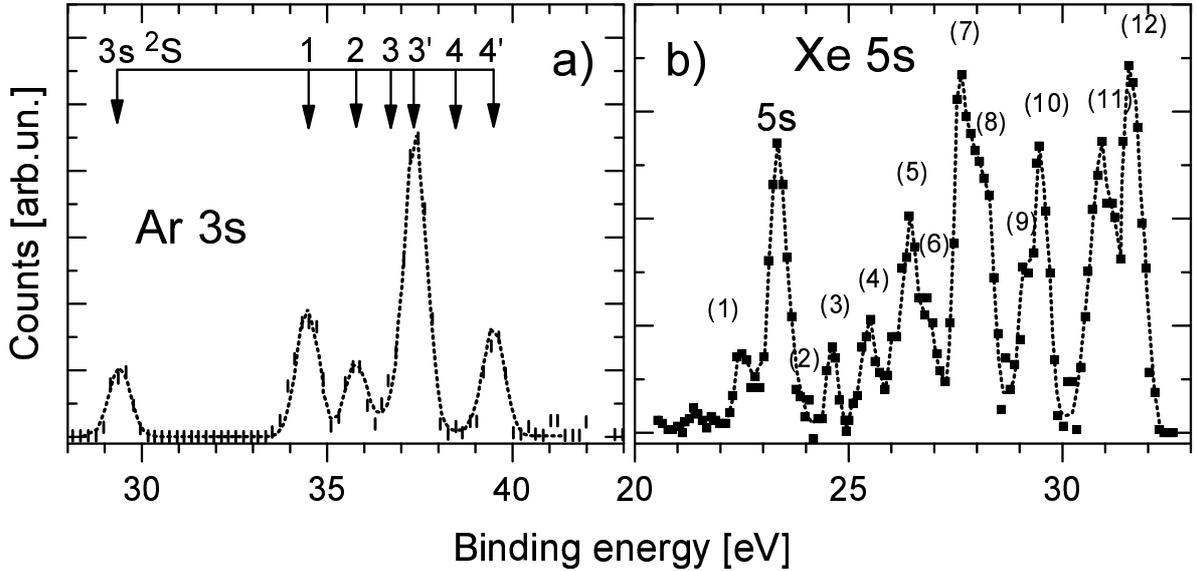}
\end{center}
\caption{Experimental {subvalence} $ns$ PES of Ar and Xe in the vicinity of the Cooper minimum a) $3s$ PES of Ar at $\hbar\omega=43$ eV adapted from \citep{adam85}. b) $5s$ PES of Xe at $\hbar\omega=33$ eV adapted from \citep{fahlman84}.}%
\label{fig:XeAr_nsPES_Coop}%
\end{figure}

\begin{figure}[ptb]
\begin{center}
\includegraphics[width=0.95\textwidth]{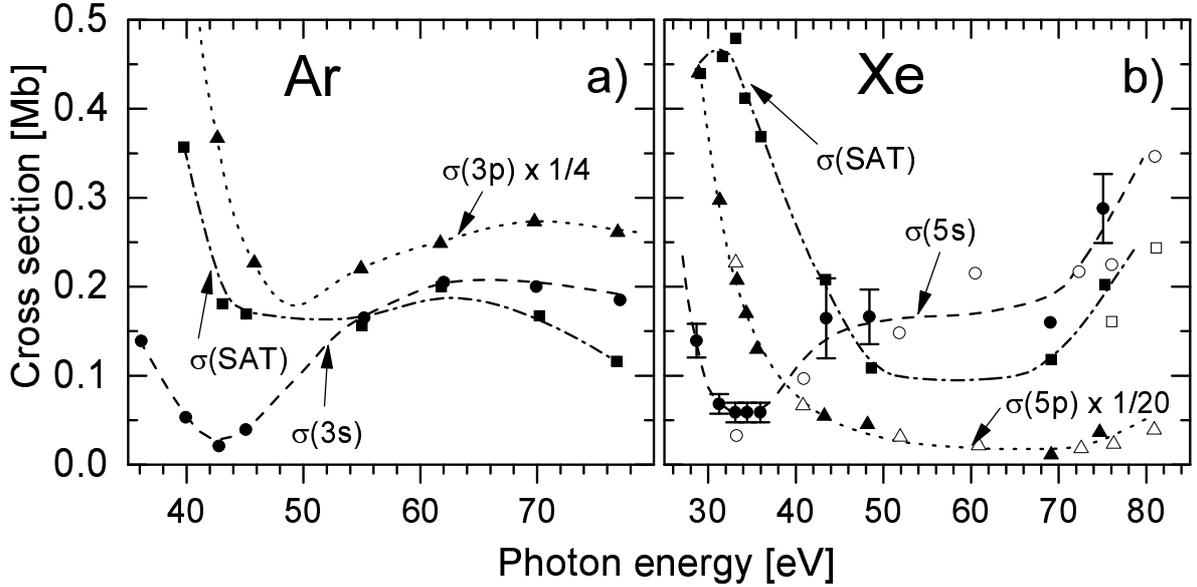}
\end{center}
\caption{Experimental photoionization cross sections for the main $ns$ levels
and for the sum of satellites (SAT) for Ar a) and Xe b). Figures a) and b)
adapted from \citep{adam85} and \citep{fahlman84}, respectively.}%
\label{fig:ArXeSatDep}%
\end{figure}

\cite{spears74} presented the $ns$ XPES for Ar, Kr and Xe excited by Al
$K_{\alpha}$ (1486.6 eV), Mg $K_{\alpha}$ (1253.6 eV) and Zr $M_{\zeta}$
(151.7 eV) radiation at a resolution of about 1.3 eV, revealing the dependence
of correlation satellite intensities on the exciting-photon energies. Several
years later, utilizing synchrotron radiation for excitation of PES,
\cite{adam78a} confirmed that the dependence of the relative intensities of
both main and satellite lines on the exciting-photon energy $\hbar\omega$ can
be very strong. Measuring the $3s$ PES of Ar, they found that decreasing
$\hbar\omega$ from 77 eV to 43 eV decreased the ratio $I_{3s~MAIN}/I_{3p}$ by
a factor of 17.5 and simultaneously increased the ratio $I_{3s~SAT}/I_{3p}$ by
a factor of 7.5. Thus, the ratio $I_{3s~SAT}/I_{3s~MAIN}$ changed by more than
two orders of magnitude (!).

The exciting-photon energy dependence of correlation satellites has been
revisited by \cite{adam85} for the $3s$ PES of Ar and by \cite{fahlman84} for
the $5s$ PES of Xe. The observed spectra for selected energies are shown in
Fig.~\ref{fig:XeAr_nsPES_Coop}. Later on, an investigation by \cite{becker88}
performed at higher energy resolution showed that the satellites shown in
Fig.~\ref{fig:XeAr_nsPES_Coop}a have a more complex structure. Fig.
\ref{fig:ArXeSatDep} shows the energy dependence of the main line intensity in
comparison with the integrated intensity of satellites. One can see that in
the vicinity of the $\sigma_{ns}(\omega)$ minimum, the $I_{ns~SAT}%
/I_{ns~MAIN}$ ratio is substantially larger then unity. The $\sigma
_{np}(\omega)$ depicted in the same figure also exhibit minima. These
minima {are of Cooper type, i.e. of} single-electron nature, whereas the minimum in $\sigma_{ns}(\omega)$ is connected with intershell correlations.

Theories existed at that time \citep{wendin77,martin78,dyall82b,smid83}
considering the intensity of correlation satellites as originating from the
main $ns^{2}np^{6}\dashrightarrow ns\ \varepsilon p$ transition only (cf.
pathway (a) in scheme (\ref{eq:5sXe-abs})) and, therefore, could not explain
the observed intensity behaviour in the threshold region. However, intershell
and ISCI correlations described by pathways (\ref{eq:5sXe-abs}b,c) are very
strong in the threshold region \citep{sukhorukov85a}. Moreover, the
$ns^{2}np^{6}\rightarrow ns^{2}np^{4}5d\ \varepsilon^{\prime}d\dashrightarrow
ns^{2}np^{4}n\ell\varepsilon^{\prime\prime}\ell^{\prime}$ and $ns^{2}%
np^{6}\dashrightarrow ns^{2}np^{5}\ \varepsilon^{\prime}d\rightarrow
ns^{2}np^{4}n\ell\varepsilon^{\prime\prime}\ell^{\prime}$\ channels result in
photoionization of the non-spherical $ns^{2}np^{4}n\ell(^{2}P,~^{2}P^{o}%
,~^{2}D)$ terms in addition to the $ns^{2}np^{4}n\ell(^{2}S)$ terms,
substantially increasing the intensity of correlation satellites in the
threshold region and explaining the discussed energy dependence.

\begin{figure}[ptb]
\begin{center}
\includegraphics[width=0.95\textwidth]{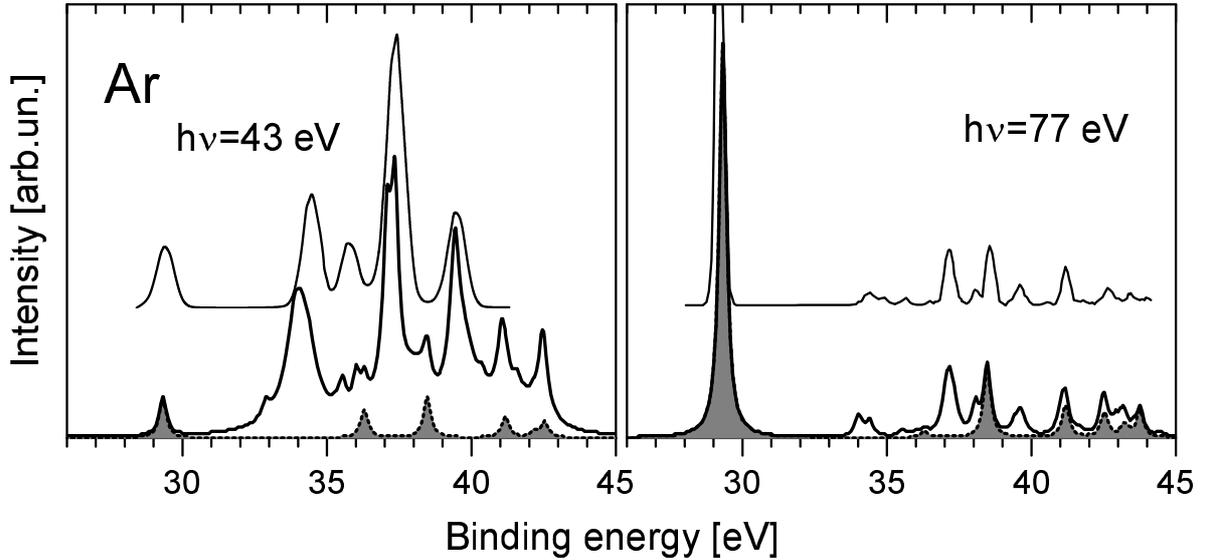}
\end{center}
\caption{Theoretical $3s$ XPES of Ar excited by synchrotron radiation
adapted from \citep{sukhorukov92}. Experimental spectra are shown by thin
lines for $\hbar\omega=43$ eV and $\hbar\omega=77$ eV from \cite{adam85} and
from \cite{kossmann87}, respectively. In the computed spectra
\citep{sukhorukov92}, the correlation satellites connected with the spherical
$^{2}S$ term are shaded.}%
\label{fig:Sukhorukov92_Ar3sPES}%
\end{figure}

The $3s$\ PES of Ar computed with taking into account the intershell and ISCI
correlations by \cite{sukhorukov92} are shown in
Fig.~\ref{fig:Sukhorukov92_Ar3sPES}. In this figure, the satellites connected
with the spherical $3s^{2}3p^{4}n\ell(^{2}S)$ terms are shaded. One can see
that the intensity of correlation satellites at small exciting-photon energies
mainly stems from the non-spherical $3s^{2}3p^{4}n\ell(LS)$ terms. The
dependence of the $3s$ PES of Ar and $4s$ PES of Kr on the exciting-photon
energies has been computed and tabulated by \cite{sukhorukov94a} and turned
out to be in good agreement with experiment.

\begin{figure}[ptb]
\begin{center}
\includegraphics[width=0.9\textwidth]{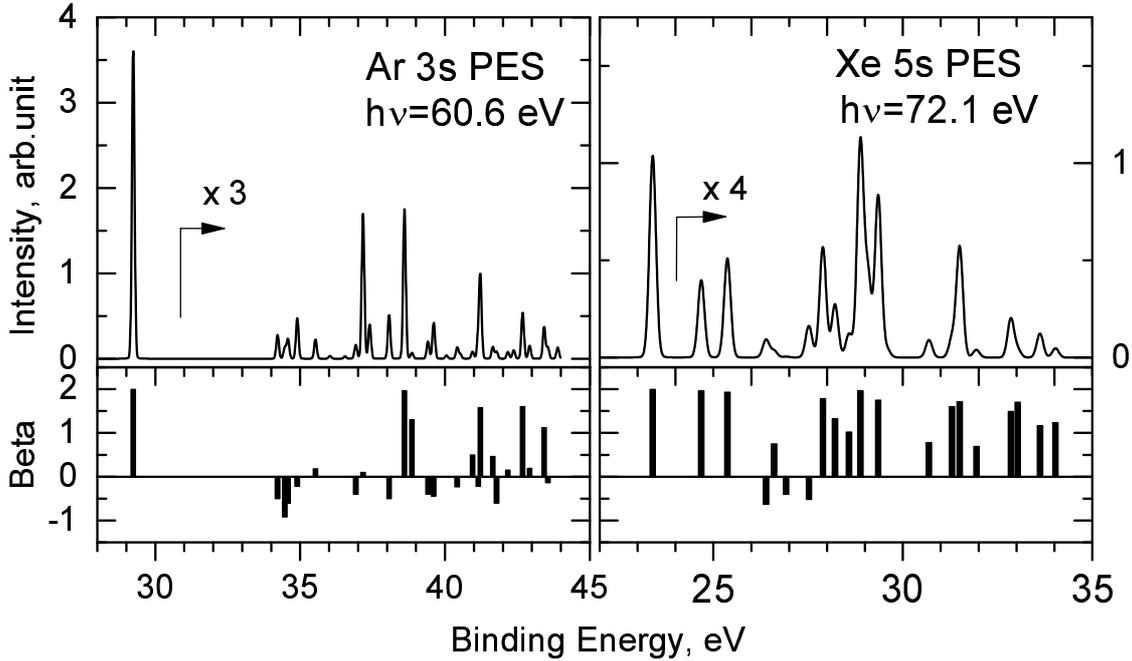}
\end{center}
\caption{Photoelectron spectra and angular distribution parameters for the
$3s$ Ar ($\Delta E=105$ meV) and $5s$ Xe ($\Delta E=200$ meV) satellites.
Figure adapted from \citep{krause92}.}%
\label{fig:Krause92_S_bet}%
\end{figure}

The possibility to obtain experimentally information about the symmetry of
satellite states has been illustrated by \cite{krause92} who measured the $ns$
PES of all \emph{Rg} at different angles, determining the parameter of angular
distribution of photoelectrons. The $ns$ PES of Ar and Xe from
\citep{krause92} are depicted in Fig.~\ref{fig:Krause92_S_bet}. This figure
illustrates that the satellite structure of the $3s$ Ar and $5s$ Xe PESs,
looking qualitatively similar, has different $\beta$ parameters and, hence,
the respective ionic levels have different symmetry.

\subsection{Correlation satellites in (e, 2e) spectra of rare gases}

Discussing the correlation satellites we should mention their observation by
(e, 2e) spectroscopy (see the basis of this method in \citep{bethe30} and
reviews of \cite{inokuti71,inokuti78,mccarthy76}). The correlation satellites
of Ar accompanying excitation of the $3s$ shell have been observed and
interpreted as the $3p^{2}-nd\varepsilon\ell$ excitations by \cite{weigold75}
who used for the interpretation optical data and the calculation of
\cite{luyken72,luyken72a}. The energy resolution of the ($e,~2e$) spectra is
much lower than the resolution of the ($\gamma,~e$) spectra, which is
illustrated in Fig.~\ref{fig:Ar3sPES-e2e} where the state-of-the-art ($e,~2e$)
and ($\gamma,~e$) $3s$ spectra of Ar are compared. However, in case the energy
resolution is not very decisive in the experiment, the ($e,~2e$) spectroscopy
is widely used for the measurement of the energy dependence of the partial
photoionization cross sections. For instance, the $3s$ photoionization cross
section of Ar has been measured by \cite{tan78}; the dependence of the
integrated intensity of correlation satellites on the exciting energy has been
measured for the $4s$ spectrum of Kr and for the $3s$ spectrum of Ar by
\cite{fuss81} and by \cite{avaldi89}, respectively.

\cite{mccarthy78} drew attention to the fact that ($e,~2e$) and ($\gamma,~e$)
spectra result in different spectroscopic factors for the $3s$ spectrum of Ar.
In the following decade, there were several papers where this phenomenon has
been investigated experimentally and theoretically
\citep{mitroy84,brion86,svensson87,avaldi89}. Finally, \cite{amusia85} showed
that taking into account many-electron correlations could explain the observed
difference. However, a detailed calculation of the ($e,~2e$) spectra that
takes into account many-electron correlations of type (\ref{eq:5sXe-abs}) has
not been performed so far.

\section{Resonance structure in the main line and satellite production cross
sections}

\subsection{Experimental investigations}

In the mid eighties, it became clear that photoionization of the subvalence
$ns$ shell of the \emph{Rg} is a complex many-electron process accompanied by
excitation of correlation satellites being sometimes more intense than the
main $ns$\ line (see section~\ref{sec:3sPES_Cooper}). Experimental
measurements resulted in \textquotedblleft smooth\textquotedblright\ PICS
$\sigma(ns)$ of the main line (see section~\ref{sec:PICS}) being rather scarce
for the satellites PICS $\sigma(np^{4}m\ell)$.

Calculations of the $\sigma(np^{4}m\ell)$ PICS for satellites were practically
absent. \cite{silfvast86} computed the PICS for the Ar odd $3p^{4}np$
satellites production considering the $3p^{6}\dashrightarrow3p^{4}%
np\varepsilon(s/d)$ transition of the `shake' type. The influence of
resonances on the computed PICS had not been included. \cite{wijesundera87}
presented theoretical results for the PICS of the Ar even $3p^{4}nd$
satellites including Rydberg series $3p^{4}ndmp$ converging to the respective
thresholds. They obtained that the $3p^{4}4dmp$ and $3p^{4}5dmp$ resonances
result in prominent Fano profiles in the $\sigma(3p^{4}3d)$ satellite cross
section. Later on, calculations of such type have been extended to the main
line $\sigma(3s)$ and satellites $\sigma(3p^{4}np)$ \citep{wijesundera89}. In
all cases, taking into account doubly-excited states resulted in pronounced
structure in the main line and satellite production cross sections.

The energy resolution of conventional photoelectron spectroscopy at this time
was not sufficient to observe the PICS of individual satellite levels.
Therefore, \cite{hall89} applied for this purpose the photoelectron
spectrometer described by \cite{king87} which collects electrons with energies
$\varepsilon$\ between 0~eV and 1~eV. This spectrometer operated in both the
threshold mode ($\varepsilon\approx0$~eV) providing a resolution of about
$\Delta E=10$~meV and in the synchronous mode providing the partial ionization
cross section of the specific ionic state. The spectrometer has been applied
to measure the relative PICS of Ne \citep{hall91}, Ar \citep{hall89}, Kr and
Xe \citep{hall90} with an energy resolution of about $\Delta E=50$~meV and
exciting-photon energies of about 1~eV above the threshold of each
investigated individual \emph{Rg}~II level.

\begin{figure}[tb]
\begin{center}
\begin{minipage}[t]{0.47\textwidth}
\begin{center}
\includegraphics[width=\textwidth]{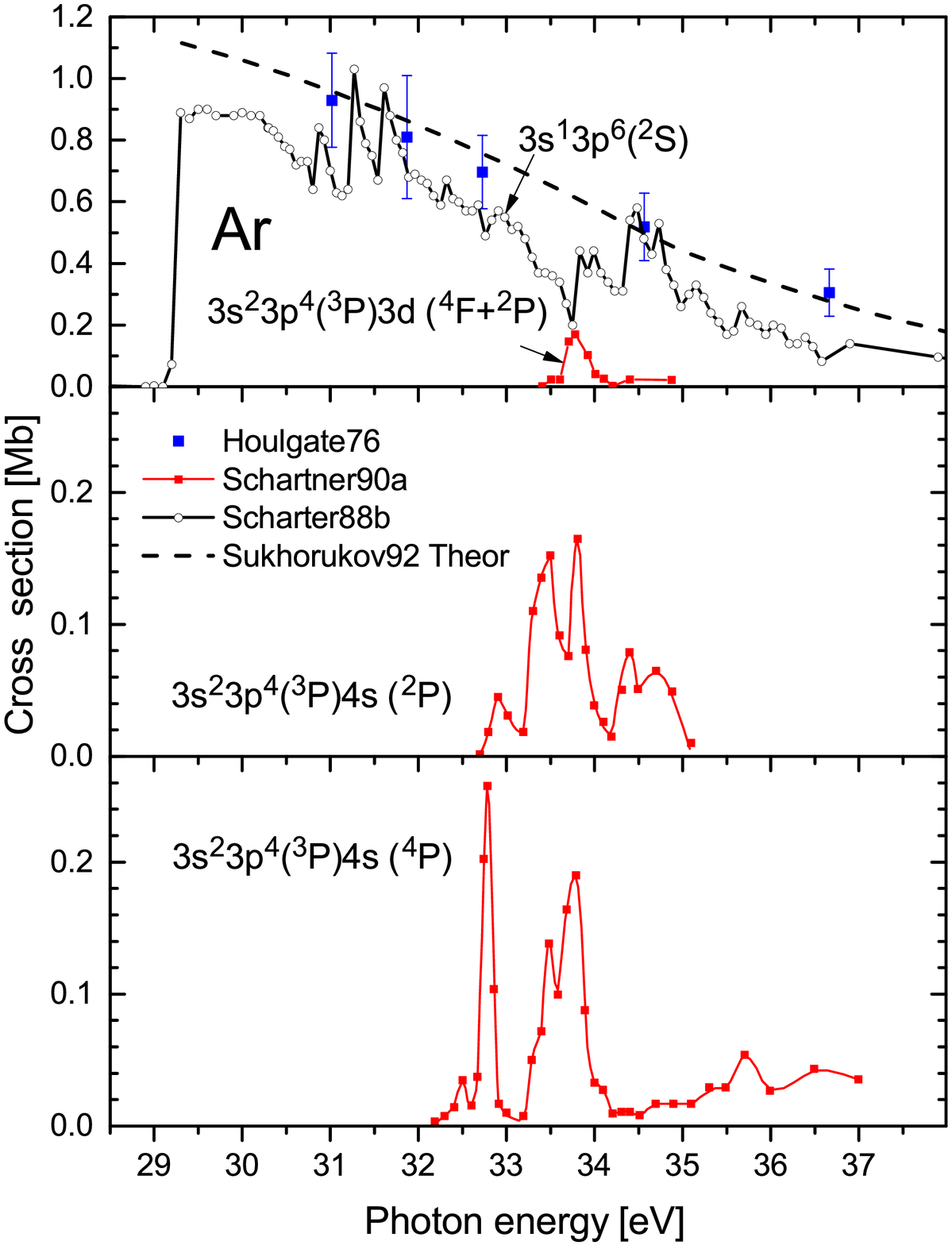}
\caption{Partial PICS of the $3s$ main and some $3p^{4}n\ell$ satellite
levels of Ar. Adapted from
\citep{schartner88a,schartner90,sukhorukov92}.}\label{fig:PIFS_Ar3s_3lay}%
\end{center}
\end{minipage} \hfill\begin{minipage}[t]{0.47\textwidth}
\begin{center}
\includegraphics[width=\textwidth]{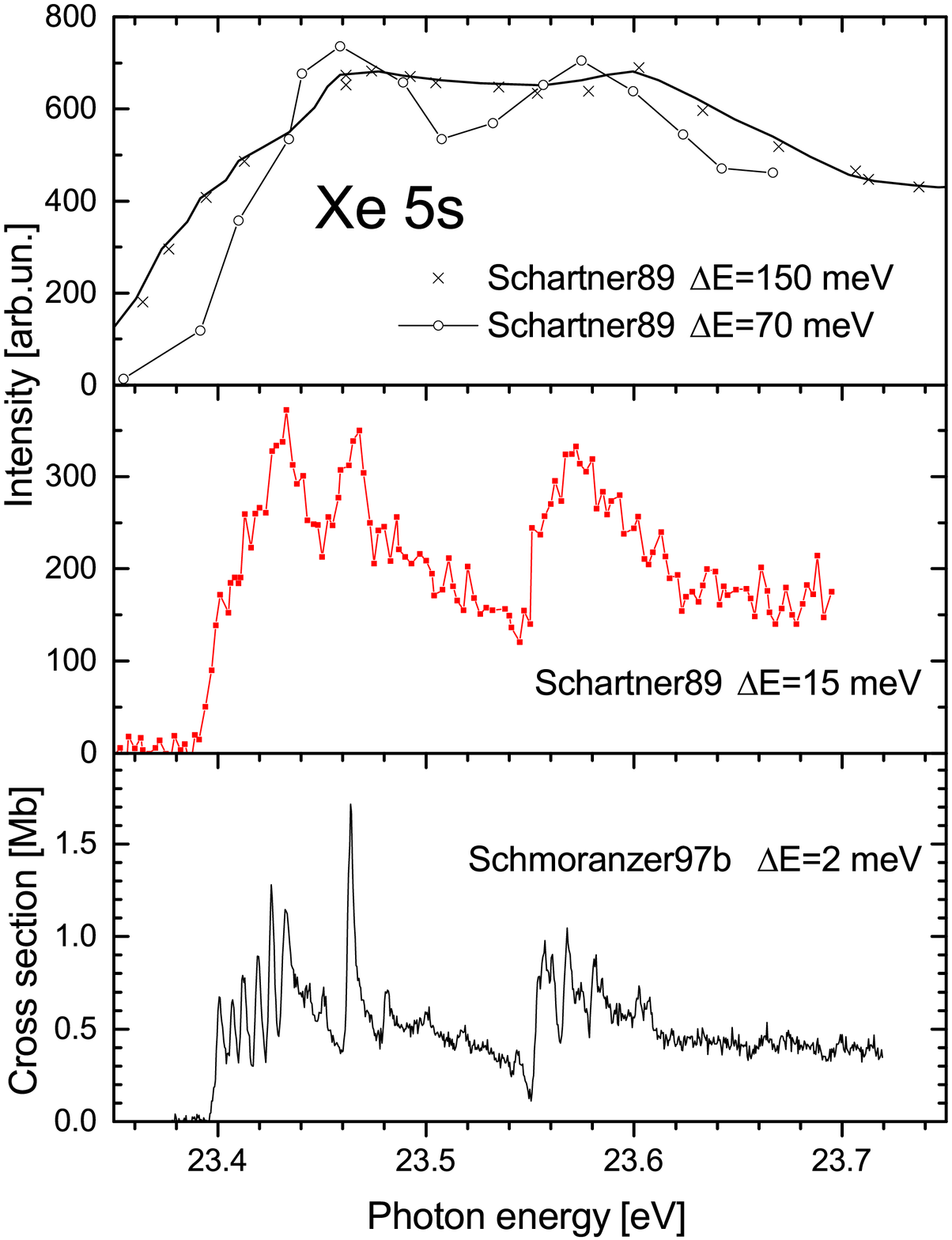}
\caption{Partial PICS of the $5s$ main level of Xe. Adapted from
\citep{schartner89,schmoranzer97}.}\label{fig:PIFS_Xe5s_3lay}\end{center}
\end{minipage}
\end{center}
\end{figure}

\cite{wills89} designed a photoelectron spectrometer which recorded PES
varying the exciting-photon energies in a broad energy interval (up to 20~eV
above the respective threshold) by steps of about 30~meV and {achieved} an
energy resolution of $\Delta E=50-100$~meV. The authors obtained relative PICS for individual ionic levels (the authors call them `constant ionic state spectra') for Ne
\citep{wills90}, Ar \citep{wills89}, and Xe \citep{wills90a}. The PICS were
obtained for more then 10 satellite levels for each \emph{Rg}~II ion and for
the main $ns$ and $np$ levels. In the case of Xe, $\sigma(5p_{j})$ has also
been specified by the total angular momentum $j$ of the $5p$ electron \citep{wills90a}.

In order to observe resonance structures in the cross sections of individual
ionic levels the photon-induced fluorescence spectroscopy, PIFS (see
section~\ref{sec:PIFSexp}), appeared to be the most suitable and effective
tool. \cite{schartner88,schartner88a} applied PIFS to measure the relative
PICS of the main line $\sigma(3s)$ and satellite $\sigma(3p^{4}(^{3}%
P)4s~^{4}P)$ for Ar~II in the exciting-photon energy interval from the $3s$
threshold ($\hbar\omega=29.24$~eV) {up to} $\hbar\omega=40$~eV at a resolution of $\Delta E=150$~meV. The relative $\sigma(5s)$ and $\sigma(5p^{4}6s/5d~LS)$
PICS have been measured by \cite{schartner89} from the $5s$ threshold
($\hbar\omega=23.40$~eV) {up to} $\hbar\omega=24.3$~eV at a resolution of $\Delta E=70-150$~meV (near the $5s$ threshold region, the resolution was set to
$\Delta E=15$~meV). The Ar PICS has been put on absolute scale by
\cite{schartner90} and later by \cite{mobus93}. The absolute
$\sigma(2s)$ and $\sigma(2p^{4}(^{3}P)3s~^{4}P,^{2}P)$ for Ne~II have been
measured by \cite{schartner90a} in the energy interval {between} $\hbar\omega=48.48$ {and} $51$~eV. For Kr, relative PICS for the main $4s$ level and for some groups of the $4p^{4}n\ell$ satellites have been measured by
\cite{schmoranzer90} from the $4s$ threshold ($\hbar\omega=27.51$~eV) up to
$\hbar\omega=31.0$~eV.

Some representative experimental cross sections illustrating the ability of
the PIFS technique are depicted in Figs.~\ref{fig:PIFS_Ar3s_3lay},
\ref{fig:PIFS_Xe5s_3lay}. The satellite cross sections for Ar are from
\citep{schartner90}and the cross section for the main $3s$ line shown in the
upper panel of Fig.~\ref{fig:PIFS_Ar3s_3lay} is from \citep{sukhorukov92}. In
the same panel, cross sections of \cite{houlgate76} are depicted. Within the
error bars, the cross sections of \cite{houlgate76} and of \cite{sukhorukov92}
coincide. However, the data obtained by PIFS clearly show a rich resonance
structure, whereas the data of \cite{houlgate76} may suggest a smooth dependence.

Cross sections for the main $5s$ line of Xe \citep{schartner89} have been
measured at three different resolutions $\Delta E=150~\mathrm{meV}%
,\ 70~\mathrm{meV},\ 15~\mathrm{meV}$ (see Fig.~\ref{fig:PIFS_Xe5s_3lay}). We
note that the typical resolution in cross section measurements by threshold
spectroscopy \citep{hall89} or by photoelectron spectroscopy \citep{wills89}
was only about $\Delta E=100~\mathrm{meV}$. Moreover, the absence of cross
sections measured directly at the ionic thresholds indicates that those
spectroscopic methods {are difficult to be applied at threshold due to the necessity of detecting very slow electrons}. Further
improvements of the PIFS measurement technique allowed \cite{schmoranzer97} to
{achieve} a resolution below $\Delta E=2~\mathrm{meV}$ and to observe a very impressive resonance structure directly at the $5s$ threshold of Xe (see the
lowest panel of Fig.~\ref{fig:PIFS_Xe5s_3lay}). This structure stems from the
$(^{3}P)5d~^{4}D_{J}5f_{j}$ resonances which become visible due to their
interaction with the $(^{3}P)5d~^{2}D_{5/2}6p_{3/2}$, $(^{3}P)5d~^{4}%
D_{1/2}8p_{3/2}$, and $(^{3}P)6s~^{2}P_{3/2}9p_{1/2}$ resonances. Note that
this threshold structure has been observed later neither by photoelectron
\citep{alitalo01} nor by threshold \citep{yoshii07} spectroscopy although the
resolution in the latter work has been quoted as $\Delta E=5~\mathrm{meV}$.

\subsection{Screening of the Coulomb interaction}

\label{sec:Coulomb_screening}

Photoionization of the subvalence shells is a prototypical object to study the
Coulomb interaction in many-electron systems. Indeed, many-electron
correlations determine not only energy levels due to DPES (see
section~\ref{sec:CorrSat}) but also photoionization cross sections due to
intershell correlations (see sections~\ref{sec:Intershell}, \ref{sec:PICS}).
Furthermore, those correlations are also influenced (usually reduced) by
higher-order correlations. In this section, we review some physical phenomena
where the reduction of the Coulomb interaction (Coulomb screening) plays a
substantial role and can be confirmed by PIFS measurements.

\subsubsection{Energy level calculations}

Reduced values of the pertinent Coulomb
integrals have been determined in earlier work by least-squares technique
{fitting between calculated and measured  energy levels. In this way} Coulomb integrals for Ne \citep{luyken71}, Ar \citep{luyken72}, and Xe \citep{hansen87} {have been determined}.

The \emph{ab initio} technique described in section \ref{sec:theory_functions}
has been widely used in calculations of energy levels of the $Rg$~II ions near
the subvalence threshold. Energy levels stemming from $ns^{1}np^{6}$ and
$ns^{2}np^{4}m\ell$ configurations have been investigated for Ar ($n=3$) in
\citep{sukhorukov92,lagutin99,kammer06}, for Kr ($n=4$) in
\citep{schmoranzer93,sukhorukov94,lagutin94,demekhin05,sukhorukov07}, and for
Xe ($n=5$) in \citep{lagutin96,schmoranzer97,ehresmann10}. All these papers
contain also original measurements performed by PIFS.

The same technique has been applied to calculate energy levels of the excited
$Rg$~I states including the doubly-excited $ns^{2}np^{4}n\ell$ states and
before-$ns$-threshold states in Ar \citep{lagutin94,lagutin99,kammer06}, Kr
\citep{demekhin05,sukhorukov07}, and Xe \citep{ehresmann10,sukhorukov10}.

\begin{table}[ptb]
\caption{{Average} ionization potentials of some Xe levels in eV. Interactions between the states listed in the table are excluded. Adapted from
\citep{lagutin96}.}%
\label{tab:XeEcorr}%
\vspace{2ex}
\begin{tabular}
[c]{rcccc}\hline\hline
Configuration & $IP_{PF}$ & $IP_{c}$ & $IP$ & $IP~^{a)}$\\
& (equation (\ref{eq:IPpf})) & (equation (\ref{eq:IPc})) & (equation
(\ref{eq:IP})) & \\\hline
$5s^{1}$ & 26.28 & --0.06 & 26.22 & 26.30\\
$5p^{4}$ & 32.54 & 2.13 & 34.67 & \\
$5p^{4}6s$ & 23.84 & 1.29 & 25.13 & 25.15\\
$7s$ & 28.19 & 1.82 & 30.01 & 29.89\\
$8s$ & 29.91 & 2.02 & 31.93 & 31.71\\
$5p^{4}6p$ & 25.98 & 1.52 & 27.50 & 27.50\\
$7p$ & 29.01 & 1.94 & 30.95 & \\
$8p$ & 30.31 & 2.07 & 32.38 & \\
$5p^{4}5d$ & 24.90 & 1.08 & 25.98 & 25.99\\
$6d$ & 28.65 & 1.66 & 30.31 & 30.33\\
$7d$ & 30.13 & 1.92 & 32.05 & 31.98\\
$8d$ & 30.91 & 2.04 & 32.95 & \\
$9d$ & 31.35 & 2.09 & 33.44 & \\\hline\hline
\end{tabular}
\par
\vspace{2ex} {\footnotesize $^{a)}$ Fitted ionization potentials from the
calculation of \cite{hansen87}}\end{table}

The technique to calculate the energy levels with respect to the ground state
comprises the computing of {average} PF energies, diagonal corrections to the energy levels and their multiplet splitting, and computing corrections to the
Slater integrals determining the multiplet splitting and configuration
interaction. The PF energy of the excited configuration $K$ relative to the ground
state, having the meaning of an ionization potential, is computed as the difference
between the respective PF total energies $E_{PF}$ as%

\begin{equation}
IP_{PF}(K)=E_{PF}(K)-E_{PF}(0). \label{eq:IPpf}%
\end{equation}

Correlational corrections $E_{C}$ to the PF energies were computed according
to equation (\ref{eq:theor_Vii}), resulting in correlational corrections to
the energy level:%

\begin{equation}
IP_{C}(K)=E_{C}(K)-E_{C}(0) \label{eq:IPc}%
\end{equation}
and determining the energy level as:%

\begin{equation}
IP(K)=IP_{PF}(K)+IP_{C}(K). \label{eq:IP}%
\end{equation}

The PF energies $E_{PF}$ entering equation (\ref{eq:IPpf}) have the order of
several thousands of Ry, whereas the correlation energies amount to several
dozens of eV. Typical corrections and diagonal matrix elements are presented
for some Xe configurations in Table~\ref{tab:XeEcorr}. Corrections computed by
\citep{lagutin96} agree with the fitted results of \cite{hansen87} fairly well
(see last column of Table~\ref{tab:XeEcorr}).

\begin{table}[ptb]
\caption{Correlational corrections $\chi(ns^{1}np^{6};ns^{2}np^{4}nd)$ for the
Slater integrals $R^{1}(npnp;nsnd)$ ($n=4$ for Kr~II) and ($n=5$ for Xe~II),
in per cent and with opposite sign to the matrix element $^{a)}$.}%
\label{tab:KrXeDPES}%
\vspace{2ex}
\begin{tabular}
[c]{rrrr}\hline\hline
\multicolumn{2}{c}{Kr $^{b)}$} & \multicolumn{2}{c}{Xe $^{c)}$}\\
Intermediate & Correction & Intermediate & Correction\\
configuration &  & configuration & \\\hline
$4s^{1}4p^{4}4d^{2}$ & 9.8 & $5s^{1}5p^{4}5d^{2}$ & 14.8\\
$4s^{1}4p^{4}\{d\}\{d\}$ & 11.5 & $5s^{1}5p^{4}\{d\}\{d\}$ & 15.7\\
$\{f\}\{f\}$ & 1.1 & $\{f\}\{f\}$ & 1.2\\
$\{d\}\{g\}$ & 1.1 & $\{d\}\{g\}$ & 1.6\\
$\{f\}\{h\}$ & 0.4 & $\{f\}\{h\}$ & 0.7\\
$3d^{9}4s^{1}4p^{5}4d\{f\}$ & 5.1 & $4d^{9}5s^{1}5p^{5}5d\{f\}$ & 14.3\\
$\{p\}$ & 0.4 & $\{p\}$ & 1.2\\
$3d^{9}4s^{2}4p^{5}\{f\}$ & 1.5 & $4d^{9}5s^{2}5p^{5}\{f\}$ & 4.0\\
other & 5.1 & other & 5.9\\
Sum & 36.0 & Sum & 59.4\\\hline\hline
\end{tabular}
\par
\vspace{2ex} {\footnotesize $^{a)}$ Summation and integration over the channel
AOs in brackets was performed. \newline$^{b)}$ Data derived from
\citep{schmoranzer93}. \newline$^{c)}$ Data derived from \citep{lagutin96}.}\end{table}

Corrections to the Slater integrals were computed introducing scaling factors
$\chi\left(  \alpha,\alpha^{\prime}\right)  $ (\ref{eq:theor_x}). The
$\chi(ns^{1}np^{6};ns^{2}np^{4}nd)$ for the Slater integrals $R^{1}%
(npnp;nsnd)$ computed for Kr~II ($n=4$) and Xe~II ($n=5$) applying the
technique of \cite{judd67,lindgren86} are listed in Table~\ref{tab:KrXeDPES}.
The Xe~II correction is noticeably larger than the Kr~II correction due to the
increased influence of correlations including high orbital angular momenta.
The screening factors computed via equation~(\ref{eq:theor_x}) are 1.59 and
1.36 for Xe~II and Kr~II, respectively (for Ar~II this constant equals 1.24).
We note that computed screening factors agree fairly well with results
obtained by the least-squares technique. For instance, in Xe~II, as computed
by \cite{lagutin96}, the coefficient reducing the $R^{1}(5p5p;5s5d)$ integral
is 1.59, whereas the fitted value of \cite{hansen87} is 1.57.

Taking into account the reduction of the Coulomb interaction resulted in an
average accuracy of the energy level calculation of $\delta E=70$~meV as it
was estimated by \cite{sukhorukov94}. This accuracy and a refined analysis of
the wave functions allowed \cite{sukhorukov94} to change the assignments of 18
ionic levels relative to \citep{moore71}. The accuracy of $\delta E=70$~meV is
worse than that obtained for Ar~II by \cite{luyken72} ($\delta E=20$~meV) and
for Xe~II by \cite{hansen87} ($\delta E=32$~meV) by the least-squares fit.
This difference can be connected with the inclusion in
\citep{luyken72,hansen87} of nonlinear Trees's corrections to the energy
levels \citep{racah52,trees52,rajnak63} containing, e.g., terms as $\alpha
L(L+1)$.

\subsubsection{Lifetimes of the subvalence $ns$ vacancies}

The interaction between the $ns^{1}np^{6}$ and $ns^{2}np^{4}m\ell$
configurations results in complex wave functions of $Rg$ with the subvalence
$ns$ vacancy (see section~\ref{sec:CorrSat}). Moreover, this configuration
mixing strongly changes the results obtained using a simple central-field
calculation. \cite{lawrence69}, who measured lifetimes of some Ar, Cl, and S
ions by using a pulsed-electron technique, has found that the measurements result
in values by more than one order of magnitude larger than obtained by the
central-field calculation. This discrepancy can be seen in
Table~\ref{tab:lives} (cf. measurement of \cite{lawrence69} in 1st line and
calculation of \cite{lauer99} within the PF approach in the 4th line) and had been
attributed to DPES, revealed by \cite{minnhagen63}.

\begin{table}[ptb]
\caption{Comparison of experimental lifetimes $\tau_{ns}$ [ns] with computed
ones in different approximations. Derived from \cite{lauer99}.}%
\label{tab:lives}%
\vspace{2ex}
\begin{tabular}
[c]{rrrr}\hline\hline
& Ar & Kr & Xe\\\hline
Experiment & 4.8$\pm$0.1 $^{a)}$ & 0.33$\pm$0.04 $^{b)}$ & 34.4$\pm$0.6
$^{c)}$\\
MCHF $^{d)}$ & 4.98 & 6.58 & 9.5\\
Pulsed SR $^{e)}$ & 4.684$\pm$0.019 & 30.78$\pm$0.17 & 35.93$\pm$0.20\\
PF $^{f)}$ & 0.18 & 0.15 & 0.19\\
PF+DPES $^{f)}$ & 2.54 & 3.01 & 10.77\\
CIPF $^{f)}$ & 3.08 & 5.5 & 36.89\\\hline\hline
\end{tabular}
\par
\vspace{2ex} {\footnotesize $^{a),~b),~c)}$ data measured by pulsed electrons
\citep{lawrence69}, beam foil \citep{irwin76}, and pulsed synchrotron
\citep{rosenberg78} techniques, respectively. \newline$^{d)}$ Calculation of
\cite{hansen77}. \newline$^{e)}$ Measurement of \cite{lauer99} \newline$^{f)}$
Calculation of \cite{lauer99}.}\end{table}

\cite{luyken72} computed the lifetime of the Ar $3s$ vacancy by using fitted
wave functions and found strong cancellation between the transition amplitudes
describing the $3s^{1}3p^{6}\dashrightarrow3s^{2}3p^{5}$ and $3s^{1}%
3p^{6}\longleftrightarrow3s^{2}3p^{4}nd\dashrightarrow3s^{2}3p^{5}$ pathways.
Taking into account this interference resulted in a deviation of the computed
value by only 25\% from the measured one by \cite{lawrence69}, $\tau
_{3s}=4.8(1)$~ns.

\cite{irwin76} applied the beam-foil technique to measure the lifetime
$\tau_{4s}$\ of the Kr $4s$ vacancy. The obtained value of $\tau_{4s}%
=0.33(4)$~ns was unexpectedly small in comparison with data obtained by
\cite{lawrence69} for Ar and several years later by \cite{rosenberg78} for Xe
who used pulsed synchrotron radiation (SR) and obtained $\tau_{5s}%
=34.4(6)$~ns. \cite{rosenberg78} noted that their result for $\tau_{5s}$ was
about three times larger than that predicted by multiconfiguration
Hartree-Fock (MCHF) calculation of \cite{hansen77} $\tau_{5s}=9.5$~ns. After
the measurement of \cite{rosenberg78}, \cite{hansen79} revisited the
calculation of the Xe $\tau_{5s}$ using wave functions fitted via experimental
energy levels of Xe~II \citep{hansen78}. They used several approximations
resulting in $\tau_{5s}=37.1$~ns in the best case. This value agrees well with
$\tau_{5s}=34.4\pm0.6$~ns measured by \cite{rosenberg78}. In case of Kr,
$\tau_{4s}$ computed by \cite{hansen79} was 20 times larger than that measured
by \cite{irwin76}.

\begin{figure}[ptb]
\begin{minipage}[t]{0.48\textwidth}
\begin{center}
\includegraphics[width=\textwidth]{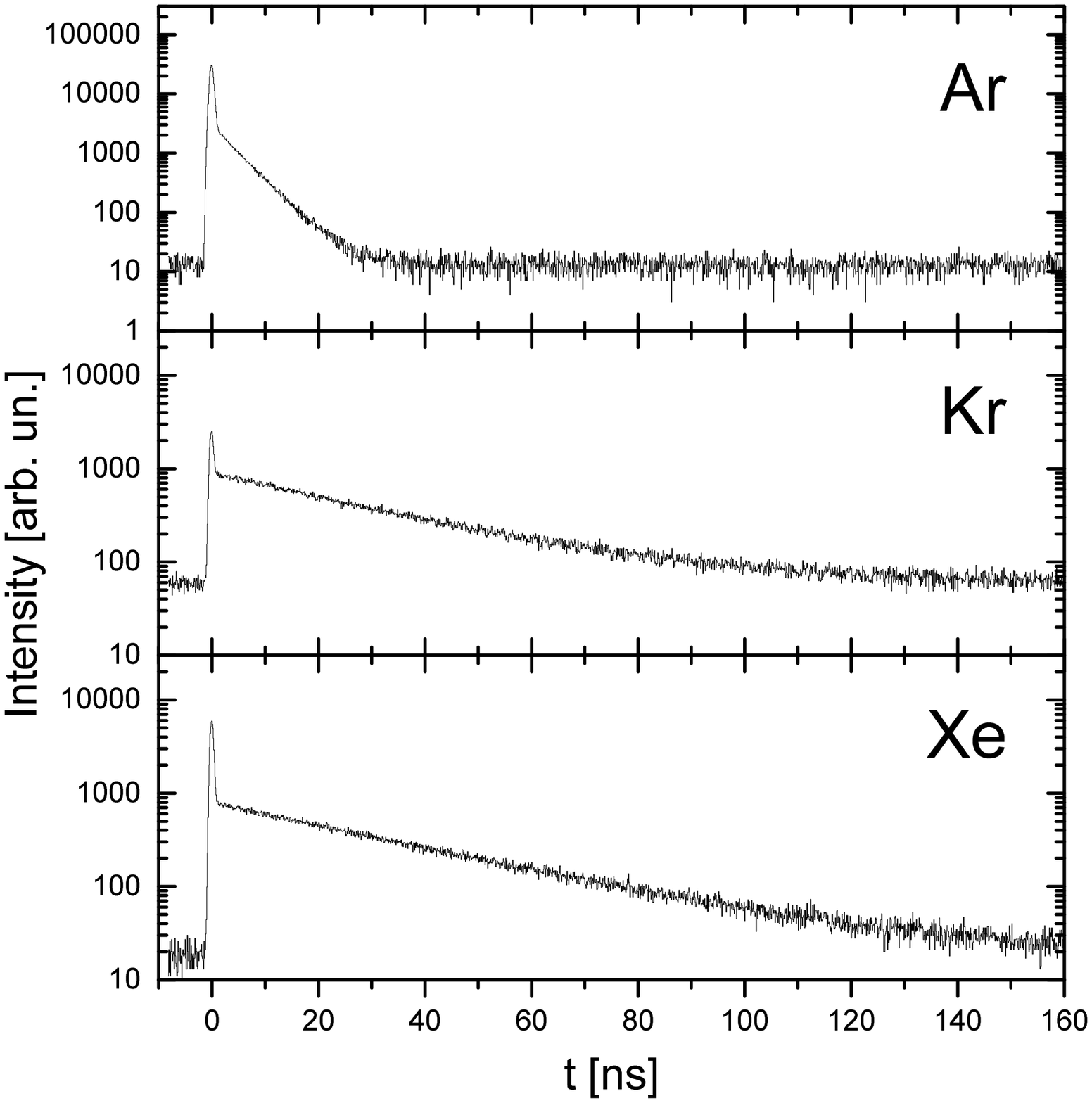}
\caption{Measured fluorescence decay curves at a gas pressure of 0.100
mbar. At $t=0$~ns, the incident photon pulse is observed which is caused by
reflections inside the target cell. Adapted from
\citep{lauer99}.}\label{fig:Life_example}\end{center}
\end{minipage} \hfill\begin{minipage}[t]{0.48\textwidth}
\begin{center}
\includegraphics[width=\textwidth]{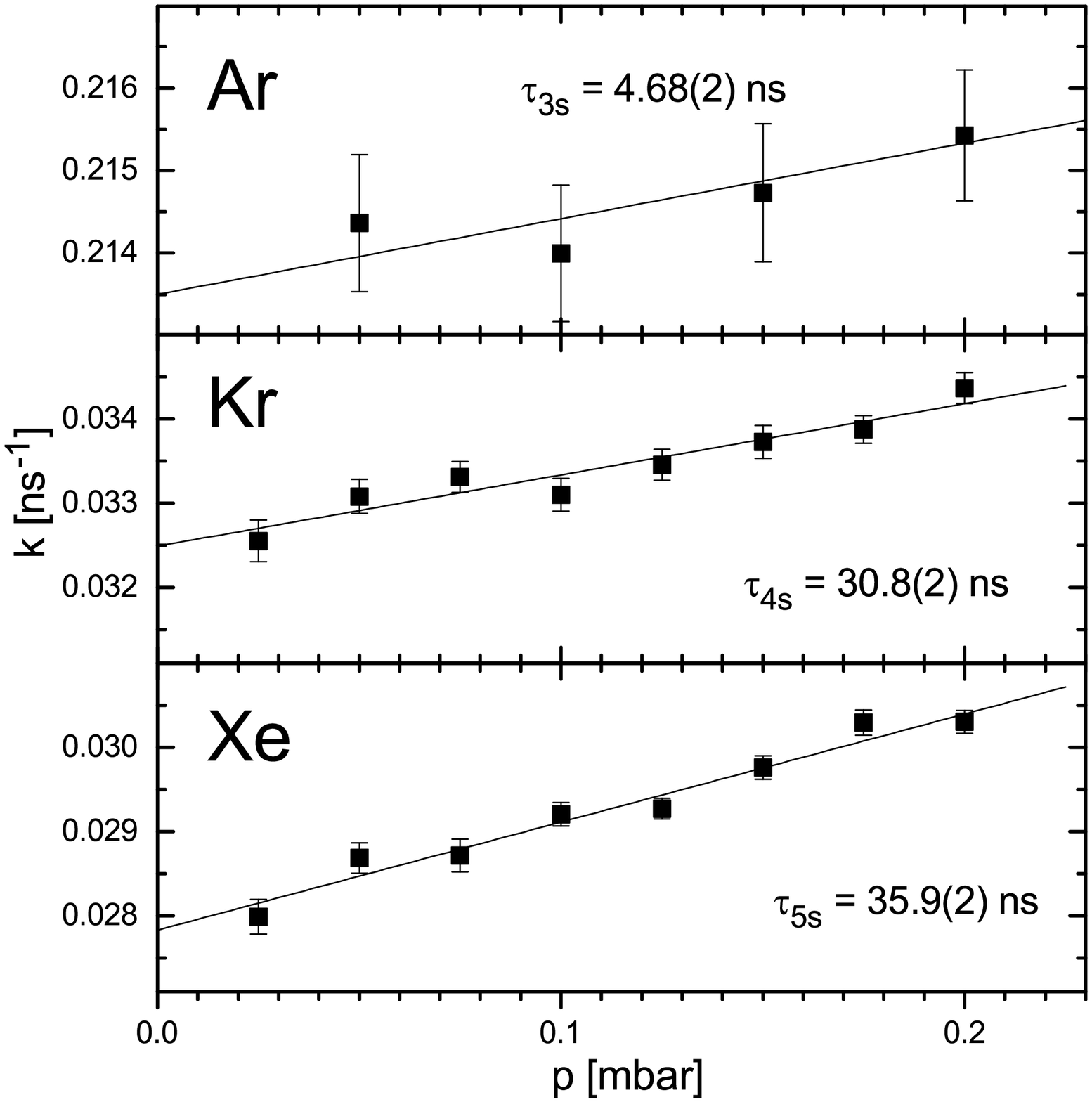}
\caption{Stern-Volmer plot for the
decay of the $ns$ vacancies. The error bars represent the
one-standard-deviation fit uncertainties. Adapted from
\citep{lauer99}.}\label{fig:Life_ALLdata}\end{center}
\end{minipage}
\end{figure}

A refined experiment has been performed by \cite{lauer99} who used pulsed
monochromatized SR and PIFS to measure lifetimes of the subvalence $ns$
vacancies for Ar, Kr, and Xe. They analyzed the time dependence of the
fluorescence intensity after the exciting SR pulse of 0.7~ns FWHM using the
single-photon start-stop technique \citep{schmoranzer84,Lauer98a}. Typical
fluorescence decay curves are depicted in Fig.~\ref{fig:Life_example} where
the trace of the exciting pulse is observed at $t=0$. The energy of exciting
photons was chosen between the respective $ns$ threshold and the first
$np^{4}m\ell$ satellite with FWHM$\simeq60$~meV. The measured decay curves
were used to derive the lifetimes at each gas pressure and corrected for the
collisional de-excitation using Stern-Volmer plots \citep{schmoranzer84}.
Results of measurements are listed in the 3rd line of Table~\ref{tab:lives}.
One can see excellent agreement between the lifetimes of \cite{lauer99} and
previous experimental data for Ar and Xe, whereas for Kr a difference by about
two orders of magnitude exists.

\cite{lauer99} also computed $\tau_{ns}$ in different approximations according
to the schemes (\ref{eq:lives_m},\ref{eq:lives_s}):%

\begin{equation}%
\begin{array}
[c]{cccc}%
\mathbf{main} &
\begin{array}
[c]{cc}
& 4s^{1}4p^{6}%
\end{array}
& \dashrightarrow &
\begin{array}
[c]{cc}%
4s^{2}4p^{5} &
\end{array}
\\
& \updownarrow &  & \updownarrow\\
\mathbf{corr} & \left.
\begin{array}
[c]{cr}%
3d^{9}4s^{2}4p^{5}\left\{  p/f\right\}  & -1.6\\
4s^{1}4p^{5}\{p\} & -0.1
\end{array}
\right\}  &  & \left\{
\begin{array}
[c]{rc}%
-1.9 & 3d^{9}4s^{1}4p^{6}\left\{  p/f\right\} \\
-20.1 & 4s^{1}4p^{5}\left\{  s/d\right\}
\end{array}
\right.
\end{array}
\label{eq:lives_m}%
\end{equation}
where the designations correspond to scheme~(\ref{eq:Ar-scheme}). A
calculation taking into account the pathway \textbf{main} (\ref{eq:lives_m}) only results in
too short lifetimes (see line PF in Table~\ref{tab:lives}). Destructive
interference between pathways \textbf{main} (\ref{eq:lives_m}) and \textbf{sat} (\ref{eq:lives_s}) strongly decreases the transition moment, increasing the lifetimes (see line PF+DPES in
Table~\ref{tab:lives}). Both results are in accord with findings of
\cite{luyken72,hansen79}.%

\begin{equation}%
\begin{array}
[c]{cccc}%
\mathbf{sat} &
\begin{array}
[c]{cc}
& 4s^{2}4p^{4}4d
\end{array}
& \dashrightarrow &
\begin{array}
[c]{cc}%
4s^{2}4p^{5} &
\end{array}
\\
& \updownarrow &  & \updownarrow\\
\mathbf{corr} & \left.
\begin{array}
[c]{cr}%
3d^{9}4s^{2}4p^{5}\left\{  p/f\right\}  & -0.6\\
4s^{1}4p^{5}\{p\} & -0.1
\end{array}
\right\}  &  & \left\{
\begin{array}
[c]{rc}%
-0.9 & 3d^{9}4s^{2}4p^{4}4d\left\{  p/f\right\} \\
-17.2 & 4s^{2}4p^{3}4d\left\{  s/d\right\} \\
+4.0 & 4s^{2}4p^{4}\left\{  p/f\right\} \\
-7.6 & 4s^{1}4p^{5}4d
\end{array}
\right.
\end{array}
\label{eq:lives_s}%
\end{equation}

In addition to {interactions and effects considered by} those authors, \cite{lauer99} {considered} the screening
of the Coulomb interaction. Correlations contributing strongly to the main and
satellite transitions are shown in schemes (\ref{eq:lives_m}) and
(\ref{eq:lives_s}). The values near each correlation are the contributions in per
cent to the transition amplitude $\langle$\textbf{main}$\rangle$ and $\langle
$\textbf{sat}$\rangle$ in schemes (\ref{eq:lives_m}) and (\ref{eq:lives_s}),
respectively. {Corresponding} lifetimes computed with taking into account all pathways are listed in Table~\ref{tab:lives} in line CIPF. One can see that the Coulomb screening significantly improved the agreement between theory and experiment for Xe.

A long-scale calculation of lifetimes for the Ar~II main and manyfold
satellite levels has been made by \cite{hibbert87,hibbert94} who applied an
\emph{ab initio} CI approximation to this problem. For Xe II, lifetimes of
some satellite levels have been computed by \cite{ehresmann10}.

In conclusion, lifetimes of the subvalence $ns$ vacancies provide a very
sensitive test of many-electron theories because of {frequently occurring} strong {destructive} interference. Albeit the physics {behind the radiationless decay of the subvalence vacancies} seems clear, up to now there are no calculations providing a
good agreement between theory and experiment for $\tau_{ns}$ of all rare gases.

\subsubsection{Cross section calculations}

\label{sec:corrsat_PICS}

When it became clear that the threshold photoionization of the subvalence $Rg$
shells is strongly influenced by intershell correlations, numerous theoretical
and experimental investigations of this process appeared (see
Section~\ref{sec:PICS} and Figs.~\ref{fig:Ar3sPICS_Theor}%
,\ref{fig:Ar3sPICS_Exp}) which sometimes gave different results. A precise
measurement of the Ar $\sigma_{3s}$ \citep{sukhorukov92,mobus93} revealed a
rich resonance structure of $\sigma_{3s}$ near the $3s$ threshold. The cross
sections agreed well with previous data of \cite{houlgate76} which
fortuitously were located in the peaks of resonances but in average were lower
than the computed $\sigma_{3s}$ (see upper left panel of
Fig.~\ref{fig:PIFS_Ar3s_3lay}).

\begin{figure}[ptb]
\begin{minipage}[t]{0.48\textwidth}
\begin{center}
\includegraphics[width=\textwidth]{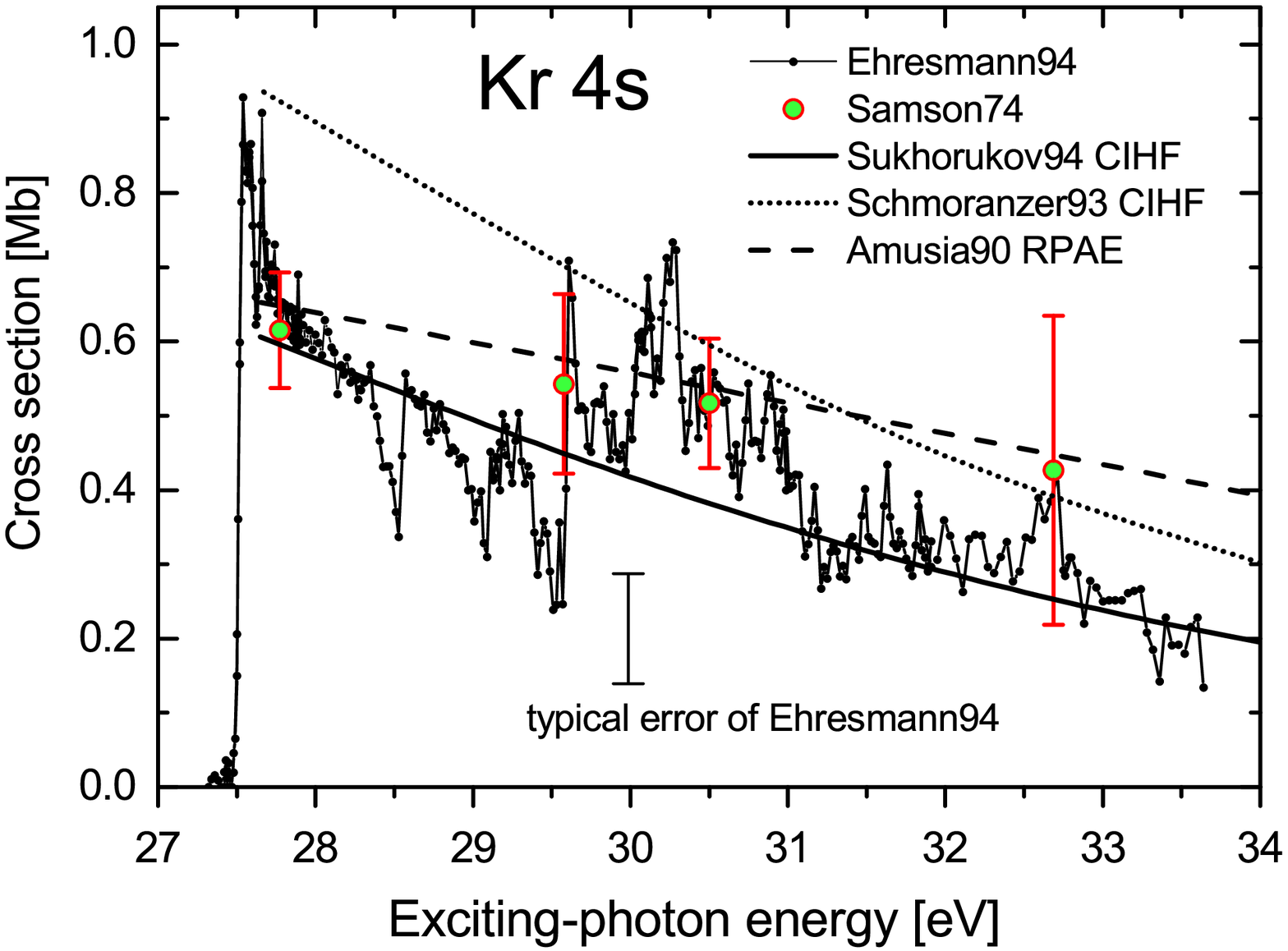}
\caption{Absolute Kr $\sigma_{4s}$ photoionization cross section in the near-threshold region. Experiment from \citep{samson74,ehresmann94a}. Theory from \citep{amusia90a,schmoranzer93,sukhorukov94}. Adapted from
\citep{ehresmann94a}.}\label{fig:Ehresmann94_Kr_4smain_PICS}\end{center}
\end{minipage} \hfill\begin{minipage}[t]{0.48\textwidth}
\begin{center}
\includegraphics[width=\textwidth]{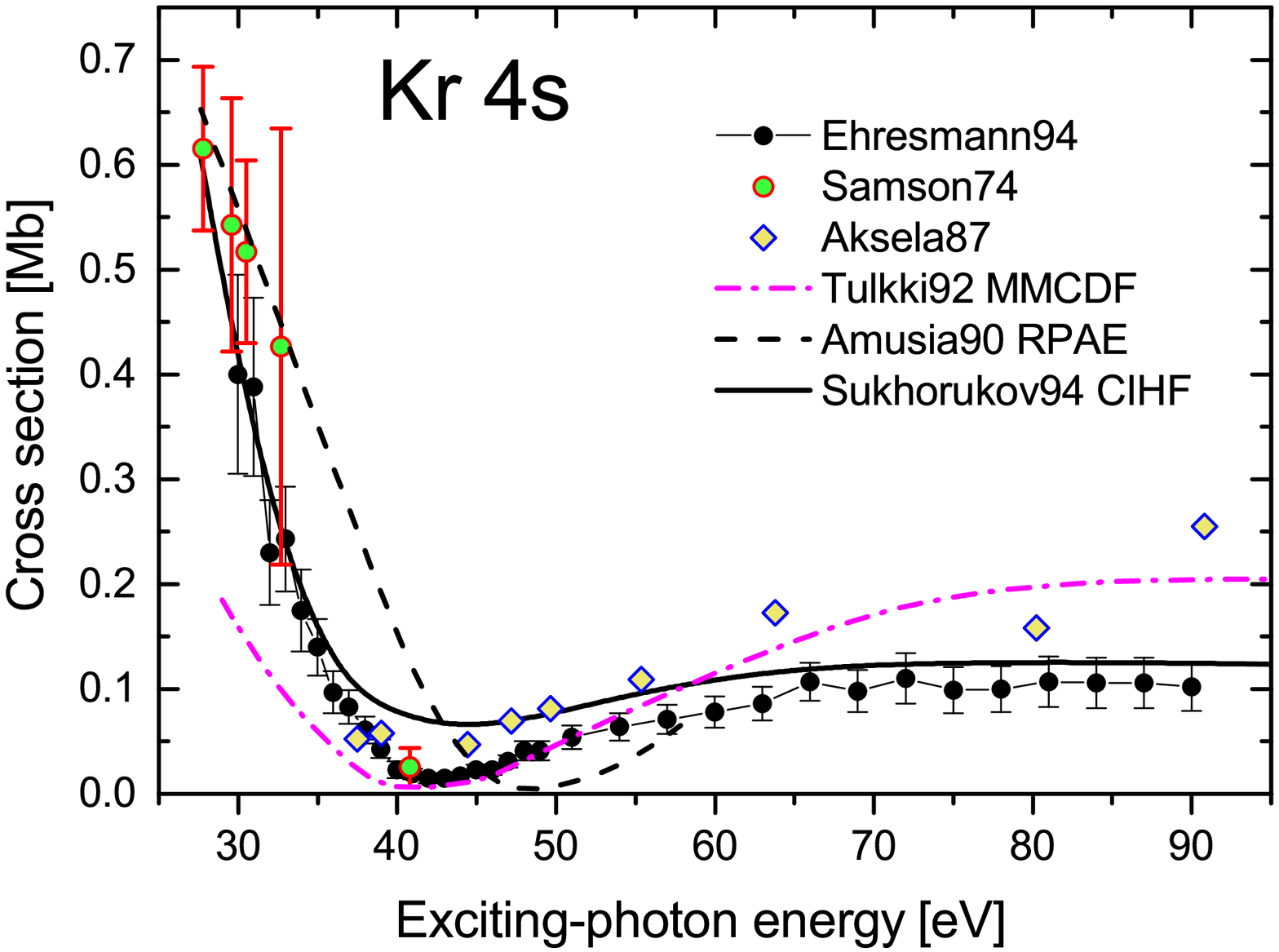}
\caption{Absolute Kr $\sigma_{4s}$ photoionization cross section in the extended region. Experiment from \citep{samson74,aksela87,ehresmann94a}. Theory from \citep{amusia90a,tulkki92,sukhorukov94}. Adapted from
\citep{ehresmann94a}.}\label{fig:Ehresmann94_Kr_4s_long_PICS}\end{center}
\end{minipage}
\end{figure}

A similar situation has been found by \cite{ehresmann94a} in the case of the
Kr $\sigma_{4s}$ (see Fig.~\ref{fig:Ehresmann94_Kr_4smain_PICS}). One can see
that the values of \cite{samson74} lie on the peaks of resonances revealed in
\citep{ehresmann94a} and in average are lower than those computed by
\cite{schmoranzer93}. In the high exciting-photon energy region, $\sigma_{4s}$
of \cite{ehresmann94a} turned out lower than the measurement of
\cite{aksela87} and the calculation of \cite{tulkki92} due to unknown reasons
(see Fig.~\ref{fig:Ehresmann94_Kr_4s_long_PICS}).

\begin{table}[ptb]
\caption{Correlation corrections for the screening coefficient $\chi
(np^{5}\varepsilon^{\prime}d;ns^{1}\varepsilon p)$, in per cent and with
opposite sign to the matrix element $\left\langle ns^{1}\varepsilon
p~^{1}P\left\vert H^{ee}\right\vert np^{5}\varepsilon^{\prime}d~^{1}%
P\right\rangle $ ($n=4$ and $n=5$ for Kr and Xe, respectively.) $^{a)}$}%
\label{tab:KrXeRPAE}%
\vspace{2ex}
\begin{tabular}
[c]{rrrr}\hline\hline
Intermediate & Correction & Intermediate & Correction\\
configuration &  & configuration & \\\hline
$4s^{1}4p^{4}\varepsilon p\varepsilon^{\prime}d\{d\}$ & 7.5 & $5s^{1}%
5p^{4}\varepsilon p\varepsilon^{\prime}d\{d\}$ & 5.9\\
$4s^{1}4p^{5}\varepsilon^{\prime}d\{d\}$ & 6.3 & $5s^{1}5p^{5}\varepsilon
^{\prime}d\{d\}$ & 5.8\\
$3d^{9}4p^{5}\varepsilon p\{f\}$ & 1.0 & $4d^{9}5p^{5}\varepsilon p\{f\}$ &
2.5\\
$3d^{9}4s^{1}\varepsilon^{\prime}d\{f\}$ & 0.0 & $4d^{9}5s^{1}\varepsilon
^{\prime}d\{f\}$ & 1.7\\
$3d^{9}4s^{1}4p^{5}\varepsilon p\varepsilon^{\prime}d\{p\}$ & 0.4 &
$4d^{9}5s^{1}5p^{5}\varepsilon p\varepsilon^{\prime}d\{p\}$ & 0.6\\
$\{f\}$ & 2.7 & $\{f\}$ & 5.5\\
$4s^{1}4p^{5}\{s\}\{d\}$ & 3.6 & $5s^{1}5p^{5}\{s\}\{d\}$ & 2.7\\
$\{p\}\{p\}$ & 0.3 & $\{p\}\{p\}$ & 0.4\\
$\{p\}\{f\}$ & 0.4 & $\{p\}\{f\}$ & 0.5\\
$\{d\}\{d\}$ & 2.1 & $\{d\}\{d\}$ & 1.7\\
$\{d\}\{g\}$ & 0.3 & $\{d\}\{g\}$ & 0.4\\
other & 2.4 & other & 3.1\\
Sum & 30.6 & Sum & 27.0\\\hline\hline
\end{tabular}
\par
\vspace{2ex} {\footnotesize $^{a)}$ The values of $\varepsilon$ and
$\varepsilon^{\prime}$ are equal to 0.01 Ry and 1.0 Ry, respectively.}
\par
{\footnotesize $^{b)}$ Summation and integration over the channel AOs in
braces were performed.}\end{table}

\cite{sukhorukov94} computed the screening of the Coulomb interaction
determining the intershell correlations. A typical result of this calculation
is listed in Table~\ref{tab:KrXeRPAE} for Kr (second column). In this
column, the partial decreasing coefficients $\chi(4p^{5}\varepsilon^{\prime
}d;4s^{1}\varepsilon p)$ for the matrix element $\left\langle 4s^{1}%
\varepsilon p~^{1}P\left\vert H^{ee}\right\vert 4p^{5}\varepsilon^{\prime
}d~^{1}P\right\rangle $ with energies of the continuum electrons
$\varepsilon=0.01$ Ry and $\varepsilon^{\prime}=1.0$ Ry are listed. For other
energies, the partial decreasing coefficients are slightly different but the
resulting coefficient {changes only slightly} \citep{sukhorukov94}.

The Coulomb screening for the intershell correlation in Xe had been computed
by \cite{lagutin96}. This result is listed in the fourth column of
Table~\ref{tab:KrXeRPAE}. One can see that partial channels connected with
electrons having high orbital angular momentum are larger in the case of Xe
than in Kr. However, the resulting decreasing coefficient changes {only slightly}, in contrast to the case shown in Table~\ref{tab:KrXeDPES} for the DPES.

Fig.~\ref{fig:Ehresmann94_Kr_4smain_PICS} shows that the photoionization cross
section $\sigma_{4s}$ of Kr computed with the `screened' intershell
interaction (or correlational decrease) fits very well the `averaged'
near-threshold cross section. The far region of $\sigma_{4s}$ exhibits also
good agreement between the theory of \cite{sukhorukov94} and the experiment of
\cite{ehresmann94a} (see Fig.~\ref{fig:Ehresmann94_Kr_4s_long_PICS}). In the
region {close to} the Cooper minimum, the $\sigma_{4s}$ cross section computed within the non-relativistic CIHF approach is too large in comparison with
experiment. Good agreement between measured and computed $\sigma_{4s}$ in the
whole energy interval has been achieved by \cite{lagutin98} after taking into
account relativistic effects within the configuration-interaction PF (CIPF)
approach (see also Fig.~\ref{fig:5sRESkr} and comments there). The CIPF
approach provides also good overall agreement between theoretical and
experimental $\sigma_{5s}$ of Xe (see Fig.~\ref{fig:5sRESxe}).

\subsection{Extended investigations of the resonance structure in PICSs}

The $4p^{4}n\ell n^{\prime}\ell^{\prime}$ doubly excited states have been
included in the calculation of the cross sections for excitation of the main
$4s^{1}4p^{6}$ and satellite $4p^{4}n\ell$ levels of Kr by \cite{lagutin94}.
This calculation allowed to describe qualitatively the rich resonance
structure in PICSs of the main line $\sigma_{4s}$ and in some satellite levels
of Kr~II observed by \cite{schmoranzer93} at a resolution (bandwidth of the
exiting radiation) of $\Delta E=150$~meV. However, the integrated cross
sections for the satellite production computed by \cite{lagutin94} exceeded
the measured ones by a factor of 4.

In order to clarify the mechanisms determining main line and satellite
production, a series of combined experimental and theoretical investigations
have been performed. Experimentally, the PIFS method at high resolution was
used with an extension allowing the simultaneous measurement of both the
partial photoionization (PICS) and the photoabsorption (PACS) cross sections.
The calculation was carried out applying the CIPF technique.

\begin{figure}[ptb]
\begin{minipage}[t]{0.48\textwidth}
\begin{center}
\includegraphics[width=\textwidth]{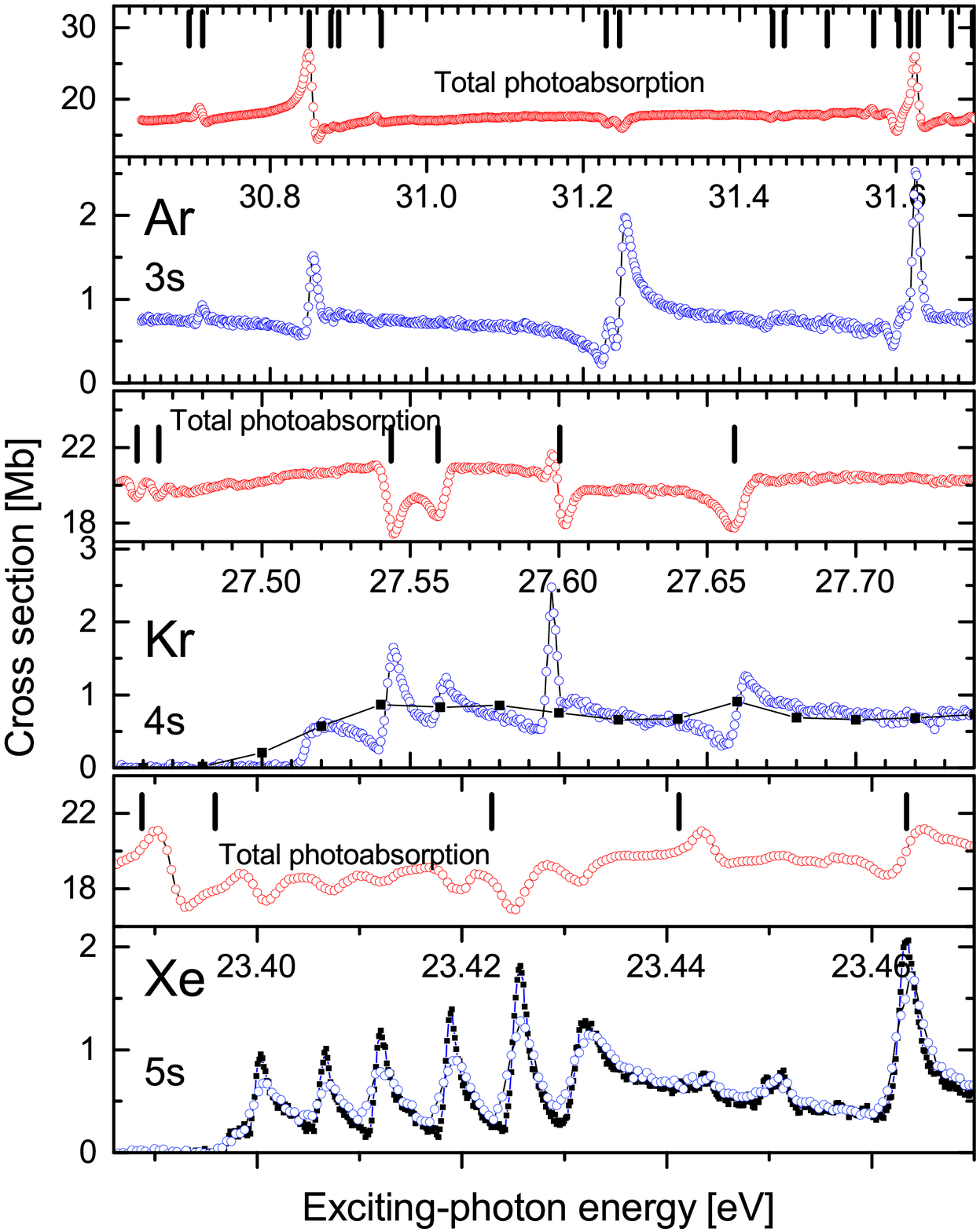}
\caption{Experimental PACS and $\sigma_{ns}$ PICSs of
Ar, Kr and Xe atoms on absolute scale. The energy positions of resonances measured by
\cite{madden69} for Ar and by \cite{codling72} for Kr and Xe are drawn in as
bars. \textbf{Ar:} data from \citep{lagutin99} at 4.8 meV resolution. \textbf{Kr:} data from
\citep{schmoranzer01a} (open circles: 3 meV resolution) and data from
\citep{ehresmann94a} (solid squares: 25 meV resolution). \textbf{Xe:} data from
\citep{schmoranzer01a} (open circles: the total and partial $5s$ PICS at a 2~meV resolution;
solid squares: the $5s$ PICS at a 1.0~meV resolution); Adapted from
\citep{schmoranzer01a}.}\label{fig:High_res_Rg}\end{center}
\end{minipage} \hfill\begin{minipage}[t]{0.48\textwidth}
\begin{center}
\includegraphics[width=\textwidth]{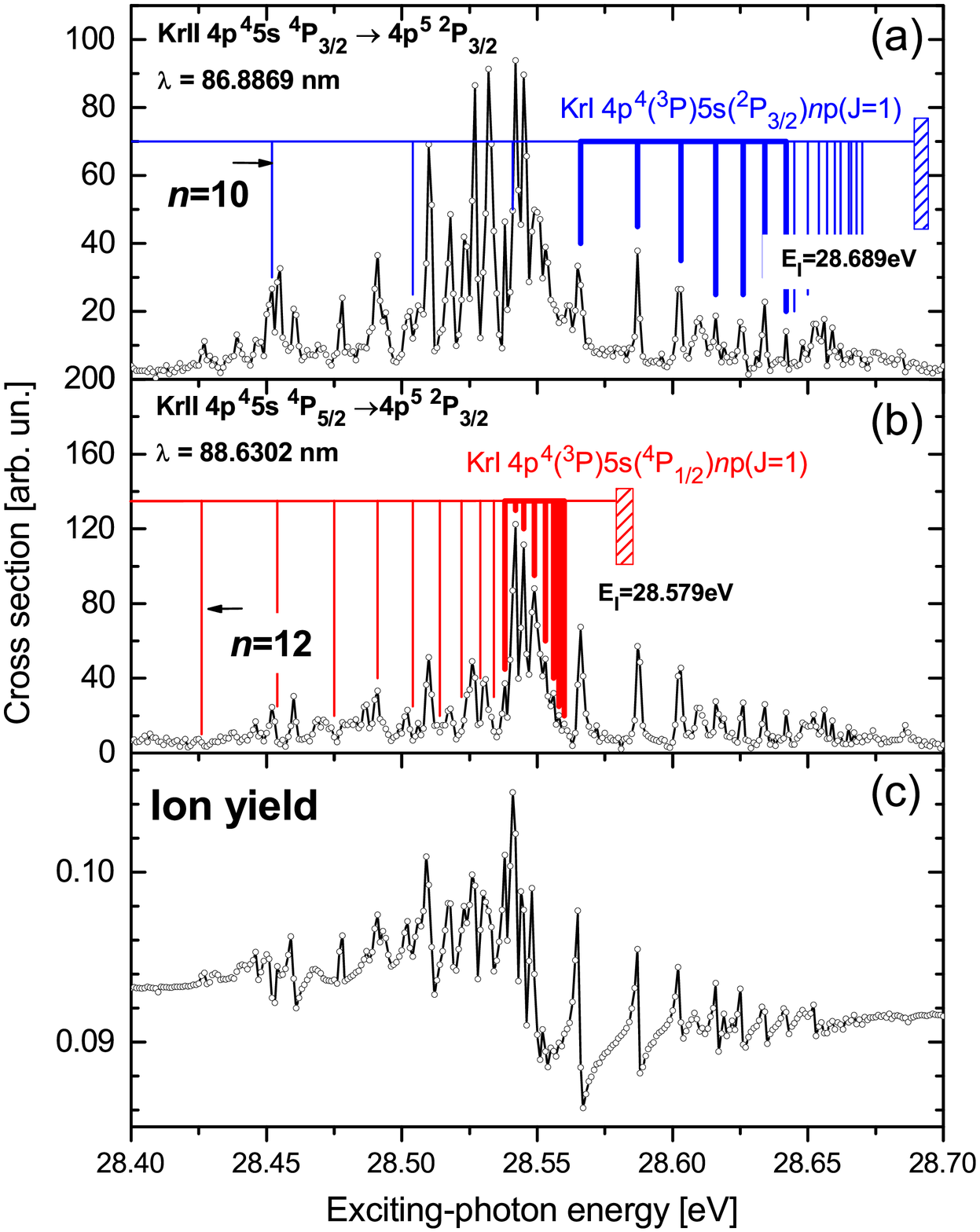}
\caption{Relative PICS of KrII (a) and (b) and photoion yields (c) in the
exciting-photon energy range between 28.4 eV and 28.7 eV. Peaks obviously
belonging to one Rydberg series were marked by bold solid bars and used to
determine the quantum numbers of the states of the series and their quantum
defect. The energies of the peaks marked by thin solid bars were calculated
using the Rydberg formula (\ref{eq:Ry}) and the determined quantum numbers and quantum defect. The energetic thresholds to which the Rydberg series converge are indicated by
the left boundary of the hatched boxes \citep{striganov68}. The two PICS are
on the same relative intensity scale. Adapted from
\citep{ehresmann04}.}\label{fig:High_res_Kr}\end{center}
\end{minipage}
\end{figure}

\cite{lauer98} measured $\sigma_{3s}$ and the PACS of Ar at a resolution of
$\Delta E=4.8$~meV for exciting-photon energies $30.65\leq\hbar\omega
\leq31.75$~eV exceeding the $3s$ threshold $IP_{3s}=29.24$~eV. Using the same
setup, the near-threshold cross sections for the main $ns$ lines
simultaneously with the PACS have been measured by \cite{schmoranzer01a} at
resolutions of $\Delta E=3$~meV for Kr and $\Delta E=1$~meV for Xe. The
results of these measurements are depicted in Fig.~\ref{fig:High_res_Rg}. One
can see that the high resolution allowed to clearly observe the Fano
profiles both in the PACS and in the $\sigma_{ns}$ PICS which had not been
seen in earlier measurements (cf. $\sigma_{4s}$ PICS of Kr in the middle panel
of Fig.~\ref{fig:High_res_Rg}).

Experimental cross sections of the main $4s$ line and some of the lowest
$4p^{4}(^{3}P)4s~^{2S+1}P_{J}$ satellite levels have been measured at a
resolution of $\Delta E=1.7$~meV by \cite{ehresmann04} in the region of
$28.40\leq\hbar\omega\leq28.70$~eV. The achieved resolution enabled clearly to
observe Rydberg series which were analyzed using the well known formula:%

\begin{equation}
E_{n\ell}=E_{thr}-\frac{\mathrm{Ry}}{\left(  n-\mu_{\ell}\right)  ^{2}}
\label{eq:Ry}%
\end{equation}
connecting the energy of resonance, $E_{n\ell}$, with its principal quantum
number $n$, the quantum defect $\mu_{\ell}$ of the $\ell$-series and the
respective threshold energy $E_{thr}$. Results for two Rydberg $np$ series
converging to the $4p^{4}\left(  ^{3}P\right)  5s~^{4}P_{1/2}$ and
$4p^{4}\left(  ^{3}P\right)  5s~^{2}P_{3/2}$\ thresholds are depicted in
Fig.~\ref{fig:High_res_Kr} together with positions of resonances computed with
equation (\ref{eq:Ry}). Extended measurements in a broader exciting-photon
energy range ($27.80\leq\hbar\omega\leq29.45$~eV) were presented in
\citep{sukhorukov07} together with a refined analysis of the observed Rydberg
series, performed on the basis of the CIPFCP calculation. This analysis for
the $4p^{4}\left(  ^{3}P\right)  5s~^{2}P_{3/2}np$ and $4p^{4}\left(
^{3}P\right)  5s~^{4}P_{1/2}np$ Rydberg series resulted in quantum defects
$\mu_{p}=2.740(60)$ and $\mu_{p}=2.763(72)$, respectively, which is in good
agreement with the value of $\mu_{p}=2.767$ calculated within the CIPFCP
approximation for the electron configuration $4p^{4}\ 5s\ np\ (n\geq10)$.

Cross sections for the main $3s$ line of Ar ($32.50\leq\hbar\omega\leq
33.00$~eV) and for the three lowest satellite levels $3p^{4}(^{3}P)3d~^{4}D$,
$3p^{4}(^{3}P)4s~^{2,4}P$\ ($32.45\leq\hbar\omega\leq33.05$~eV) have been
measured at a resolution of $\Delta E=3.4$~meV by \cite{kammer06}. The Rydberg
series converging to the $3p^{4}(^{3}P)4s~^{2}P_{3/2}$ and $p^{4}%
(^{3}P)4s~^{2}P_{1/2}$ thresholds were used by \cite{kammer06} for calibration
of the exciting-photon energy scale with an estimated accuracy of a few meV.

Measurements of the fluorescence cross sections for the main $5s$ line and
eight satellite lines of Xe have been performed by \cite{ehresmann10} and
consisted of two experiments. One experiment aimed to get the overview
fluorescence cross section in a broad interval ($23\leq\hbar\omega\leq28$~eV)
of the exciting-photon energies at a comparatively low resolution of $\Delta
E=20$~meV. Another experiment was focused on detailed features in a narrow
energy interval ($25.25\leq\hbar\omega\leq25.50$~eV) with extremely high
resolution of $\Delta E=1.8$~meV. The cross sections computed by
\cite{ehresmann10} within the CIPF approach exhibited good agreement with the
experimental results, especially in integrated form.

\subsubsection{Interaction between resonances through autoionizing
continua}

The reason for the large difference between computed and measured integrated
cross sections for satellite production has been clarified by
\cite{demekhin05} and \cite{petrov05}. These authors applied the following
scheme to reveal the physical mechanisms determining the main line
and satellite production in the $4s$ photoionization of Kr:%

\begin{equation}%
\begin{array}
[c]{ccc}%
\begin{array}
[c]{cc}%
\left(  \mathbf{0}\right)  & 4s^{2}4p^{6}\\
& ground~state
\end{array}
& \dashrightarrow & \left\{
\begin{array}
[c]{cc}%
4s^{2}4p^{5}\varepsilon\left(  s/d\right)  & \left(  4\mathbf{p}\right) \\
4s^{1}4p^{6}\left(  n/\varepsilon\right)  p & \left(  4\mathbf{s}\right)
\end{array}
\right. \\
\updownarrow &  & \updownarrow\\
\left.
\begin{array}
[c]{cc}%
\left(  \mathbf{a}\right)  & 4s^{2}4p^{4}\left\{  s/d\right\}  ~\left(
n/\varepsilon\right)  (s/d)\\
\left(  \mathbf{b}\right)  & 4s^{1}4p^{5}n(s/d)~\left(  n^{\prime}%
/\varepsilon\right)  (p/f)\\
\left(  \mathbf{c}\right)  & 4s^{2}4p^{4}\left\{  p/f\right\}  ~\left(
n^{\prime}/\varepsilon\right)  (p/f)
\end{array}
\right\}  &  & \left\{
\begin{array}
[c]{cc}%
4s^{2}4p^{4}n(s/d)~\varepsilon(p/f) & \left(  \mathbf{sat}\right) \\
4s^{2}4p^{4}n(s/d)~n^{\prime}(p/f) & \left(  \mathbf{2ex}\right)
\end{array}
\right. \\
\mathbf{ISCI} &  & \mathbf{FISCI}%
\end{array}
\label{eq:Kr-scheme}%
\end{equation}
where the designations are the same as in (\ref{eq:Ar-scheme}): the
dashed and double-side arrows denote electric-dipole and Coulomb interaction,
respectively; ISCI and FISCI denote the initial and final state configuration
interaction, respectively; summation and integration are performed over the
channels shown in braces, etc. Besides the correlations indicated in
(\ref{eq:Kr-scheme}), all Coulomb interactions were screened as described in
section~\ref{sec:Coulomb_screening}. Interaction between the FISCI continuum
channels entering the scheme (\ref{eq:Kr-scheme}) was included via the
K-matrix technique.

Scheme (\ref{eq:Kr-scheme}) describes the following many-electron correlations:

\begin{description}
\item[(i)] \emph{intrashell correlations} are described by pathway
$\left\langle \mathbf{0\longleftrightarrow a\dashrightarrow}4\mathbf{p}%
\right\rangle $ and change the $\left\langle \mathbf{0\dashrightarrow
}4\mathbf{p}\right\rangle $ transition amplitude describing the
photoionization of the $4p$ shell via accounting for the correlative motion of
electrons inside the same $4p$ shell.

\item[(ii)] \emph{intershell correlations} change strongly the $\left\langle
\mathbf{0\dashrightarrow}4\mathbf{s}\right\rangle $ transition amplitude via
the pathway $\left\langle \mathbf{0\dashrightarrow}%
4\mathbf{p\longleftrightarrow}4\mathbf{s}\right\rangle $, i.e. photoionization
of the $4s$ shell, via the transition through the $4p$ shell (so-called
`borrow-intensity' mechanism).

\item[(iii)] \emph{dipole polarization of electron shells}. This correlation
is described by the interaction between the configuration $4s^{1}4p^{6}$ and
the channel $4s^{2}4p^{4}n/\varepsilon(s/d)$. The influence of such kind of
interaction was already discussed for Xe (see scheme~(\ref{eq:5sXe-abs}))
\end{description}

Cross sections for the satellite production borrow intensity mainly from the
$4p$ photoionization $\left\langle \mathbf{0\dashrightarrow}4\mathbf{p}%
\right\rangle $ via the $\left\langle \mathbf{0\dashrightarrow}%
4\mathbf{p\longleftrightarrow sat}\right\rangle $, $\left\langle
\mathbf{0\mathbf{\dashrightarrow}}4\mathbf{s\longleftrightarrow sat}%
\right\rangle $ and $\left\langle \mathbf{0\mathbf{\dashrightarrow}%
4\mathbf{p}\longleftrightarrow}4\mathbf{s\longleftrightarrow sat}\right\rangle
$ pathways. The influence of the doubly-excited states on the direct $4s$ main
and satellite photoionization is described by the $\left\langle
4\mathbf{s\longleftrightarrow2ex}\right\rangle $ and $\left\langle
\mathbf{sat\longleftrightarrow2ex}\right\rangle $ interactions, respectively,
and manifests itself in a strong modulation of the photoionization cross sections.

The calculation performed by \cite{petrov05} clarified the many-electron
correlations providing the correct strength of the `borrow-intensity'
mechanism. It turned out that the integrated intensity of satellites is
strongly dependent on two factors: (i) relaxation of the $4s$ and $4p$ atomic
orbitals and (ii) interaction between resonances through the autoionizing
continua. Including these effects substantially decreases the computed
fluorescence cross sections. For instance, the relaxation decreases the
computed cross sections integrated within the exciting-photon energy region
(27.48--28.68~eV) by factors of 1.33, 2.44, and 2.27 for the $4p^{4}%
(^{3}P)5s~^{4}P_{J}$ $J=5/2$, $J=3/2$, and $J=1/2$ levels, respectively.
Taking into account the interaction between continua results in an additional
decrease of the integrated intensity by factors of 3.00, 2.27, and 3.30, respectively. The
resulting decreasing factors (4.0, 5.5, and 7.5) provide agreement between
computed and measured cross sections within 10~\%. Similar results have been
obtained by \cite{demekhin05} in the energy interval (27.90--29.93~eV).

\begin{figure}[ptb]
\begin{center}
\includegraphics[width=0.9\textwidth]{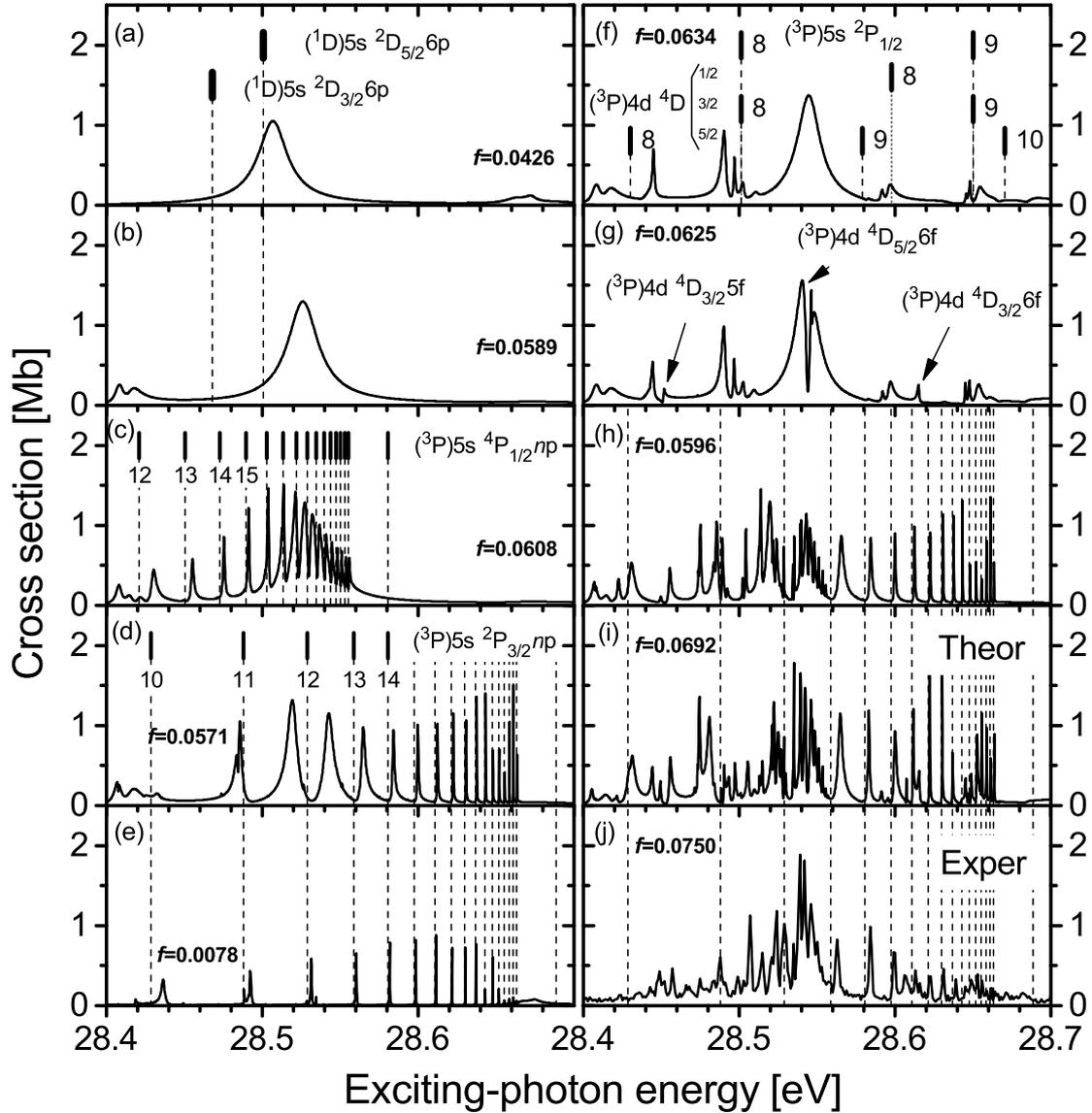}
\end{center}
\caption{Cross sections for the production of the $4p^{4}(^{3}P)5s~^{4}P$
satellite computed in different approximations in the region around the
$4p^{4}(^{1}D)5s~^{2}D_{J}~6p$ resonances. The experimental cross section is
shown in panel (j) for comparison. The integrated cross section
\emph{\textbf{f}} [Mb$\times$eV] is shown in each panel. $(a)$ Resonance
$(^{1}D)5s~^{2}D_{5/2}~6p$ is included, $(b)=(a)+(^{1}D)5s~^{2}D_{3/2}~6p$
resonance, $(c)=(b)+(^{3}P)5s~^{4}P_{1/2}~np$ series, $(d)=(b)+(^{3}%
P)5s~^{2}P_{3/2}~np$ series, $(e)\ (^{3}P)5s~^{2}P_{3/2}\ np$ series is
included only, $(f)=(b)+$ low-number $np$ series, $(g)=(f)+nf$ series,
$(h)=(i)-nf$ series,$(i)=(g)+(^{3}P)5s~^{4}P_{1/2},^{2}P_{3/2}~np$ series. All
computed cross sections are shown with the natural broadening only. Figure
adapted from \citep{sukhorukov07}.}%
\label{fig:Sukhorukov_07_Kr}%
\end{figure}

In detail, the interaction between different resonances including several
Rydberg series has been investigated by \cite{sukhorukov07} applying (\ref{eq:Kr-scheme}). In their calculations, the authors removed all
resonances from the {exciting-photon energy} region $28.40~\textrm{eV}\leq\hbar\omega\leq28.70~\textrm{eV}$ and included them in the
following step-by-step. The results of the calculations in each step are
depicted in Fig.~\ref{fig:Sukhorukov_07_Kr}. In order to monitor the changes
in the oscillator strengths, the cross sections \emph{\textbf{f}} integrated
over the interval $28.40~\textrm{eV}\leq\hbar\omega\leq28.70~\textrm{eV}$ and having the
dimension of $[\mathrm{Mb\times eV}]$ are listed in each panel of this figure.

The following approximations in the calculation of the cross sections were used:

\begin{description}
\item[(a)] The $(^{1}D)5s\ ^{2}D_{5/2}\ 6p_{3/2}$ resonance only was included
in the region $\Delta\hbar\omega=28.40~\div28.70$~eV.
Fig.~\ref{fig:Sukhorukov_07_Kr}(a) shows that this broad resonance promotes
the intensity from the $4\mathrm{p}\rightarrow\varepsilon\ell$%
\ photoabsorption via the $0\dashrightarrow4p\rightarrow(^{1}D)5s\ ^{2}%
D_{5/2}\ 6p_{3/2}$ transition providing practically all the intensity in the
investigated region and is, therefore, called a \emph{promoter}. The small
intensity at 28.67~eV originates from the resonances which are outside the
investigated region before solving the secular equation.

\item[(b)] In addition to the approach (a), the $(^{1}D)5s\ ^{2}%
D_{3/2}\ 6p_{3/2,1/2}$ resonances were included. These resonances do not
contribute much to the oscillator strength: it is increased by 38\%. The
$(^{1}D)5s\ ^{2}D_{3/2}\ 6p_{3/2,1/2}$ resonances shift the \emph{promoter}
and redistribute intensity within the investigated region.

\item[(c)] In addition to the approach (b), the $\left(  ^{3}P\right)
5s~^{4}P_{1/2}np$ Rydberg series was included. The interaction between the
$\left(  ^{3}P\right)  5s~^{4}P_{1/2}np$ Rydberg series and the
\emph{promoter} results in quite a regular structure. The integrated cross
section remains practically unchanged. On an enlarged scale (not shown in the
figure), one can see the well known `sign-reversal' behaviour of the $q$
parameter \citep{connerade88} around the resonance with an energy of 28.532~eV.
This resonance which can be {considered as an} \emph{intruder resonance} or as a `remaining \emph{promoter}' increases the amount of resonances connected
with the $\left(  ^{3}P\right)  5s~^{4}P_{1/2}np$ Rydberg states by one.

\item[(d)] In addition to the approach (b), the $\left(  ^{3}P\right)
5s~^{2}P_{3/2}np$ Rydberg series was included. The `sign-reversal' behaviour
of the $q$ parameter can be seen around the \emph{intruder resonance} with the
energy of 28.542~eV (between the Rydberg resonances $n=12$ and $n=13$) which can
be interpreted as the `remaining \emph{promoter}'.

\item[(e)] In order to illustrate the `own' intensity of the $\left(
^{3}P\right)  5s~^{4}P_{1/2}np$ and $\left(  ^{3}P\right)  5s~^{2}P_{3/2}np$
Rydberg series, they were included in the calculation excluding the
\emph{promoter}. Fig.~\ref{fig:Sukhorukov_07_Kr}(e) illustrates that (i) the
`own' intensity is 7 times smaller than the promoted intensity; (ii) the major
part of the `own' intensity stems from the $\left(  ^{3}P\right)
5s~^{2}P_{3/2}np$ Rydberg series, whereas the $\left(  ^{3}P\right)
5s~^{4}P_{1/2}np$ Rydberg series is not seen within the ordinate scale of
Fig.~\ref{fig:Sukhorukov_07_Kr}(e).

\item[(f)] In addition to (b), the `low-number' $np$ Rydberg
series were included. This approach shows the role of the `low-number'
($n=8,9,10$)\ Rydberg states stemming from the $\left(  ^{3}P\right)
5s~^{2}P_{1/2}$ and $\left(  ^{3}P\right)  4d~^{4}D_{J}$ levels in the
formation of the investigated structure. These states increase the oscillator
strength by 8\% and yield some features around 28.44~eV, 28.50~eV, 28.59~eV,
and 28.65~eV. In the experimental spectrum depicted in panel (j), these
features are clearly seen as a bell-shaped `background' below the resolved
resonances, e.g., at 28.61~eV and 28.65~eV.

\item[(g)] In addition to (f), the $nf$ Rydberg series were
included. Including the $nf$ series results in two small resonances at
28.452~eV and 28.615~eV and the pronounced profiles of the resonances stemming
from the $\left(  ^{3}P\right)  4d~^{4}D_{5/2}6f_{5/2,7/2}$ doubly-excited
states. In order to see whether it is possible to observe these features
experimentally, the calculation (h) was performed.

\item[(h)] In this approach, the $nf$ Rydberg series were excluded from the
most precise approach (i). The features connected with the $\left(
^{3}P\right)  4d~^{4}D_{3/2}5f$\ and $\left(  ^{3}P\right)  4d~^{4}D_{3/2}6f$
doubly-excited states are difficult to identify in the experimental spectrum
(j), whereas the pronounced profile stemming from the $\left(  ^{3}P\right)
4d~^{4}D_{5/2}6f_{5/2,7/2}$ doubly-excited states can be seen there.
\end{description}

In conclusion, the combined experimental and theoretical investigations shed
light on the main mechanisms determining the photoionization of the rare gas
atoms in the vicinity of the subvalence $ns$ threshold. A correct
interpretation of the observed photoionization cross section requires to take
into account intershell and intrashell correlations, dipole polarization of
the $np$ shell by the $ns$ vacancy, Coulomb screening, relaxation of electron
shells upon the $ns$ vacancy, and interchannel interaction. In addition,
near-threshold photoionization of the main $ns$ and satellite levels requires
to take into account numerous doubly excited states. The inclusion all of
these interactions in the calculation allows to describe many of the observed
spectral features, including integrated cross sections and line shapes in many
cases.

\subsubsection{Photoionization of rare-gas atoms between the $ns$ and $np$
thresholds}

The amount of pronounced resonances observed in the $\sigma_{np}$ cross
sections of $Rg$ between the valence $np$ and subvalence $ns$ thresholds is
much smaller than that observed in the main $ns$ and $np^{4}n^{\prime}\ell$
satellite photoionization cross sections described in the previous section.
Autoionizing resonances lying between the $np$ and $ns$ thresholds in $Rg$
represent a very important phenomenon allowing to precisely study different
types of many-electron correlations. The interaction between $ns^{1}np^{6}mp$
and $ns^{2}np^{5}\varepsilon\left(  s/d\right)  $ continua results in well
known interference profiles described by \cite{fano61} (see
equation~(\ref{eq:cross_shore})). The parameters of those autoionizing
resonances are very sensitive to the approximations used in calculations and
allow to draw conclusions about the adequacy of the approach.

Autoionizing resonances in the total photoionization cross sections of He, Ne,
and Ar have `Beutler-Fano' shape, as it has been observed by
\cite{madden63} who used synchrotron radiation as excitation source and
presented a densitometer trace for the He photoionization. Later on, the
observation of the $ns\dashrightarrow mp$ resonances has been revisited
several times. \cite{samson63} presented photoionization cross sections
measured for Ar and Kr using a high-voltage repetitive spark discharge in Ar
as excitation source.

Refined measurements of the $ns\dashrightarrow mp$ resonances have been
performed by \cite{codling67} for Ne, by \cite{madden69} for Ar (see
Fig.~\ref{fig:Madden69_3p}), and by \cite{ederer71,codling72} for Kr and Xe.
The measurements of Codling and Madden contain the above-$ns$-threshold
region, too (see Fig.~\ref{fig:Madden69_2ex}). The `window' type of the main
autoionizing resonances observed in Ar, Kr, and Xe indicate that the imaginary
part prevails in the transition amplitude.

The first calculation of the $ns\dashrightarrow mp$ resonances line shapes has
been performed by \cite{kelly73} for Ar within the many-body perturbation
theory. In this calculation, they took into account the intrashell
correlations and neglected the DPES. Nevertheless, they obtained a fairly good
agreement between the calculated and the measured total $3p$ PICS of
\cite{madden69}. It can be assumed that the used approximation would result in
a poor description of the $\sigma_{3s}$ PICS above the Cooper minimum (see
Fig.~\ref{fig:RPAE_3s_Ar}) and of the correlation satellites. Thus, the
calculation of \cite{kelly73} can serve as an example that an approximate
calculation can give a good description of a \emph{part} of the experiment only.

\cite{burke75} applied the R-matrix theory to describe the $ns\dashrightarrow
mp$ autoionizing resonances and their parameters $\Gamma_{m}$, $q_{m}$, and
$\rho_{m}^{2}$ entering equation (\ref{eq:cross_shore}). They found that
taking into account many-electron correlations substantially changes the
resonance parameters, even changing the sign of the quality parameter $q$.
Good agreement between measured and computed values of resonance parameters
has been obtained. A large number of many-electron correlations included in
the calculation took into account to a large extent the DPES and the Coulomb
screening. However, a detailed analysis of the influence of each correlation
on the computed quantities is absent in this paper. Also \cite{burke75} did
not discuss the shift of the computed cross section $\sigma_{3s}$ in
comparison with the measured one (see Fig.~\ref{fig:Ar3sPICS_Theor}). It can
possibly be connected with the use of a semiempirical core polarization
potential in the AO calculation of these authors or with the {used approximation accounting for} DPES.

\cite{wijesundera89} applied many-body perturbation theory to compute the
influence of single-electron excitations ($3s\dashrightarrow mp$) and of
doubly excited states on the photoionization of the $3p$ and $3s$ shells
and of the $3s^{2}3p^{4}n\ell$ satellites. They were the first who showed the
importance of including doubly excited states in the calculation of the
near-$3s$-threshold photoionization of Ar. However, the computed line shapes
of the $3s\dashrightarrow mp$ resonances in the $3p$ photoionization did not
reproduce the experimental data well. Supposedly, this is due to the fact that
the authors did not fully take into account Coulomb screening (at least,
the configuration $3s^{1}3p^{4}\{d\}\{d\}$, which largely determines the
Coulomb screening (see, e.g., Table~\ref{tab:KrXeDPES}), is absent in the list
of electron configurations included in the calculation).

Photoelectron spectroscopy provided a new possibility to measure the shape
of the autoionizing $ns\dashrightarrow mp$ resonances in the partial PICS
specified by the total angular momenta of the core $np_{j}$ vacancy.
\cite{samson68} were the first who presented {experimental} partial PICS $\sigma_{np_{j}}$ for Ar, Kr, and Xe. The resolution of the photoelectron spectrometer in
this first experiment was rather low but, nevertheless, allowed the authors to
{observe} that some resonances have a `peak' shape in the $\sigma_{np_{1/2}}$ {cross section} and
a `window' shape in the $\sigma_{np_{3/2}}$ {one} (or vice versa). Thus, these resonances, being strong in $\sigma_{np_{1/2}}$ and $\sigma_{np_{3/2}}$,
cancel each other and result in only a weak resonance in the total
photoionization cross section. Later on, autoionizing resonances of such kind
have been called `mirroring' resonances.

The angular distributions of photoelectrons in the region of the
$3s\dashrightarrow4p$ resonance in Ar and of the $5s\dashrightarrow6p$
resonance in Xe have been measured by \cite{codling80}. \cite{ederer82}
measured the shape of the $ns\dashrightarrow mp$ resonances in Kr and Xe and
{determined} partial cross sections at a resolution of $\Delta E=100$~meV. The angular distribution parameter of the photoelectrons $\beta_{5p_{j}}$,
specified by the total angular momentum of the $5p$ vacancy, has also been
measured in this work. Analyzing the experimental data, \cite{ederer82}
pointed out the necessity to take into account doubly excited
$np^{4}n^{\prime}\ell^{\prime} n^{\prime\prime} \ell^{\prime\prime}$
resonances stemming from the correlation satellite $np^{4}n^{\prime}%
\ell^{\prime}$ levels.

\cite{wills89} used photoelectron spectroscopy to measure the
$ns\dashrightarrow mp$ autoionizing resonances in an extended exciting-photon
energy range at a resolution of $\Delta E\simeq50$~meV. They presented results
for Ar \citep{wills89}, Xe \cite{wills90a}, and Ne \cite{wills90}, which for
Xe were specified by the total angular momentum of the $5p_{j}$ vacancy.

\cite{flemming91} measured cross sections $\sigma_{np_{j}}$ and photoelectron
angular distribution parameters $\beta_{np_{j}}$ for the energy regions
20.6---21.5~eV in Xe and 24.6---25.3 eV in Kr at a high resolution of $\Delta
E\simeq10$~meV. They clearly observed resonances connected with the doubly
excited $np^{4}n^{\prime}\ell^{\prime}n^{\prime\prime}\ell^{\prime\prime}$
states for both atoms. \cite{flemming91} analyzed the measured line shapes of
the $ns\dashrightarrow mp$ resonances in terms of the \cite{shore67a} parameters.

The relationship between the \cite{fano65} and \cite{shore67a} parameters was
investigated in great detail by \cite{sorensen94}. They also performed a
calculation of the line shapes within the MMCDF approach of \cite{tulkki92a}
neglecting, however, the DPES and Coulomb screening.

An extended calculation of the autoionizing $ns\dashrightarrow mp$ resonances
has been performed for Ar by \cite{lagutin99}, for Kr by \cite{demekhin05},
and for Xe by \cite{sukhorukov10} within the CIPFCP approximation. They took
into account a {multitude} of many-electron correlations such as: (i) intrashell and intershell correlations; (ii) DPES; (iii) Coulomb screening; (iv) interaction
between resonances stemming from the main $ns$ level and from the
$np^{4}n^{\prime}\ell^{\prime}$ satellite levels (direct and via continua);
(v) interchannel interactions. For Kr and Xe, the $j$ specific cross sections
$\sigma_{np_{j}}$ and the photoelectron angular distribution parameters
$\beta_{np_{j}}$ were also computed.

\begin{figure}[ptb]
\begin{center}
\includegraphics[width=0.80\textwidth]{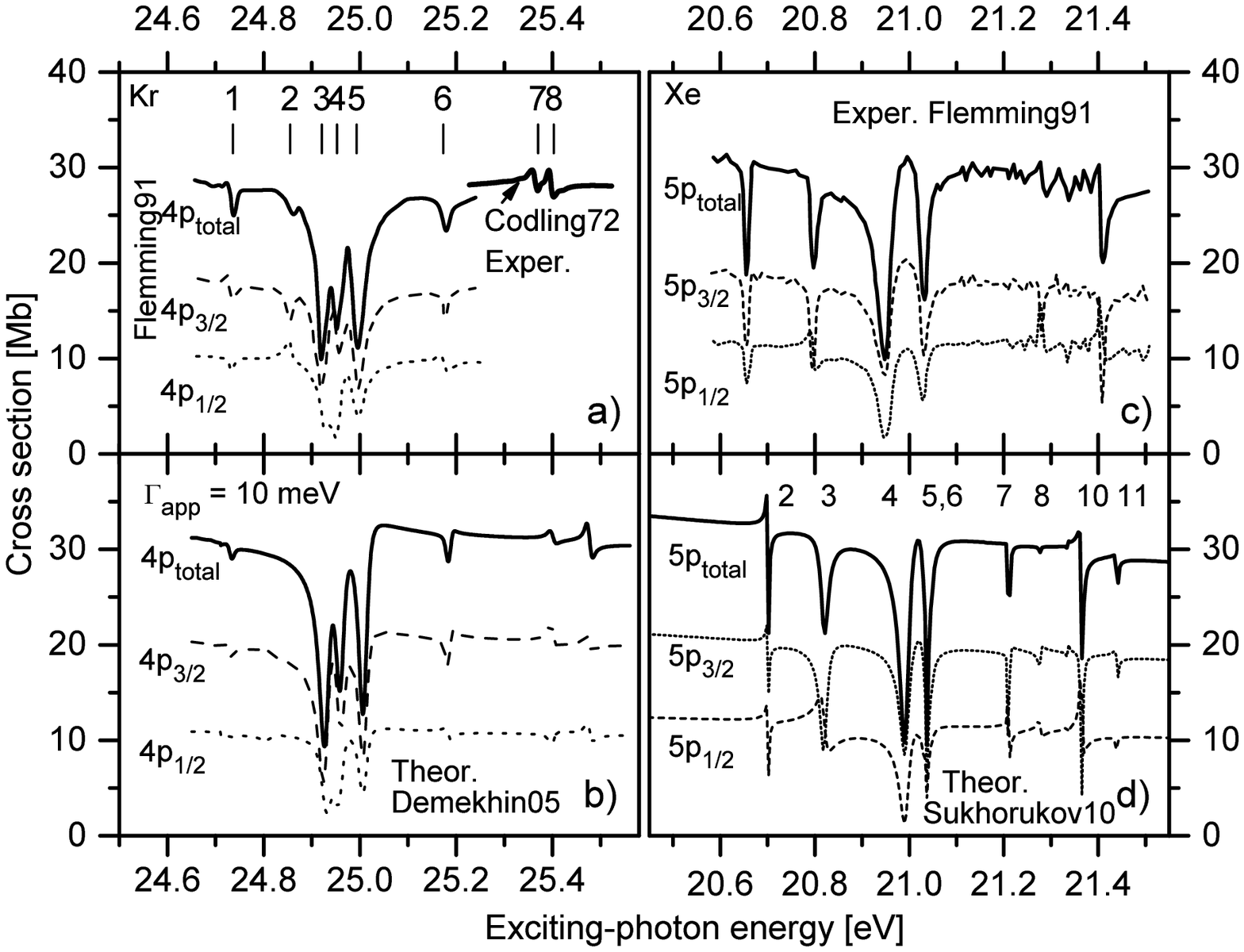}
\end{center}
\caption{Comparison between computed and measured cross sections in the
proximity of the first $ns(n+1)p$ resonance. Data for Kr and Xe are from
\citep{codling72,flemming91,demekhin05} and \citep{flemming91,sukhorukov10},
respectively. Numbering of resonances corresponds to
Table~\ref{tab:KrXe4p_lev}. Figure adapted from
\citep{demekhin05,sukhorukov10}.}%
\label{fig:5p_KrXe}%
\end{figure}

The partial and the total $np$ photoionization cross sections computed by
\cite{demekhin05} for Kr ($n=4$) and by \cite{sukhorukov10} for Xe ($n=5$) are
compared with the respective measurements of \cite{flemming91} in
Fig.~\ref{fig:5p_KrXe} for the energy region around the first
$ns\dashrightarrow mp$ resonances. In both cases, the computed cross sections
are in good agreement with the measured ones.

The identification of resonances is listed in Table~\ref{tab:KrXe4p_lev}. In
the first two columns, the numbering of resonances corresponding to
Fig.~\ref{fig:5p_KrXe} and the computed resonance energies, respectively, are
listed. The 3rd column shows the largest coefficient $| b_{m}^{(i)}|^{2}$ (in
per cent) in the linear combination of the doubly excited states
(\ref{eq:wf_disc}). The 4th column shows the largest coefficient $\left\langle
\alpha LSJ\left\vert E_{c}J\right.  \right\rangle ^{2}$ (in per cent) in the
core wave function~(\ref{eq:core_WF}) and the $\alpha LSJ$ quantum numbers of
the core. It is possible to indicate the genealogy of the
main and satellite levels in the majority of cases, especially in Kr. In case
of the $ns\dashrightarrow mp$ resonances, the percentage of the dominating
component in the doubly excited state hardly exceeds $30-40$\% due to the
strong interaction between the doubly excited states stemming from the main
and satellite levels. This makes the identification of the autoionizing
resonances in Kr and Xe to a large extent conventional.

\begin{table}[ptb]
\caption{Identification of some first resonances in Kr and Xe {below} the $ns$ threshold (computed energies are listed).}%
\label{tab:KrXe4p_lev}%
{\small
\begin{tabular}
[c]{ccccccccccc}\hline\hline
\multicolumn{5}{c}{Kr from \citep{demekhin05}} &  & \multicolumn{5}{c}{Xe from
\citep{sukhorukov10}}\\\cline{1-5}\cline{7-11}%
no & $E$ [eV] & \% & \multicolumn{2}{c}{Doubly excited state $^{b)}$} &  &
no & $E$ [eV] & \% & \multicolumn{2}{c}{Doubly excited state $^{b)}$%
}\\\cline{4-5}\cline{10-11}
&  & $^{a)}$ & Ion state & $np$ &  &  &  & $^{a)}$ & Ion state &
$np$\\\cline{1-5}\cline{7-11}%
1 & 24.732 & 29 & 67\% ($^{3}$P)5s $^{4}$P$_{3/2}$ & 5p$_{3/2}$ &  & 1 &
20.319 & 31 & 87\% ($^{3}$P)6s $^{4}$P$_{5/2}$ & 6p$_{3/2}$\\
2 & 24.811 & 52 & 64\% 4s $^{2}$S$_{1/2}$ & 5p$_{1/2}$ &  & 2 & 20.692 & 33 &
47\% ($^{3}$P)6s $^{2}$P$_{3/2}$ & 6p$_{1/2}$\\
3 & 24.929 & 17 & 96\% ($^{3}$P)5s $^{4}$P$_{5/2}$ & 5p$_{3/2}$ &  & 3 &
20.819 & 75 & 55\% (5s) $^{2}$S$_{1/2}$ & 6p$_{1/2}$\\
4 & 24.960 & 36 & 67\% ($^{3}$P)5s $^{2}$P$_{3/2}$ & 5p$_{1/2}$ &  & 4 &
20.992 & 65 & 55\% (5s) $^{2}$S$_{1/2}$ & 6p$_{3/2}$\\
5 & 25.008 & 23 & 64\% 4s $^{2}$S$_{1/2}$ & 5p$_{3/2}$ &  & 5 & 21.032 & 33 &
47\% ($^{3}$P)6s $^{2}$P$_{3/2}$ & 6p$_{3/2}$\\
6 & 25.185 & 24 & 93\% ($^{3}$P)5s $^{4}$P$_{1/2}$ & 5p$_{3/2}$ &  & 6 &
21.036 & 42 & 60\% ($^{3}$P)5d $^{4}$D$_{1/2}$ & 6p$_{1/2}$\\
7 & 25.400 & 11 & 93\% ($^{3}$P)5s $^{4}$P$_{1/2}$ & 5p$_{1/2}$ &  & 7 &
21.211 & 24 & 69\% ($^{3}$P)6s $^{4}$P$_{1/2}$ & 6p$_{1/2}$\\
8 & 25.477 & 30 & 67\% ($^{3}$P)5s $^{2}$P$_{3/2}$ & 5p$_{3/2}$ &  & 8 &
21.277 & 28 & 80\% ($^{3}$P)5d $^{4}$D$_{5/2}$ & 6p$_{3/2}$\\
&  &  &  &  &  & 9 & 21.336 & 58 & 56\% ($^{3}$P)6s $^{4}$P$_{3/2}$ &
6p$_{1/2}$\\
&  &  &  &  &  & 10 & 21.364 & 31 & 60\% ($^{3}$P)5d $^{4}$D$_{3/2}$ &
6p$_{3/2}$\\
&  &  &  &  &  & 11 & 21.439 & 52 & 60\% ($^{3}$P)5d $^{4}$D$_{1/2}$ &
6p$_{3/2}$\\\hline\hline
\end{tabular}
} \vspace{2ex}
\par
{\footnotesize $^{a)}$ The percentage of the largest component in the doubly
excited state.}
\par
{\footnotesize $^{b)}$ Before the quantum numbers of the ion state, the
percentage of its largest component is shown.}\end{table}

\begin{figure}[ptb]
\begin{minipage}[t]{0.48\textwidth}
\begin{center}
\includegraphics[width=\textwidth]{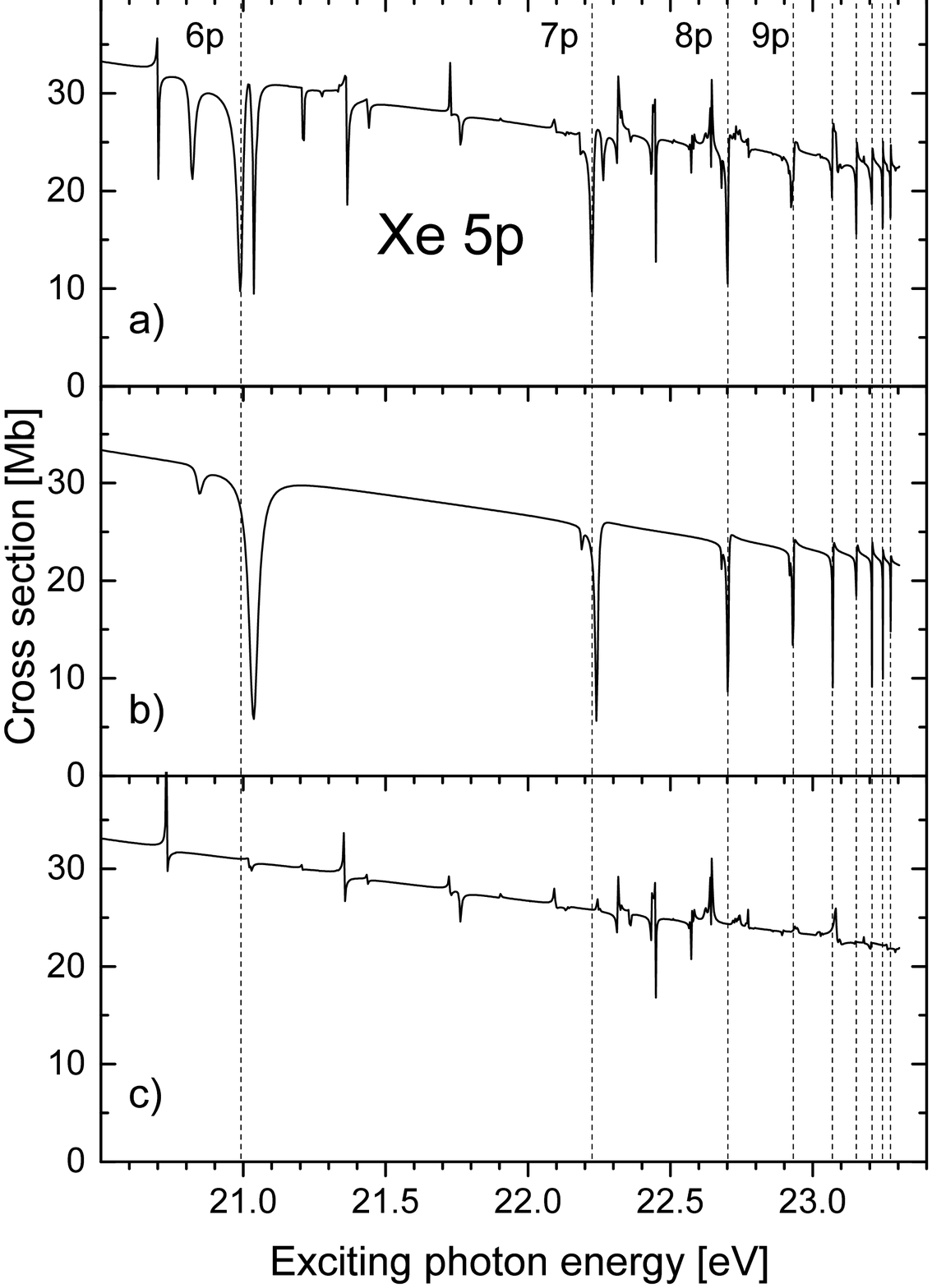}
\caption{Comparison between the photoionization cross section
computed with full inclusion of `main' and `satellite' Rydberg
series (a) with the cross section where only `main' (b) or `satellite'
(c) Rydberg series are included. Vertical dashed lines show the position of the unperturbed $5s\dashrightarrow mp$ series converging to the $5s$ threshold.
Adapted from \citep{sukhorukov10}.}\label{fig:5p_INV}\end{center}
\end{minipage} \hfill\begin{minipage}[t]{0.48\textwidth}
\begin{center}
\includegraphics[width=\textwidth]{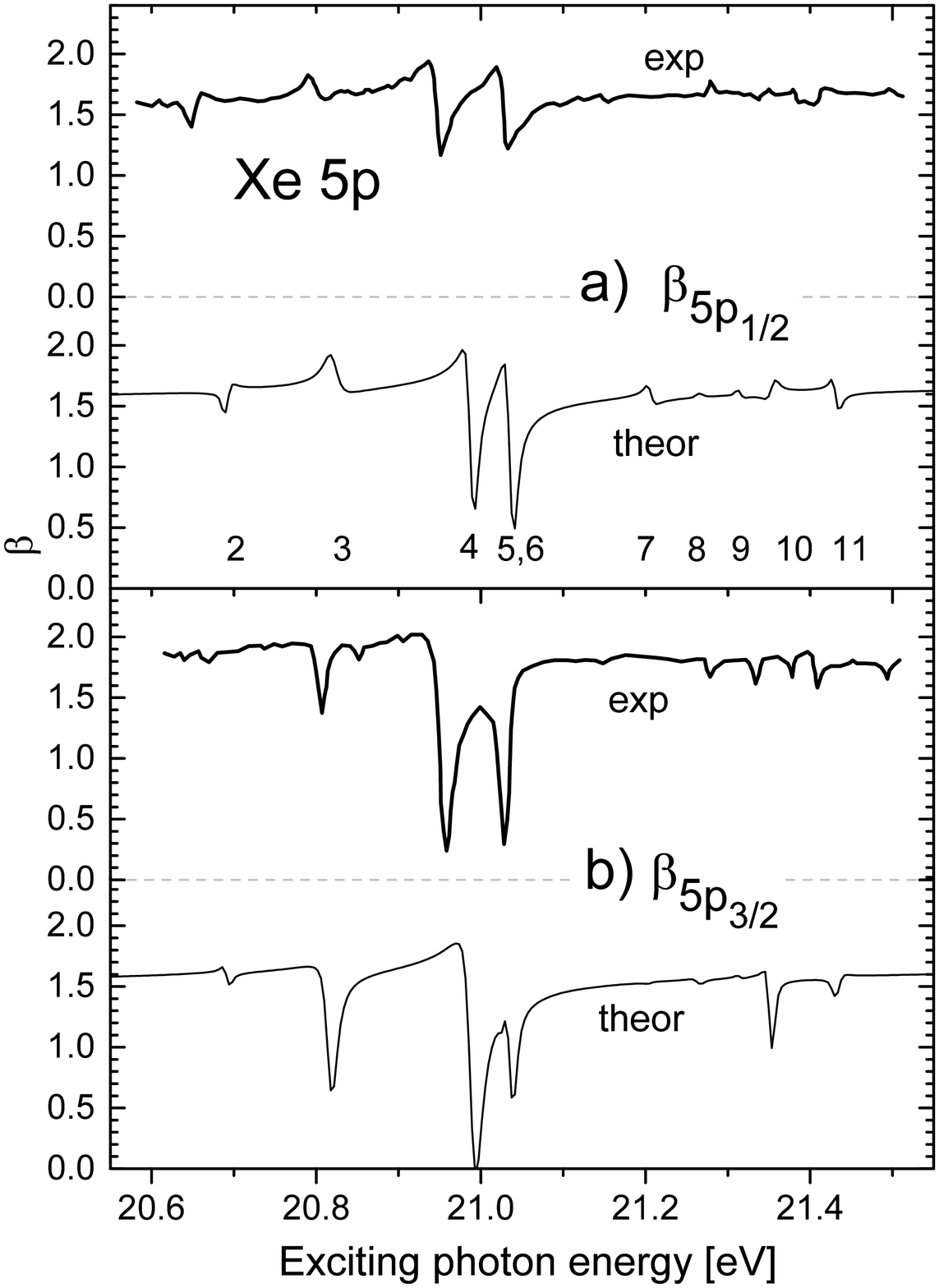}
\caption{Comparison between measured \citep{flemming91} (thick lines)
and computed \citep{sukhorukov10} (thin lines) partial $5p$ photoionization cross
sections of Xe. Numbering of resonances corresponds to
Table~\ref{tab:KrXe4p_lev}.
Adapted from \citep{sukhorukov10}.}\label{fig:5p_BET}\end{center}
\end{minipage}
\end{figure}

\cite{sukhorukov10} studied the formation of a resonant structure in the Xe
$5p$ photoionization. For this purpose, they performed two model calculations.
In the first one, the $5s^{1}mp$ Rydberg series was taken into account only,
whereas the second one included the Rydberg series converging to the $5s^{2}
5p^{4} mp$ satellite thresholds only. The results of this study are depicted
in Fig.~\ref{fig:5p_INV}b and Fig.~\ref{fig:5p_INV}c, respectively.
Fig.~\ref{fig:5p_INV}a shows the result of the calculation where all Rydberg
series are taken into account.

Fig.~\ref{fig:5p_INV}b illustrates that taking into account the $5s^{1}mp$
Rydberg series forms the most prominent details in the total $\sigma_{5p}$
cross section, with the most intense components belonging to the
$5s^{1}mp_{3/2}$ series. Rydberg series originating from the satellite ionic
states result in relatively small resonances (see Fig.~\ref{fig:5p_INV}c).
Taking into account the interaction between the `main' and `satellite' Rydberg
series leads to their strong mixing, which manifests itself in a change of the
energies and intensities of resonances, and results in a good agreement
between theory and experiment (see the right part of Fig.~\ref{fig:5p_KrXe}).

To verify the accuracy of the {calculated} wave functions \cite{demekhin05} and \cite{sukhorukov10} computed the
photoelectron angular distribution parameters $\beta_{4p_{j}}$ for Kr and
$\beta_{5p_{j}}$ for Xe, respectively. For Xe, the computed $\beta_{5p_{j}}$
is compared with the measured one in Fig~\ref{fig:5p_BET}. The good agreement
between theory and experiment {demonstrates the adequacy} of the
CIPFCP approach for describing the atomic photoionization.

\subsubsection{Isoelectronic sequences}
\label{sec:Iso}

In the previous sections, it was illustrated how changing intraatomic
interactions in the calculation of photoionization cross sections {clarifies the major correlative effects}. An interesting series of papers has been published by
\cite{kampen97}, \cite{neogi03}, and \cite{yeates04} who found a way to change
the relationship between intraatomic interactions \emph{experimentally}. To
achieve this goal, they investigated the K$^{+}$ and Ca$^{2+}$ ions,
isoelectronic to Ar \citep{kampen97}, and the Rb$^{+}$, Sr$^{2+}$
\citep{neogi03}, and Y$^{3+}$ \citep{yeates04} ions, isoelectronic to Kr. To
measure the photoabsorption of these ions, the Dual Laser Plasma (DLP)
technique was used.

The DLP technique uses two plasmas created by lasers. One laser operated with
pulses of $\sim1$ J in 10~ns and produced the plasma column consisting of the
ions of interest (the sample) at about 15~eV ($150\ 000$~K). The temperature
of the expanding plasma decreases with time, lowering the dominant ion stage.
The second laser operating in a similar regime produced the second plasma
column serving as a source of the VUV exciting photons. Setting a suitable
time delay between the operation of the sample source and of the source of the
exciting photons, one can investigate the photoabsorption of the selected ions
using the spectrometer. The energy resolution in the PACS measurements varies
from $\Delta E=15$~meV at $\hbar\omega=30$~eV to $\Delta E=50$~meV at
$\hbar\omega=80$~eV. Here we gave the salient details of the DLP technique
only, while its full description can be found in the exhaustive reviews of
\cite{costello91} and \cite{kennedy94,kennedy04}.

The investigation of the $3p$ photoabsorption of the Ar-like ions Ar--K$^{+}%
$--Ca$^{2+}$ by \cite{kampen97} revealed dramatic changes in the shape of the
$3s\dashrightarrow np$ resonances. The $q$ parameter of the first
$3s\dashrightarrow4p$ resonance in K$^{+}$, $q=0.4(1)$, changed the sign with
respect to $q=-0.249(3)$ of Ar \citep{sorensen94}. In Ca$^{2+}$, this
resonance `lost' oscillator strength and became less intense than the
resonances with $n=5,6,7$, having a peak shape. The calculation performed by
\cite{kampen97} explained this behaviour of the resonance shape by the impact
of the many-electron correlations, emphasizing the large role of
`double-electron correlations' which in the present review are called DPES correlations.

A detailed calculation of the $3p$ photoabsorption of the Ar-like ions has
been performed by \cite{lagutin99} within the CIPFCP approximation. These
authors illustrated that the strong changes of the shape of the
$3s\dashrightarrow np$ resonances are determined by the complex interrelation
between the direct and intershell transition amplitudes including their real
and imaginary parts.

\begin{figure}[ptb]
\begin{center}
\includegraphics[width=0.9\textwidth]{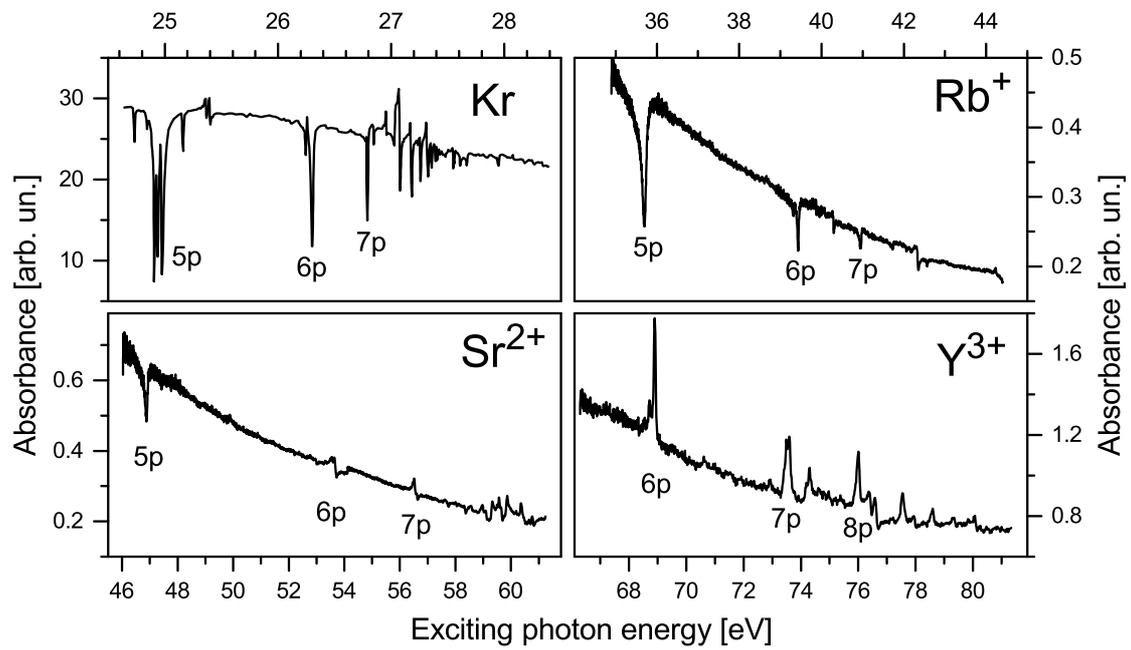}
\end{center}
\caption{Photoabsorption spectra of Kr from \citep{codling72}, Rb$^{+}$ and
Sr$^{2+}$ from \citep{neogi03} and Y$^{3+}$ from \citep{yeates04}. Note that
the first autoionizing resonance in Y$^{3+}$ is the $6p$ one because the
$4s^{1}5p$ resonance lies below the $4p^{5}$ threshold. Adapted from
\citep{neogi03,yeates04}.}%
\label{fig:Kr_like_exp}%
\end{figure}

A similar behaviour of the line shapes of the $4s\dashrightarrow np$
resonances in the Kr-like ions Kr--Rb$^{+}$--Sr$^{2+}$--Y$^{3+}$ has been
observed by \cite{neogi03} (for Rb$^{+}$, Sr$^{2+}$) and \cite{yeates04} (for
Y$^{3+}$). The results of their observation by the DLP technique are depicted
in Fig.~\ref{fig:Kr_like_exp} together with the $\sigma_{4p}$ PICS of Kr
\citep{codling72}. One can recognize several salient features in the behaviour
of the resonance line shapes: (i) the $q$ parameter varies strongly in the
sequence Kr--Rb$^{+}$--Sr$^{2+}$--Y$^{3+}$, being small in Kr ($|q|\ll1$) and
large in Y$^{3+}$ ($|q|\gg1$); (ii) in Sr$^{2+}$, the $6p$ and $7p$ resonances
exhibit too small oscillator strengths in comparison with the $5p$ one, and
(iii) $q$ parameters strongly change along the $4s\dashrightarrow np$ series.

\begin{figure}[ptb]
\begin{minipage}[t]{0.85\textwidth}
\begin{center}
\includegraphics[width=\textwidth]{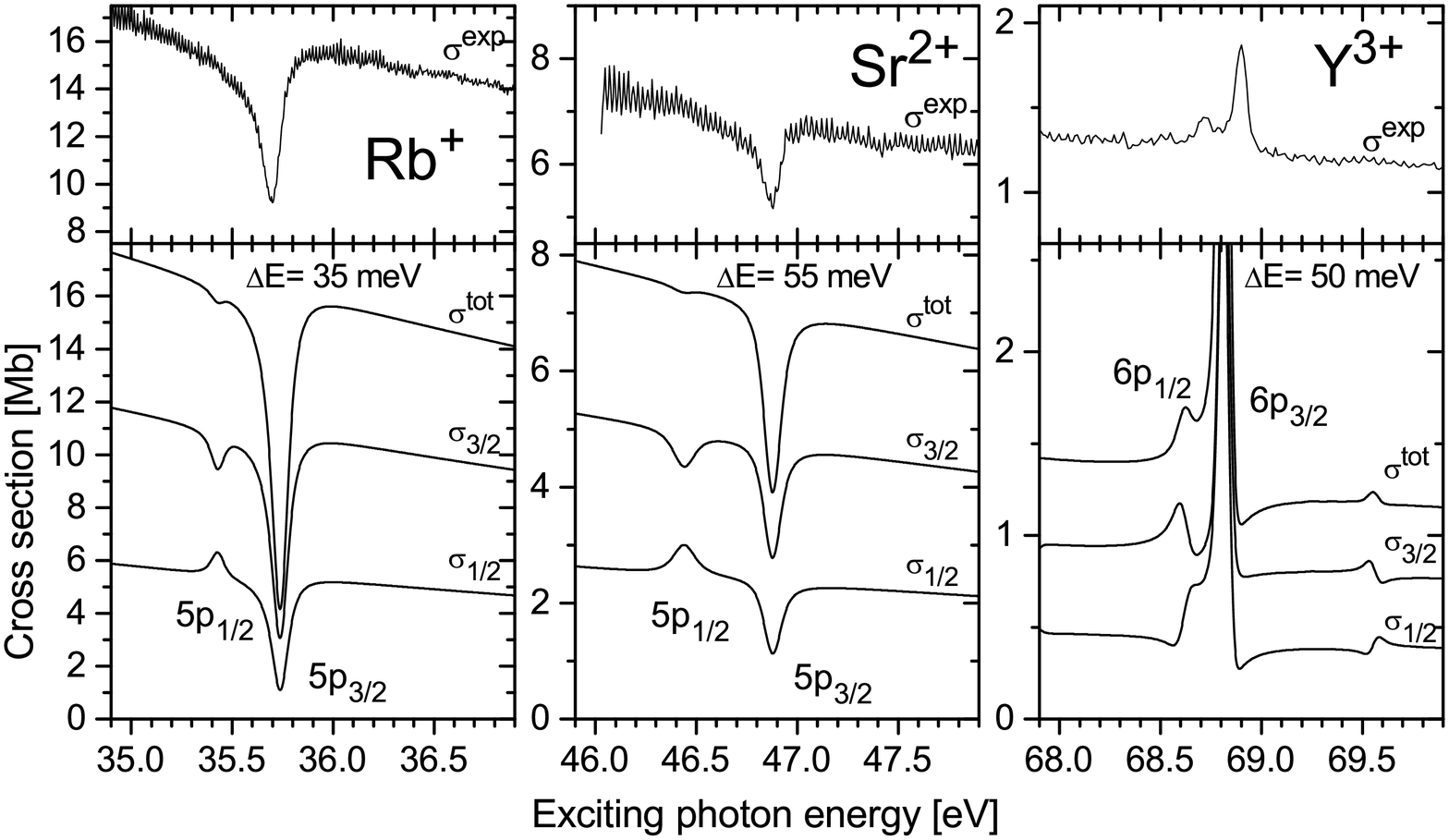}
\caption{Comparison between computed and measured (normalized) photoionization
cross sections in the region of the first $4s^{1}np$ resonance. The
$4s^{1}5p_{1/2}$ resonances in Rb$^{+}$ and Sr$^{2+}$ are cancelled in the total
cross section due to mirroring behaviour, whereas the $4s^{1}6p_{1/2}$ resonance
in Y$^{3+}$ is clearly observed in the total cross section because of the large
value of $\rho^{2}q^{2}$. The computed spectra were additionally deconvolved by an experimental function having Gaussian shape with the FWHM $\Delta E$ shown for each spectrum. Adapted from
\citep{yeates04}.}\label{fig:Kr_like_1st}\vspace{4ex}
\end{center}
\end{minipage} \hfill\begin{minipage}[t]{0.85\textwidth}
\begin{center}
\includegraphics[width=\textwidth]{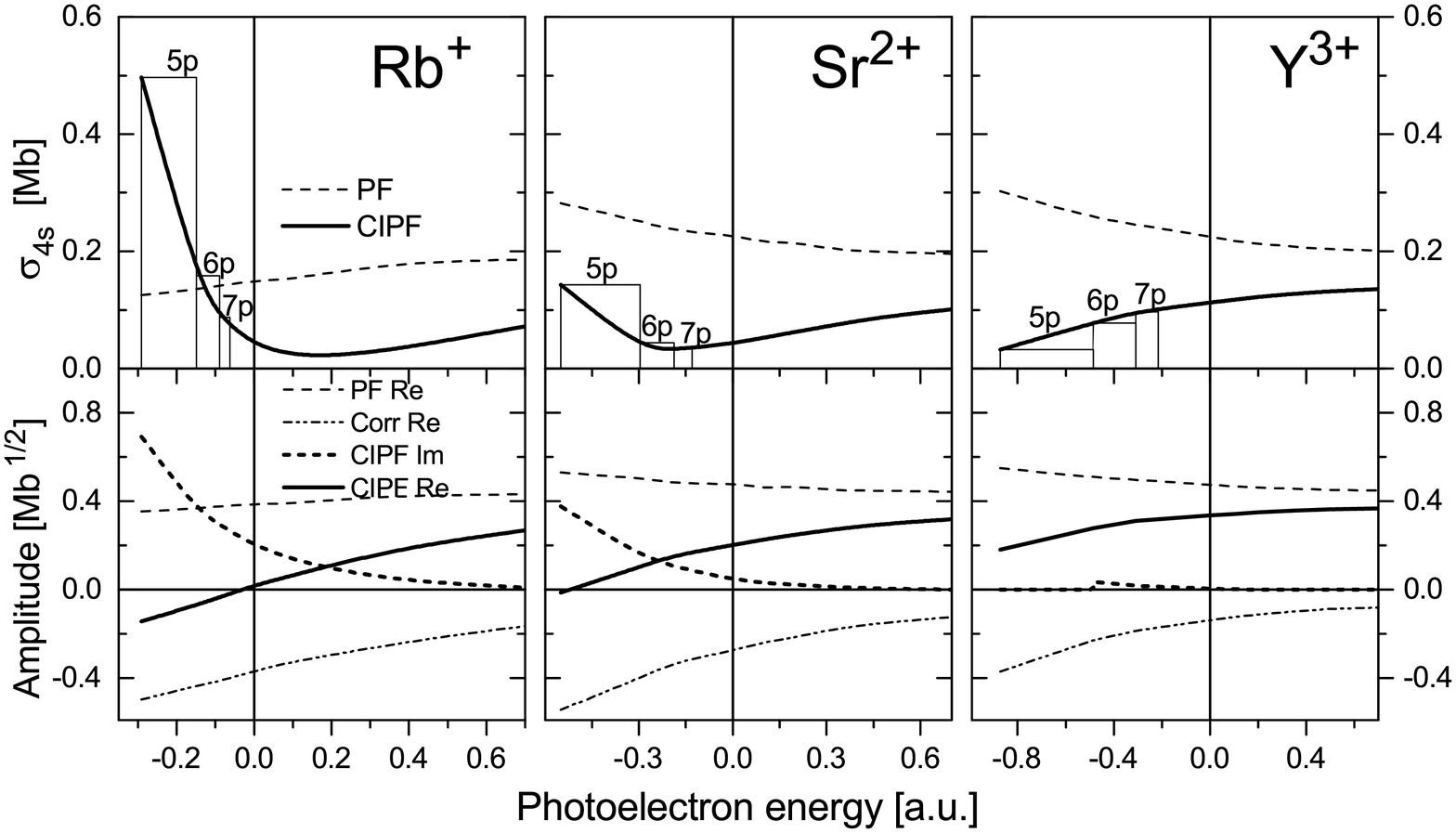}
\caption{Cross sections $\sigma_{4s}$ (upper panels) and transition amplitudes (lower panels) of the Rb$^{+}$--Sr$^{2+}$--Y$^{3+}$ isoelectronic
sequence computed within the PF and CIPF approaches. Adapted from \citep{yeates04}.}
\label{fig:Kr_like_amp}\end{center}
\end{minipage}
\end{figure}

The calculations of \cite{neogi03} and \cite{yeates04} showed that the
$4p^{4}n^{\prime}\ell^{\prime}n^{\prime\prime}\ell^{\prime\prime}$ doubly
excited states in the ions have larger energy than the first autoionizing
resonance, in contrast to Kr discussed above where they strongly
overlap with the $4s^{1}5p$ \emph{promoter}. In Rb$^{+}$--Sr$^{2+}%
$--Y$^{3+}$, the doubly excited states result in the rich structure of the
second and higher resonances. Therefore, in Fig.~\ref{fig:Kr_like_1st}, we
show results from \citep{yeates04} for the first resonance only.

In Fig.~\ref{fig:Kr_like_amp}, the partial and total $4s\dashrightarrow
n/\varepsilon p$ transition amplitudes are depicted for Rb$^{+}$--Sr$^{2+}%
$--Y$^{3+}$. For the discrete states, the amplitudes are presented according
to \citep{cooper62} as areas proportional to the respective oscillator
strengths. From this figure, one can recognize that the intershell
correlations result in large real (Corr Re) and imaginary (Corr Im$=$CIPF Im)
parts of the partial transition amplitude which are comparable to the direct
PF amplitude. The interrelation between all these amplitudes explains all the
peculiarities in the behaviour of the autoionizing resonances described above.
Indeed, (i) Y$^{3+}$ has a very small imaginary part of the transition
amplitude in comparison with Rb$^{+}$--Sr$^{2+}$ which explains the peak shape
of the observed resonances (see equation~(\ref{eq:Fano_par}) for $q$); (ii) in
Sr$^{2+}$, the $4s\dashrightarrow np$ PICS have a minimum in the vicinity of
the $6p$ and $7p$ resonances which explains their anomalously small oscillator
strength; (iii) in Sr$^{2+}$, the ratio of the imaginary and real parts of the
transition amplitude varies strongly along the $4s\dashrightarrow np$ series
which explains the respective behaviour of the $q$ parameter.

Fig.~\ref{fig:Kr_like_1st}, where the computed $\sigma_{4p}$ PICS are compared
with the measured ones, illustrates how the mirroring behavior of the cross
sections $\sigma_{4p_{1/2}}$ and $\sigma_{4p_{3/2}}$ at the $5p_{1/2}$
resonance explains its absence in the total cross section $\sigma_{4p}$ in
Rb$^{+}$--Sr$^{2+}$. In Y$^{3+}$ the $j=1/2$ component of the first
autoionizing resonance can clearly be seen in the total cross section.

In conclusion, the investigation of the isoelectronic sequences allows one to
judge about the interrelation between the single-electron and many-electron
processes in a wide range of the intraatomic interactions. This information is
extremely useful for the creation and development of the \emph{ab initio}
methods for atomic structure calculations.

\section{Alignment and orientation of ionic states}

\begin{figure}[ptb]
\begin{center}
\includegraphics[width=0.95\textwidth]{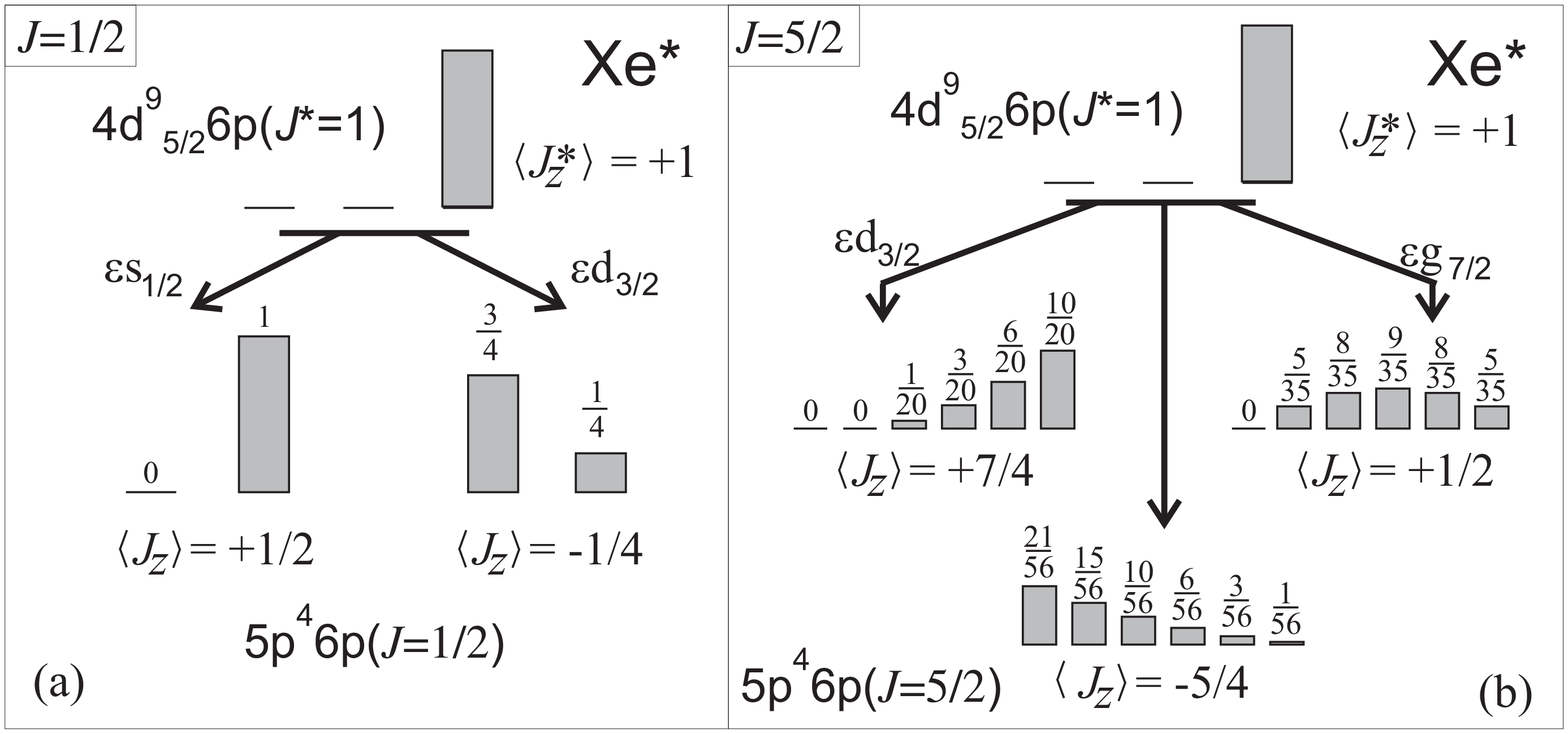}
\end{center}
\caption{Orientation transfer during the Auger decay of the $4d^{9}%
_{5/2}6p(J^{*}=1)$ state prepared by right-hand circularly polarized photons
and, therefore, 100\% populated at $M_{J^{*}}=+1$. The populations of the
magnetic sublevels of the final $5p^{4}6p$ states for $J=1/2$ and $J=5/2$ are
shown in panels (a) and (b), respectively. Bars indicate the normalized
population of the $M_{J}$ sublevels $\sigma_{M_{J}}$. Average projection of
the resulting ion angular momentum along the SR beam $\langle J_{z}
\rangle=\sum\sigma_{M_{J}}\cdot M_{J}$ is indicated for each decay mode.
Adapted from \citep{okeeffe03}.}%
\label{fig:Meyer_POP}%
\end{figure}

When polarized synchrotron radiation excites the $Rg$ atom, the magnetic
sublevels are populated non-statistically. For instance, linearly polarized
exciting radiation populates the sublevel $M_{J^*}=0$, whereas the population of the
$M_{J^*}=\pm1$ sublevels equals zero (thus, the exited state is fully aligned).
Circularly polarized light produces fully oriented states, e.g. right-hand
polarized photons excite the $M_{J^*}=+1$ sublevel only (see
Fig.~\ref{fig:Meyer_POP}).

During the decay of the excited state, the alignment or orientation is
transferred to the decay fragments and results in their anisotropic
distribution and polarization. In Fig.~\ref{fig:Meyer_POP}, the orientation
transfer computed for the process $[\text{Xe}]\dashrightarrow4d^{9}%
6p(J^{*}=1)\rightarrow5p^{4}6p(J) \varepsilon\ell_{j}$ is {schematically visualized}. The non-statistical population of the magnetic sublevels is clearly
seen. Thus, studying the angular distribution or polarization of the decay
fragments allows to get information about the spectral distribution of the
angular momenta of the, e.g. ionic states. This information provides an
excellent test of the theoretical models for interpreting experimental data
and is useful in the interpretation and disentanglement of the complex
structure of the aforementioned core-excited spectra.

In the present chapter, we intend to present the basic results obtained in the
polarization experiments by PIFS. Therefore, for the detailed theoretical description of
the photofragmentation, alignment, and orientation, we refer the reader to the
reviews of \cite{greene82}, \cite{schmidt92}, \cite{kabachnik07}, and to the
book of \cite{balashov00} where the details and references to numerous original investigations can be found.

\cite{mehlhorn68} stressed that colliding particles will align atoms with
respect to the direction of the exciting beam and, therefore, X-ray radiation
and Auger electrons should have a nonisotropic angular distribution. Studying
that angular distribution has been proposed in order to determine the relative
partial photoionization cross sections \citep{flugge72}. The basic relations
describing the alignment of ions in the ionization process and the following
angular distribution of the decay fragments have been obtained by
\cite{eichler76} and \cite{berezhko77}. For example, the angular distribution
of the Auger electrons ejected from an atom by linearly polarized photons with
energy $\omega$ can be represented by the differential cross section as:%

\begin{equation}
\frac{d\sigma(\omega)}{d\Omega}=\frac{\sigma_{tot}(\omega)}{4\pi}[1+\beta
^{e}P_{2}(\mathrm{cos}\theta)], \label{eq:beta_Kab}%
\end{equation}
where $P_{2}$ is the Legendre polynomial, $\theta$ is the angle between the
polarization vector of the exciting radiation and the direction of the ejected
Auger electron, and $\beta^{e}$ is the quantity called photoelectron
angular distribution (PAD) parameter. \cite{berezhko77} also calculated the
alignment of ions and PAD parameters for some selected cases within the HS
approach. In the following, a considerable amount of investigations were
focused on the alignment transfer during the resonant Auger (RA) effect after
excitation of a deep core electron to the first Rydberg state, e.g.
$3d\dashrightarrow5p$ in Kr or $4d\dashrightarrow6p$ in Xe, because these
processes look comparatively simple.

\subsection{Resonant Auger (RA) effect}

\label{sec:RA}

The decay of the resonantly excited Kr$^{*}$ $3d^{9}5p$ and Xe$^{*}$
$4d^{9}6p$ states has been observed for the first time by \cite{eberhardt78}
at a resolution of 0.3~eV and initiated numerous investigations. Measurements
of the RA effect were used to investigate post collision interaction
\citep{schmidt81,southworth83}, the alignment of the Xe$^{+}$ ions and angular
distribution of the Auger electrons \citep{southworth83}, and many-electron
correlations like DPES \citep{aksela86}, etc. (see, e.g., the review
\cite{armen00} and references therein).

The first measurements of the RA effect and related phenomena were performed
at rather low resolution of $\sim0.5$~eV. Better resolution of 150~meV was
used by \cite{carlson88,carlson89} who measured the Auger electron angular
distribution parameters $\beta^{e}$ in the resonantly excited Ar$^{*}$
($2p\dashrightarrow4s,5s$), Kr$^{*}$ ($3d\dashrightarrow5p,6p$), and Xe$^{*}$
($4d\dashrightarrow6p,7p$). Earlier theoretical treatments of the RA effect
were based on the stepwise (or two-step (2ST)) model which implies that the
interference between the resonant and direct ionization channels as well as
between different resonant channels is ignored. This model was used by
\cite{kabachnik91} and by \cite{hergenhahn91} to compute the spin polarization
and the angular distribution of the Auger electrons in the RA decay of Ar, Kr,
and Xe. The authors used in the calculation the core AOs obtained by applying
the DF package of \cite{grant80} and the continuum AOs computed within the
relativistic local density approach.

Considerable progress in the investigation of the RA effect has been achieved
by experiments in the Raman regime of the excitation of atoms. In this regime,
the experimental resolution $\Delta E$ is smaller than the natural width of
the intermediate resonance \citep{brown80,kivimaki93}, which makes it possible
to observe the majority of the final ionic states. The influence of
many-electron correlations in both the initial and final states of the RA
effect on the angular distribution of the Auger electrons has been
investigated by \cite{tulkki94} within the MMCDF approach (see
section~\ref{sec:suites_PC}).

The Raman regime was utilized to observe many new experimental features in the
ordinary RA spectra of Kr \citep{aksela96,jauhiainen96} and to measure
$\beta^{e}$ in the RA spectra of Kr \citep{aksela96a} and Xe
\citep{aksela96b}. In order to interpret the experiments, these authors
performed calculations using the theoretical technique of \cite{tulkki94}.

\cite{ehresmann98} applied the PIFS technique to measure the alignment of the
$5p^{4}6p\left(  E_{1}J_{1}\right)  $ Xe\thinspace II ions obtained in the
following process:%

\begin{equation}%
\begin{array}
[c]{ccccc}%
\text{\lbrack0] }(J_{0}=0)+\gamma_{ex} & \dashrightarrow & 4d_{\bar{J}}%
^{9}6p_{\bar{j}}(J=1)\;(\mathbf{R}) &  & \\
&  & \updownarrow &  & \\
&  & 5p^{4}6p\left(  E_{1}J_{1}\right)  +\varepsilon\ell j\;(\mathbf{dir}) &
\dashrightarrow & 5p^{4}6s\left(  E_{2}J_{2}\right)  +\varepsilon\ell
j+\gamma_{fl}%
\end{array}
\label{eq:2ST}%
\end{equation}
where the designations are the same as in (\ref{eq:Ar-scheme}): the
dashed and double-side arrows denote electric-dipole and Coulomb interaction,
respectively; [0] denotes the ground state of the atom and the internal
quantum numbers of atomic and ionic states are self-explanatory. All
transitions in scheme (\ref{eq:2ST}) are initiated by the exciting photon
$\gamma_{ex}$, while the fluorescence photon $\gamma_{fl}$ is registered.
The measurements for Xe were revisited at higher accuracy by \cite{lagutin00}
and a similar experiment for Kr was performed by \cite{zimmermann00}.

Different experimental setups have been applied by those authors: to measure
the alignment of the Xe$^{+}$ $5p^{4}6p\left(  E_{1}J_{1}\right)  $ states,
\cite{lagutin00} applied the technique of the `crossed undulators' (see
Fig.~\ref{fig:HS_fig7_cdrbw} and comments to it), while for the Kr$^{+}$
$4p^{4}5p\left(  E_{1}J_{1}\right)  $ states, \cite{zimmermann00} used the
`Wollaston prism' setup (see Fig.~\ref{fig:HS_fig6_polset}). In order to
compute the alignment parameters $A_{20}(E_{1}J_{1},\omega)$, \cite{lagutin00}
and \cite{zimmermann00} used the 2ST model and the CIPF approximation
according to the following formula (see, e.g. \citep{schartner07}):%

\begin{equation}
A_{20}(E_{1}J_{1},\omega)=\sum_{j(\ell)}b2(j,J_{1})\cdot\gamma_{E_{1}J_{1}%
}^{\varepsilon\ell j}(\omega) \label{eq:Schrt_A20}%
\end{equation}
where the kinematics coefficients $b2(j,J_{1})$\ can be found in
\citep{lagutin00,zimmermann00,lagutin03a}. The reduced partial cross sections
$\gamma_{E_{1}J_{1}}^{\varepsilon\ell j}(\omega)$%

\begin{equation}
\gamma_{E_{1}J_{1}}^{\varepsilon\ell j}(\omega)=\sigma_{E_{1}J_{1}%
}^{\varepsilon\ell j}(\omega)/\sigma_{E_{1}J_{1}}(\omega)
\label{eq:Schrt_red_PICS}%
\end{equation}
were computed using the cross sections $\sigma_{E_{1}J_{1}}^{\varepsilon\ell
j}(\omega)$ and $\sigma_{E_{1}J_{1}}(\omega)$ determined by equation
(\ref{eq:PICS}). The absolute values of the calculated alignment parameter
$A_{20}$ appeared to be systematically larger than the measured ones. This was
partly attributed to the disalignment of ions due to cascade transitions,
which have an estimated probability of 8.5\% in comparison with the main
process (\ref{eq:2ST}) \citep{lagutin00}. The strongest disalignment channel
$[0]\dashrightarrow4d^{9}6p\rightarrow5p^{4}7p\dashrightarrow5p^{4}%
7s\dashrightarrow5p^{4}6p$\ contributes 75\% of that total probability.
Additional disalignment of the ionic states is connected with the hyperfine
interactions \citep{meyer01}. The PAD parameters $\beta^{e}$ for the RA decay
in Kr and Xe were computed in \citep{zimmermann00,lagutin00} also.

\subsubsection{Beyond the two-step model}

\label{sec:beyond_2ST}

All the above papers used for the interpretation of the experimental data the
2ST model, treating the RA decay as two independent steps. The first step
leads to the excitation of the intermediate atomic resonances which decay
during the second one. The limitation of the 2ST model has been pointed out by
\cite{camilloni96,kukk97,saito00,defanis02}. In these papers, the parameters
for individual RA lines in the resonance excitation of Ar $2p-3d$
\citep{camilloni96}, Kr $3d-np$ \citep{kukk97} and Ne $1s-3p$
\citep{saito00,defanis02} were found to depend on the excitation energy. The
2ST model cannot explain these observations.

\cite{lagutin03a,lagutin03b} performed a combined theoretical and experimental
investigation of the RA effect in Kr. These authors measured the energy
dependence of the alignment of some selected $4p^{4}5p\left(  E_{1}%
J_{1}\right)  $ ionic states being populated via the RA decay of the
$3d_{\bar{J}}^{9}5p_{\bar{j}}(J=1)$ resonances using the crossed undulators
PIFS technique (see Fig.~\ref{fig:HS_fig7_cdrbw}). These resonances, having a
natural width of 83~meV \citep{king77,sairanen96}, were scanned within the
Raman regime at a resolution of 10~meV by steps of 5~meV.

\begin{figure}[ptb]
\begin{center}
\includegraphics[width=0.95\textwidth]{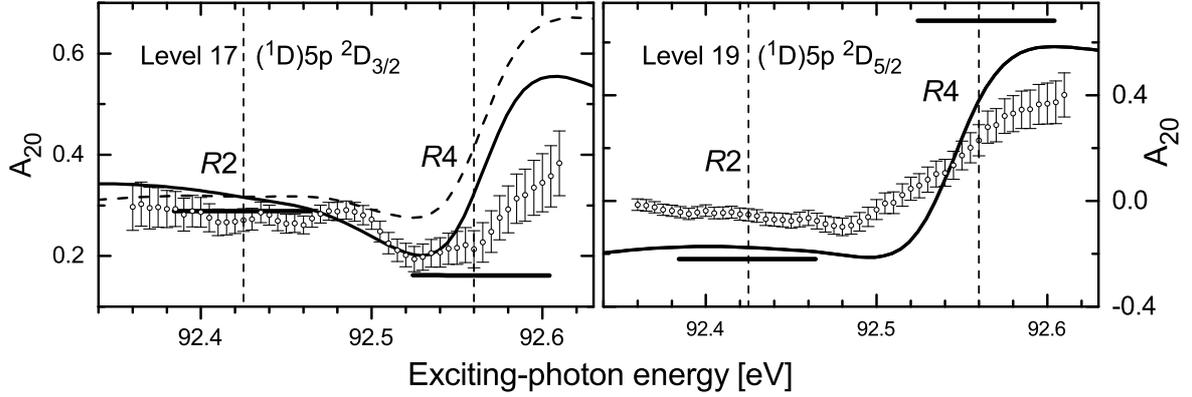}
\end{center}
\caption{Comparison between measured (open circles) and computed alignment for
two selected KrII $4p^{4}5p$ states. Solid and dashed lines: calculation with
and without the direct photoionization amplitude, respectively, in equation
(\ref{eq:Ram_ampl}). Horizontal bars with length equal to $\Gamma(R_{i})$:
$A_{20}$ computed within the stepwise model. Vertical lines mark the positions
of $R2$ and $R4$ resonances. Adapted from \citep{lagutin03b}.}%
\label{fig:Polar_PRL}%
\end{figure}

In order to compute the alignment parameters $A_{20}$,
\cite{lagutin03a,lagutin03b} applied the CIPF approach (see
section~\ref{sec:CIPFCP}) taking into account interference between the
resonant (\textbf{R}) and direct ionization (\textbf{dir})\ channels and
between different resonant channels (see scheme (\ref{eq:2ST})). In this
approximation, the transition amplitudes determining the dynamics of all
processes are given by:%

\begin{equation}
D(E_{1}J_{1},\varepsilon\ell)=\left\langle E_{1}J_{1},\varepsilon
\ell\,\left\vert \mathbf{d}\right\vert 0\right\rangle +\underset{R}{\sum}%
\frac{\left\langle E_{1}J_{1},\varepsilon\ell\,\left\vert \mathbf{H}%
^{ee}\right\vert R\right\rangle \left\langle R\left\vert \mathbf{d}\right\vert
0\right\rangle }{(\varepsilon+E_{1}-E)+\imath\Gamma(R)/2} \label{eq:Ram_ampl}%
\end{equation}
where $\mathbf{H}^{ee}$ and $\mathbf{d}$ are the Coulomb and electric dipole
operators, respectively; $\Gamma(R)$ is the total width of the resonance. The
first term entering equation (\ref{eq:Ram_ampl}) describes the direct
photoionization and the sum over $R$ in the second term includes all resonance
channels. In Kr, four resonances lie in the range of interest at the computed
energies (coinciding with the measured ones within several meV) $R1$%
(91.200~eV), $R2$(92.424~eV), $R3$(92.454~eV), and $R4$(92.565~eV). Resonances
$R1$ and $R4$ are practically pure $3d_{5/2}^{9}5p_{3/2}\,J=1$ and
$3d_{5/2}^{9}6p_{3/2}\,J=1$ basis states, respectively, whereas resonances
$R2$ and $R3$ are perceptible mixtures of the $3d_{3/2}^{9}5p_{3/2}\,J=1$ and
$3d_{3/2}^{9}5p_{1/2}\,J=1$ basis states.\ The oscillator strength for the
$R3$ resonance is 1/80 of the $R2$ resonance only, so that it was excluded
from consideration.

The results of the calculation for two selected $4p^{4}5p\left(  E_{1}%
J_{1}\right)  $ ionic states excited at energies close to the $R2$ and
$R4$\ resonances are depicted in Fig.~\ref{fig:Polar_PRL} together with the
experimental data. In this figure, the results of three calculations are
presented: (i) the 2ST results are shown as horizontal bars; (ii) the dashed
curve shows the result without taking into account direct ionization
(\textbf{dir}) (for the $4p^{4}5p\left(  ^{2}D_{5/2}\right)  $ level, the
(\textbf{dir}) pathway is forbidden by a selection rule); (iii) solid lines
represent the results of the complete calculation. One can recognize the following:

\begin{description}
\item[(a)] the interference between the (\textbf{R2}) and (\textbf{R4})
pathways only results in substantial changes of $A_{20}$ computed within the
2ST model (see the $A_{20}$ results at the $R4$ resonance);

\item[(b)] the interference between the (\textbf{R}) and (\textbf{dir})
pathways noticeably changes the computed $A_{20}$;

\item[(c)] the computed $A_{20}$ energy dependences are in good agreement
with the measured ones, resulting, however, in a little larger values. This
discrepancy can be attributed to the disalignment of ionic states which was
not taken into account {in the calculations of} \cite{lagutin03a,lagutin03b}.
\end{description}

The coherent {treatment of} the partial transition amplitudes resulted in
better agreement between computed and measured alignment parameters $A_{20}$
for the majority of the $4p^{4}5p\left(  E_{1}J_{1}\right)  $ ionic states in
Kr \citep{lagutin03a}. The remaining differences between theory and experiment
have been discussed by \cite{lagutin11} who included in the calculation
cascade transitions and four-fold excitations in the vicinity of the
$3d_{\bar{J}}^{9}5p_{\bar{j}}(J=1)$ resonances.

The Raman regime was also used by \cite{sankari07} in a combined investigation
of the RA decay of the $3d^{9}np$ resonances of Kr$^{*}$. These authors used
photoelectron spectroscopy at a resolution of 25~meV to cover the
exciting-energy range from 92.12 to 92.87 eV by 10 meV steps to measure the
population of the $4p^{4}5p\left(  E_{1}J_{1}\right)  $ ionic states and the
respective PAD parameter $\beta_{E_{1}J_{1}}^{e}$. In order to compute the
measured quantities, the authors applied the RATIP suite (see
section~\ref{sec:suites_PC}). The population of the $4p^{4}5p\left(
E_{1}J_{1}\right)  $ ionic states has also been investigated by
\cite{sankari08} using the PIFS technique. The calculation performed by these
authors includes also cascades. In both papers, the importance of
taking into account the (\textbf{R}) and (\textbf{dir}) pathways coherently
has been confirmed.

\subsection{Partial wave analysis}

\label{sec:PWA}

{In the photoionization of the closed-shell \emph{Rg}, the} cross section $\sigma_{E_{1}J_{1}}(\omega)$ for the population of the ionic state $\left\vert E_{1}J_{1}\right\rangle $ is
determined by the three partial waves $\varepsilon\ell j$
\citep{klar80,greene82} (see equation~(\ref{eq:PICS}) for details):%

\begin{equation}
\sigma_{E_{1}J_{1}}(\omega)=\sum_{\ell j}\sigma_{E_{1}J_{1}}^{\varepsilon\ell
j}(\omega). \label{eq:PICS_tot}%
\end{equation}

According to equation~(\ref{eq:PICS_tot}) one can define the dimensionless
reduced partial wave cross sections $\gamma_{E_{1}J_{1}}^{\varepsilon\ell
j}(\omega)$%

\begin{equation}
\gamma_{E_{1}J_{1}}^{\varepsilon\ell j}(\omega)=\sigma_{E_{1}J_{1}%
}^{\varepsilon\ell j}(\omega)/\sigma_{E_{1}J_{1}}(\omega) \label{eq:PICS_red}%
\end{equation}
interconnected by the normalization condition:%

\begin{equation}
\sum_{\ell j}\gamma_{E_{1}J_{1}}^{\varepsilon\ell j}(\omega)=1.
\label{eq:PICS_norm}%
\end{equation}

The three reduced partial cross sections determine also the alignment
$A_{20}(E_{1}J_{1},\omega)$ and orientation $O_{10}(E_{1}J_{1},\omega)$
parameters as \citep{klar80,greene82}:%

\begin{equation}
A_{20}(E_{1}J_{1},\omega)=\sum_{\ell j}a_{20}(j,J_{1})\cdot\gamma_{E_{1}J_{1}%
}^{\varepsilon\ell j}(\omega) \label{eq:A20_w}%
\end{equation}

\begin{equation}
O_{10}(E_{1}J_{1},\omega)=\sum_{\ell j}o_{10}(j,J_{1})\cdot\gamma_{E_{1}J_{1}%
}^{\varepsilon\ell j}(\omega). \label{eq:O10_w}%
\end{equation}
with the kinematics coefficients tabulated, e.g., in \citep{lagutin03a}.
Solving equations (\ref{eq:PICS_norm}) to (\ref{eq:O10_w}) we get%

\begin{equation}
\gamma_{E_{1}J_{1}}^{\varepsilon\ell j}(\omega)=s_{0}\left(  j\right)
+o_{1}\left(  j\right)  \cdot O_{10}(E_{1}J_{1},\omega)+a_{2}\left(  j\right)
\cdot A_{20}(E_{1}J_{1},\omega) \label{eq:inverse}%
\end{equation}
where the kinematics coefficients $o_{1}\left(  j\right)  $ and $a_{2}\left(
j\right)  $\ are tabulated, e.g. in \citep{schartner05}, and $s_{0}\left(
j\right)  =(2j+1)[(2J+1)(2J_{1}+1)]^{-1}$ ($J=1$ is the total angular momentum
of ion plus photoelectron, see \cite{schartner07}). Thus, the two observables
$A_{20}(E_{1}J_{1},\omega)$ and $O_{10}(E_{1}J_{1},\omega)$ provide sufficient
information to derive the reduced partial wave cross sections $\gamma
_{E_{1}J_{1}}^{\varepsilon\ell j}(\omega)$.

\cite{schmoranzer97a}, \cite{mentzel98}, and \cite{yenen97} obtained the
alignment parameters $A_{20}(E_{1}J_{1},\omega)$ using different setups of the
PIFS technique: \cite{schmoranzer97a} and \cite{mentzel98} applied the
`crossed undulator' setup (see Fig.~\ref{fig:HS_fig7_cdrbw}) while
\cite{yenen97} analyzed the polarization of fluorescence.
\cite{schmoranzer97a} measured $A_{20}(E_{1}J_{1},\omega)$ for the even
$4p^{4}(L_{0}S_{0})5s\ E_{1}J_{1}$ states of Kr\thinspace II at a resolution
of $\Delta E=15$~meV. \cite{yenen97} obtained the polarization of the odd
$3p^{4}(L_{0}S_{0})4p\ ^{2}P_{3/2}^{o}$ state of Ar\thinspace II at a
resolution of $\Delta E=3$~meV. \cite{mentzel98}\ reported $A_{20}(E_{1}%
J_{1},\omega)$\ measured for seven odd $3p^{4}(L_{0}S_{0})4p\ E_{1}J_{1}$
states of Ar\thinspace II at a resolution of $\Delta E=10$~meV.

\begin{figure}[ptb]
\begin{center}
\includegraphics[width=0.95\textwidth]{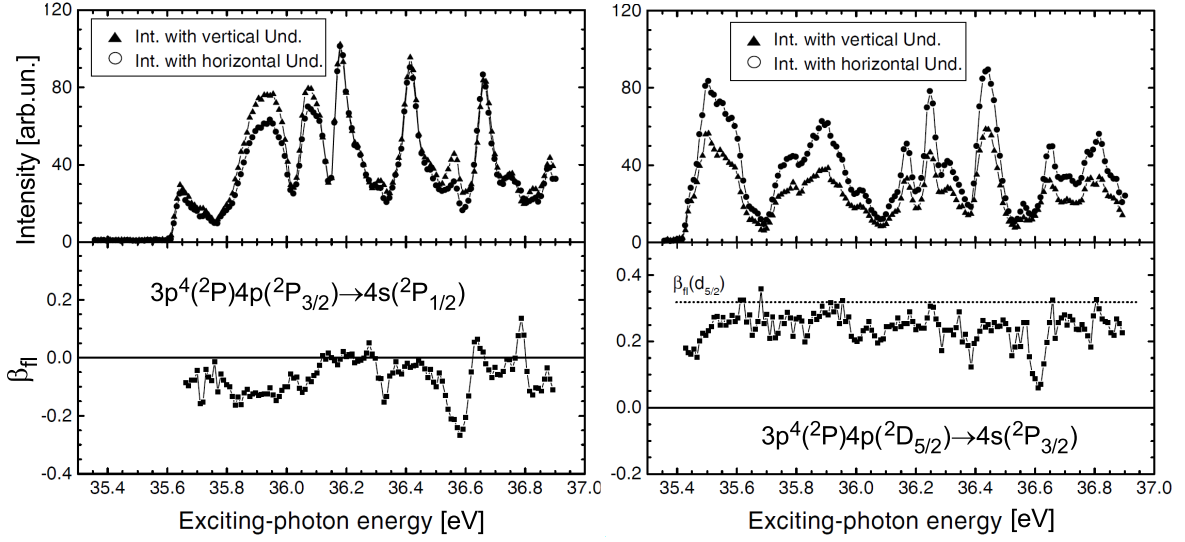}
\end{center}
\caption{Intensities $I_{ver}(\theta=90^{\circ})$, $I_{hor}(\theta=0^{\circ})$
and $\beta_{fl}$ for the $3p^{4}(^{3}P)4p(^{2}P_{3/2})\dashrightarrow
4s(^{2}P_{1/2})$ (left panel) and $3p^{4}(^{3}P)4p(^{2}D_{5/2})\dashrightarrow
4s(^{2}P_{3/2})$ (right panel) transitions. The dotted line in the right panel
indicates $\beta_{fl}(d_{5/2})$, corresponding to the decay by one partial
wave only. Adapted from \citep{mentzel98}.}%
\label{fig:Polar_Mentzel}%
\end{figure}

Typical results of the measurements of \cite{mentzel98} are shown in
Fig.~\ref{fig:Polar_Mentzel} for two selected levels. \cite{schmoranzer97a}
and \cite{mentzel98} used the intensities obtained for the undulators
operating in vertical ($I_{\perp}$) and horizontal ($I_{\parallel}$) modes,
respectively, to derive the fluorescence angular distribution parameter
$\beta2_{E_{1}J_{1}}^{J_{2}}(\omega)$ via%

\begin{equation}
\beta2_{E_{1}J_{1}}^{J_{2}}(\omega)=2\frac{I_{\parallel}-I_{\perp}%
}{I_{\parallel}+2I_{\perp}} \label{eq:b2fl_exp}%
\end{equation}
kinematically connected with $A_{20}(E_{1}J_{1},\omega)$ \citep{berezhko77} by%

\begin{equation}
\beta2_{E_{1}J_{1}}^{J_{2}}(\omega)=\alpha2_{J_{1}}^{J_{2}}\cdot A_{20}%
(E_{1}J_{1},\omega) \label{eq:b2fl}%
\end{equation}
where $J_{2}$ is the total angular momentum\ of the final state of the
fluorescence process (numerical coefficients can be found, e.g., in
\citep{lagutin03a}). We note that the alignment can also be {determined} when
the atom is ionized by circularly polarized radiation. In this case the
quantization axis coincides with the direction of the exciting radiation (as,
e.g., in \citep{mclaughlin02}) referred to as $z$- axis and $A_{20}^{z}$,
respectively. This alignment is connected with $A_{20}$ by $A_{20}
=-2A_{20}^{z}$ \citep{schmidt92}.

An advanced PIFS setup using a Wollaston prism (see
Fig.~\ref{fig:HS_fig6_polset}) has been applied by \cite{zimmermann04} to
measure the alignment parameters of the odd $2p^{4}(^{3}P)3p\ ^{4,2}%
P_{3/2}^{o},^{2}D_{5/2}^{o}$ states of Ne\thinspace II in the exciting-photon
energy range of $52~\mathrm{eV}\leq\hbar\omega\leq55.5~\mathrm{eV}$ at a
resolution of 18~meV. In addition, these authors obtained PICS for the
population of the even $2s^{1}2p^{6}\,^{2}S$ and $2p^{4}(^{1}D)3s\ ^{2}D$
states at the high resolution of 4.5~meV.

The orientation parameter $O_{10}(E_{1}J_{1},\omega)$\ for the odd
$3p^{4}(L_{0}S_{0})4p\ ^{2}P_{3/2}^{o}$ state of Ar\thinspace II has been
measured by \cite{mclaughlin02} using circularly polarized exciting photons
obtained by application of a four-reflector quarter-wave retarder to the
initial linearly polarized synchrotron radiation. Using the technique
described above \citep{yenen97}, the authors measured also $A_{20}(E_{1}%
J_{1},\omega)$. Having $A_{20},O_{10}$ and equations similar to
(\ref{eq:A20_w}), (\ref{eq:O10_w}), \cite{mclaughlin02} derived the reduced
partial wave cross sections $\gamma_{E_{1}J_{1}}^{\varepsilon\ell j}(\omega)$
in the exciting-photon energy range $35.6~\mathrm{eV}\leq\hbar\omega
\leq36.65~\mathrm{eV}$.

Calculations of the energy dependence of the alignment $A_{20}(E_{1}%
J_{1},\omega)$ and orientation $O_{10}(E_{1}J_{1},\omega)$\ parameters are
rather scarce. \cite{vanderhart99} computed $\sigma_{E_{1}J_{1}}(\omega)$ and
$A_{20}(E_{1}J_{1},\omega)$ for the $3p^{4}(L_{0}S_{0})4p\,^{2}P_{3/2}^{o}$
state of Ar\thinspace II using the R-matrix technique described earlier
\citep{vanderhart98}. Later on, the same technique has been applied for the
calculation of the orientation parameter $O_{10}(^{2}P_{3/2}^{o},\omega)$ and
of $\sigma_{E_{1}J_{1}}(\omega)$ for some selected even $3p^{4}(L_{0}%
S_{0})4s,3d\,E_{1}J_{1}$ states \citep{vanderhart02}. The calculations of
these authors qualitatively reproduced many of the experimentally observed
features. However, some of the computed features deviated (sometimes up to 0.1
eV) from the measured ones. Supposedly, this disagreement can be connected
with neglecting Coulomb screening (see section~\ref{sec:Coulomb_screening}%
) in the calculations.

The absolute partial PICSs for the even $4p^{4}(^{3}P)5s\,^{4}P_{J_{1}}$
states of Kr\thinspace II were calculated by \cite{schmoranzer97a}, allowing
these authors to compute the respective fluorescence angular distribution
parameters, too. More precise calculations of these parameters have been
performed by \cite{demekhin05} within the CIPF approximation. In this paper,
PICSs and the alignment parameters $A_{20}(E_{1}J_{1},\omega)$ for the
$4p^{4}(^{3}P)5s\,^{2}P_{J_{1}}$ and $4p^{4}(^{3}P)4d\,^{4}D_{J_{1}}$ states
were also computed and measured. The comparison of theory and experiment led
to the conclusion that core rearrangement and {coherently} taking into account
the autoionizing resonances and the autoionization continua strongly influence
the computed quantities.

\cite{schill03a} and \cite{okeeffe03,okeeffe04} performed a partial wave
analysis (PWA) of the RA decay of resonantly excited Kr and Xe using the
`sitting-on-resonance' regime. \cite{schill03a} used the setup shown in
Fig.$~$\ref{fig:HS_fig6_polset} to measure alignment and orientation of the
$4p^{4}(^{1}D)5p\,^{2}F_{7/2}$ Kr$\,$II state after the decay of the
$3d_{5/2}^{9}5p_{3/2}\,(J=1)$ resonance. Using equations (\ref{eq:PICS_norm})
to (\ref{eq:O10_w}) results in the ratios $\gamma(\varepsilon d_{5/2}%
):\gamma(\varepsilon g_{7/2}):\gamma(\varepsilon g_{9/2}%
)=3.2(5.6):1.4(3.8):96(6)\%$ which coincide with the predicted ones by
\cite{zimmermann00} $2.92:2.01:95.07$ within the error bars.

\begin{figure}[ptb]
\begin{minipage}[t]{0.48\textwidth}
\begin{center}
\includegraphics[width=\textwidth]{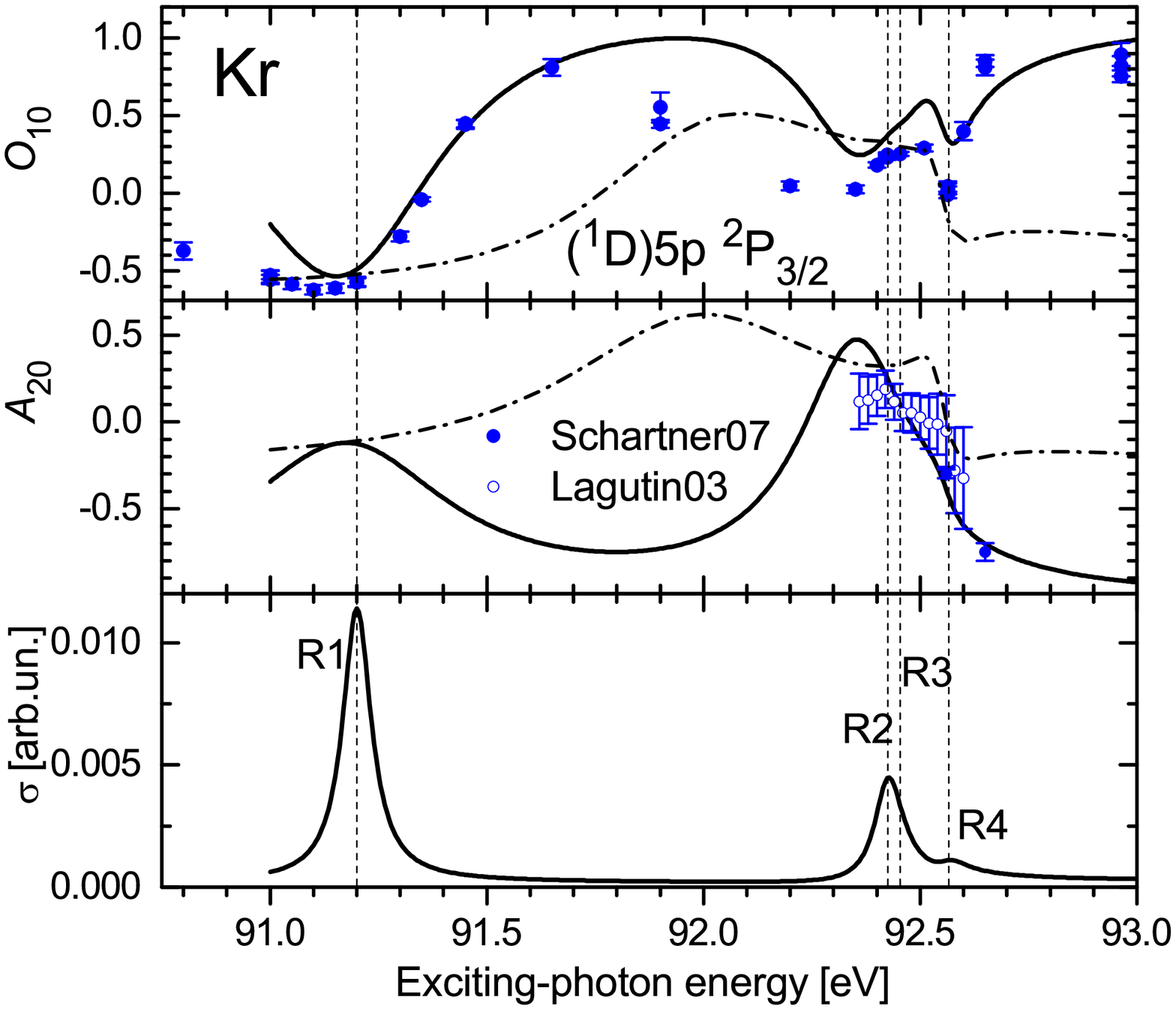}
\caption{Measured (open and full dots) and computed (full lines) orientation $O_{10}$ and alignment $A_{20}$ of the $4p^4(^1D)5p ^2P_{3/2}$ level as a function of the exciting-photon energy in the region of the $3d^9_j np$ resonances. Vertical dashed lines mark the positions of resonances. Dash-dotted curves represent the calculation without taking into account the non-resonant channel. The total cross section is shown in the bottom panel. Adapted from \citep{schartner07}.}
\vspace{4ex}
\label{fig:Polar_A2O1}\end{center}
\end{minipage} \hfill\begin{minipage}[t]{0.48\textwidth}
\begin{center}
\includegraphics[width=\textwidth]{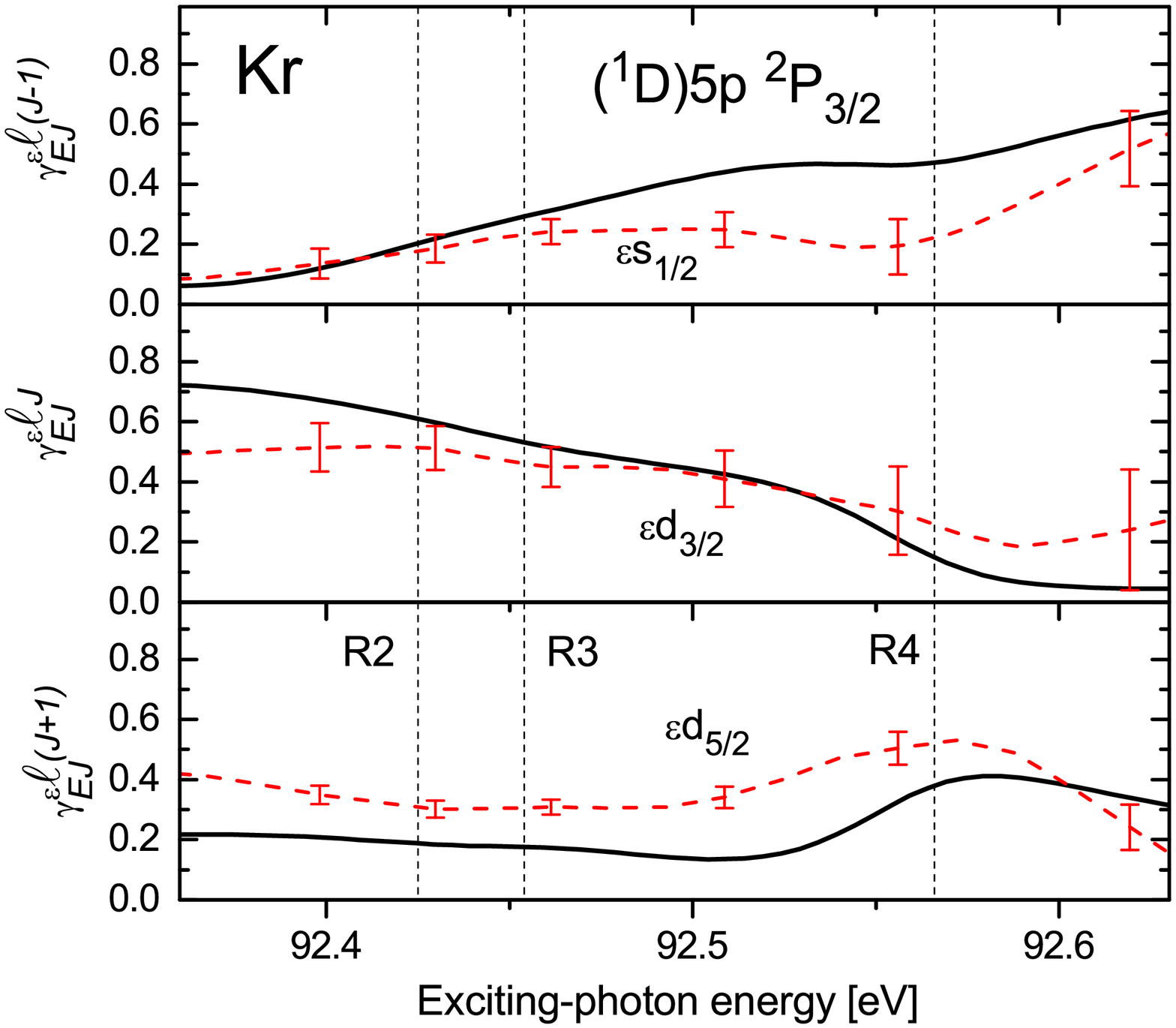}
\caption{Computed (solid curves) and measured (dashed curves with the error bars)
relative partial $\gamma_{EJ}^{\varepsilon\ell J}$ cross sections (see
equation (\ref{eq:PICS_red})) for the production of the $4p^{4}(^{1}D)5p\, ^{2}P_{3/2}$ and $4p^{4}(^{1}D)5p\, ^{2}D_{5/2}$ satellites as a function of the exciting-photon
energy in the region of the R2, R3, and R4 resonances (positions of resonances are marked by the vertical dashed lines). Adapted from
\citep{schartner07}.}\label{fig:Polar_PWA}\end{center}
\end{minipage}
\end{figure}

\cite{okeeffe03} measured the orientation of five selected $5p^{4}(L_{0}%
S_{0})6p$ Xe$\,$II states after the decay of the $4d_{5/2}^{9}6p_{3/2}\,(J=1)$
resonance and used the alignment parameters from \cite{meyer01} to perform the
PWA of this decay. A combined theoretical and experimental investigation of
the RA decay of the $4d_{5/2}^{9}6p_{3/2}\,(J=1)$ resonance with following
population of the $5p^{4}(L_{0}S_{0})6p$ Xe$\,$II states has been performed by
\cite{okeeffe04}. These authors found a good overall agreement between the
measured and computed PWA including also the calculation of \cite{lagutin00}.

The Raman regime for the PWA was used by \cite{schartner05,schartner07} who
populated the $4p^{4}(L_{0}S_{0})5p$ Kr$\,$II states via the $R1$ to $R4$
resonances ($\Gamma=83$~meV, see section~\ref{sec:beyond_2ST}) at a resolution
of $\Delta E=10$~meV. The alignment parameters $A_{20}$ and the orientation
parameters $O_{10}$ were measured for the $4p^{4}(^{1}D)5p\,^{2}F_{7/2}$ and
$4p^{4}(^{1}D)5p\,^{2}D_{5/2}$ states in \citep{schartner05} and for the
$4p^{4}(^{1}D)5p\,^{2}P_{3/2}$ state in \citep{schartner07}. These
measurements enabled the authors to determine the reduced partial wave cross
sections $\gamma_{E_{1}J_{1}}^{\varepsilon\ell j}(\omega)$\ in the
exciting-photon energy range of $90.8~\mathrm{eV}\leq\hbar\omega
\leq92.9~\mathrm{eV}$. The $A_{20}$ and $O_{10}$ results for the $4p^{4}%
(^{1}D)5p\,^{2}P_{3/2}$ state are depicted in Fig.~\ref{fig:Polar_A2O1}. In
this figure, where the results of the calculation from \cite{schartner07} are
also shown, one can see good overall agreement between theory and experiment.
The comparison between calculations performed without and with taking into
account the non-resonant channel emphasizes the importance of the latter.
Fig.~\ref{fig:Polar_PWA}, where the results of the experimental and
theoretical PWA are depicted on an enlarged scale, shows the interrelation
between partial waves passing through resonances.

\begin{figure}[ptb]
\begin{minipage}[t]{0.47\textwidth}
\begin{center}
\includegraphics[width=\textwidth]{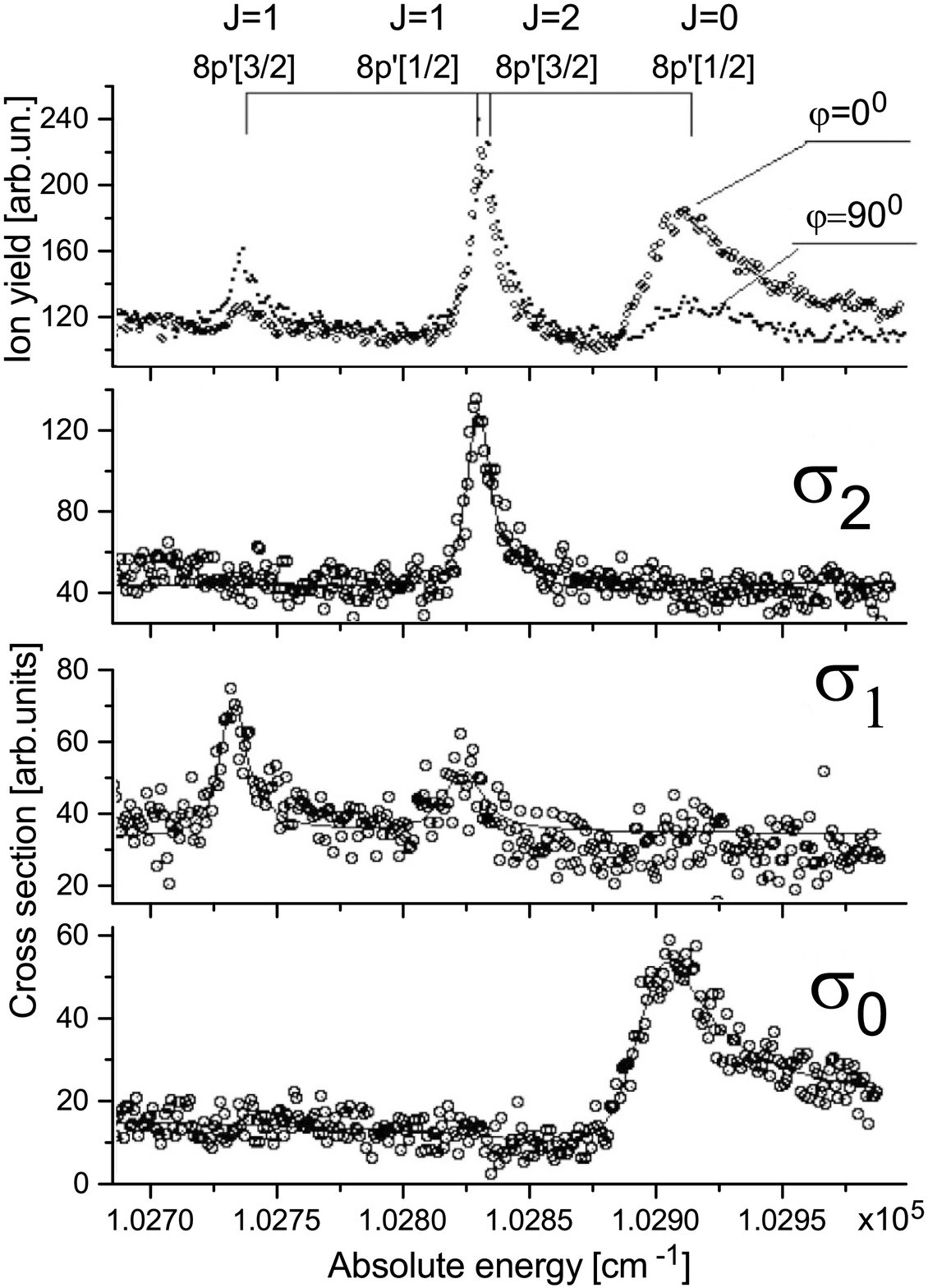}
\caption{Photoionization spectra in the region of the Xe $5p_{1/2}^{5}8p^{\prime}[K]_J$
resonances excited from the $5p_{3/2}^{5}7s[3/2]_1$ state using linearly polarized
laser and SR radiation of parallel ($\varphi=0^{\circ}$) and perpendicular
($\varphi=90^{\circ}$) relative orientation of their electric field vector (upper
panel). Partial photoionization cross sections $\sigma_0$, $\sigma_1$, and $\sigma_2$ in
the same energy region (lower panels). Adapted from
\citep{aloise05}.}\label{fig:Meyer_PICS}\end{center}
\end{minipage} \hfill\begin{minipage}[t]{0.47\textwidth}
\begin{center}
\includegraphics[width=\textwidth]{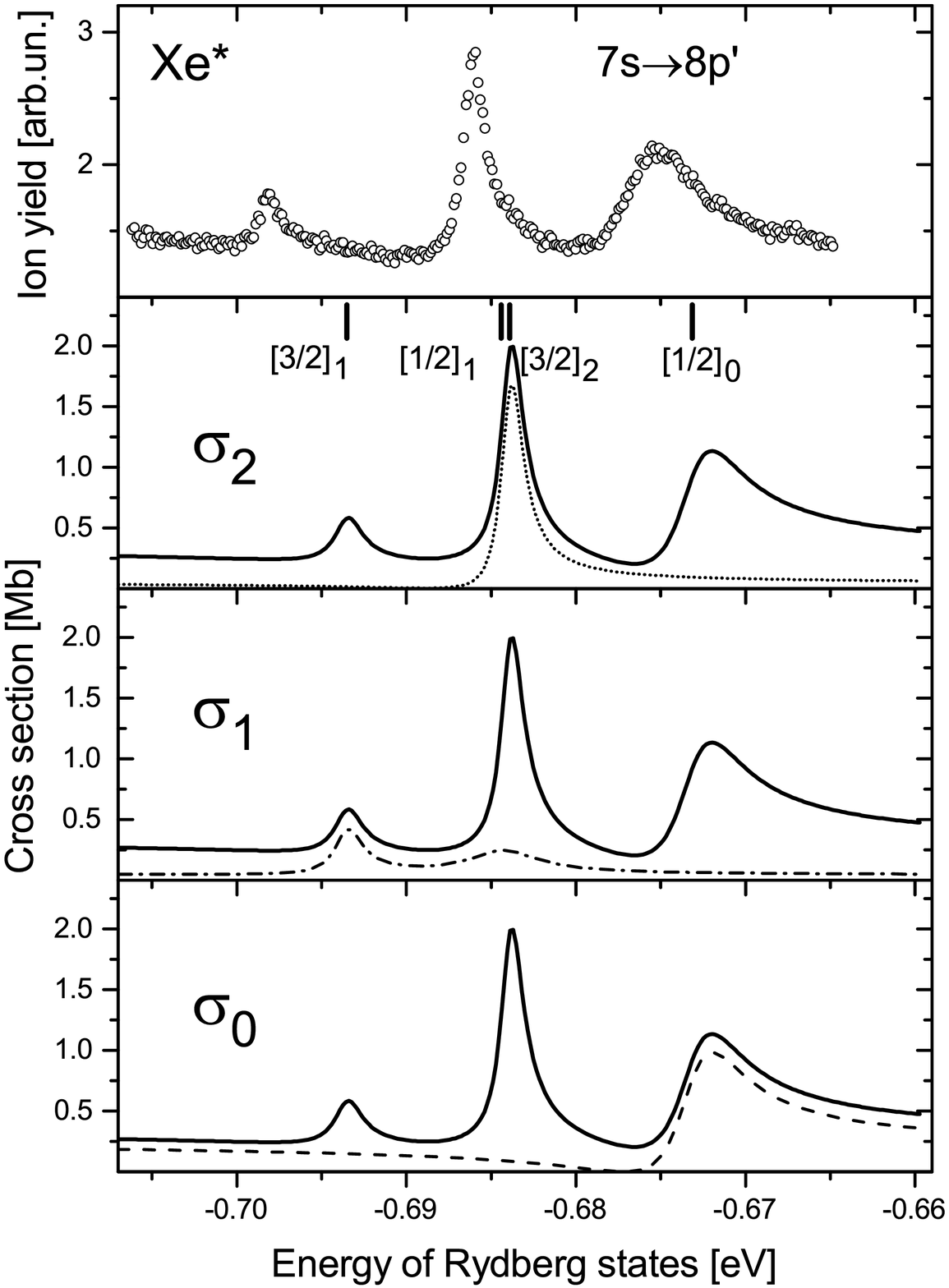}
\caption{The partial and total PICSs of Xe at the $5p_{3/2}^{5}7s[3/2]_1\dashrightarrow 5p_{1/2}^{5}8p^{\prime}[K^{\prime}]_J$ resonances. Solid lines and open circles represent the computed and measured total PICSs, respectively. Dotted, dash-dotted, and dashed lines shows the partial cross sections. Adapted from \citep{petrov06a}.}
\label{fig:Meyer_PICS_thy}\end{center}
\end{minipage}
\end{figure}

Concluding this section we mention an experimental technique which differs
from PIFS and allows to measure the partial photoionization cross sections
quantified by the total angular momentum of the final state. \cite{aloise05}
reported the experimental `two-colour' setup consisting of two radiation
sources: the synchrotron radiation (SR) was used to prepare the intermediate
resonant state while a tunable dye laser was used for the subsequent
photoionization from this state. The authors used an experimental scheme with
two collinearly counterpropagating exciting-photon beams of variable linear and
circular polarization. The photoion current was measured depending on the
varied mutual polarization between SR and laser radiation. Using linearly and
circularly polarized light makes it possible to derive linearly independent
equations connecting the photoion current measured at different polarization
as input data and the partial photoionization cross sections as unknowns.
Solving these equations results in partial PICS.

\cite{aloise05} applied the two-colour technique to study several processes
among which was the following ($jK$ coupling is used for the notation of the
atomic states):%

\begin{equation}
\text{Xe}[0]+\gamma^{SR}\dashrightarrow5p_{3/2}^{5}7s[3/2]_{1}+\gamma
^{las}\dashrightarrow\left\{
\begin{array}
[c]{c}%
5p_{3/2}^{5}\varepsilon\ell j\\
\updownarrow\\
5p_{1/2}^{5}8p^{\prime}[K]
\end{array}
\right.  ~(J=0,1,2). \label{eq:Meyer}%
\end{equation}
where the notations of arrows are the same as in (\ref{eq:Ar-scheme}). Two of
the ion yields (for the different mutual orientations of the linearly
polarized SR and laser) are depicted in the upper panel of
Fig.~\ref{fig:Meyer_PICS} as an example. In the panels below, the partial
cross sections $\sigma_{J}(\omega)$\ derived from the measured ion yields are
shown. The positions of the $5p_{1/2}^{5}8p^{\prime}[K]_{J}$ resonances are
marked on top. It is clear that the resonances $5p_{1/2}^{5}8p^{\prime}%
[K]_{J}$ affect those cross sections only that have the same value of $J$
because the Coulomb interaction is a scalar. Measured partial and total cross
sections are compared with the ones calculated by \cite{petrov06a} within the
CIPFCP approximation in Fig.~\ref{fig:Meyer_PICS_thy}. Good agreement between
all computed and measured cross sections allows to state the adequacy of both the
theoretical and experimental techniques.

\section{Summary and perspectives}

In summary, we reviewed the {work on} the dynamics of photon-induced
atomic processes over the past half century showing that many of them are
largely determined by many-electron correlations. We pursued chronologically
how our understanding of the photoionization process developed, when the
measurements initiated the development of the theory and then the development
of the theory required to reconsider the results of the experiments and to
carry out refined measurements. One of the examples is the photoionization of
the subvalence shells of rare-gas atoms. In this case, based on the
calculations, the presence of a minimum in the photoionization cross section
was predicted which visually looks like a Cooper minimum but actually has a
substantially different nature. This many-electron nature should also lead to
an anomalous angular distribution of photoelectrons. Subsequently, the
predictions were confirmed qualitatively by experiments, but the experiments
revealed also a significant discrepancy between measured and computed cross
sections. This, in turn, stimulated the development of the theory and new
methods for calculating atomic structures.

We have paid special attention to cases where the PIFS results
significantly improved the understanding of atomic processes. For example,
PIFS made it possible to reliably register a resonant structure in the
near-threshold photoionization of the rare-gas atoms, which in turn
constituted a challenge for creating new precise methods for the atomic
structure calculations. One of such methods was based on the Pauli-Fock (PF)
approximation, which allows one to take into account relativistic effects
saving the one-component structure of AOs on the one hand, and ensures high
accuracy of calculation on the other hand. The PF AOs were used to take into
account \emph{all} electron configurations differing from the state of
interest by single- and double-excitations. Strongly interacting
configurations were taken into account via the solution of the secular
equation while the other configurations were taken into account by computing
the corrections to the matrix elements of that equation (usually reducing them
and, therefore, called the Coulomb screening). Some selected excitations were
included in the calculation by means of a non-empirical core polarization (CP)
potential and following recalculation of the set of the PF AOs, thus
generating a new complete set of the PFCP AOs. Using the CIPFCP (configuration
interaction Pauli-Fock with core polarization) technique made it possible to
understand the origin of most of the features observed by PIFS. In turn, the
development of the CIPFCP approach initiated the investigation of new
interference phenomena by the PIFS method. In particular, it has been shown
that the alignment of atoms in a resonant Auger decay is significantly
affected by interference between different photoionization pathways that can
be observed experimentally.

The PIFS method allowed to reliably determine the lifetime of rare-gas atoms
with a subvalence vacancy, which is anomalously large because of the
destructive interference between the main and satellite processes. The
measured lifetimes agree reasonably with the computed ones which confirms the
adequacy of the considered mechanism. In photoelectron spectra, such an
interference causes a strong dependence of the spectra on the exciting-photon
energy in the threshold range: some well-resolved lines change their relative
intensity by more than an order of magnitude.

In a further development of PIFS, methods for measuring the alignment and
orientation of atoms and ions excited by polarized radiation and for
performing partial wave analysis have been created. The use of partial waves
in turn helps to improve the methods of the atomic structure calculations.

The ability of PIFS to study threshold processes has led to the fact that
recently the method has been used to study processes induced by photons in
molecules, clusters, and even in liquids \citep{hans17}. In particular, the
processes of neutral dissociation of molecules
\citep{liebel02,ehresmann04a,ehresmann06,demekhin10}, double photoionization
of molecules \citep{ehresmann03}, interference phenomena of the
`lifetime-vibrational interference' type
\citep{ehresmann07,demekhin08,demekhin10b}, resonant interatomic Coulombic
decay \citep{knie14,hans16} were investigated, and the relations for partial
wave analysis were proposed \citep{demekhin10c}.

In conclusion, we express the hope that PIFS will find also wide application
in the study of threshold phenomena in polyatomic systems such as molecules
and clusters, which in turn will lead to an improvement in the understanding
of molecular processes and will lead to the development of theoretical methods
for the calculation of polyatomic systems.

\section*{Acknowledgements}

We gratefully appreciate useful discussions related to this review with our
colleagues M. Amusia, U. Becker, N. Cherepkov, Ph. Demekhin, V. Ivanov, V.
Kilin, W. Mehlhorn, V. Schmidt, B. Sonntag. The special thanks by V.L.S, B.M.L., and I.D.P
are due to H. Hotop who contributed much to the creation of the CIPFCP method
by providing experimental data and fruitful discussions. V.L.S, B.M.L., and
I.D.P, acknowledge financial support by the Hessen State Initiative for the
Development of Scientific and Economic Excellence (LOEWE) in the LOEWE-Focus
project ELCH and by Deutsche Forschungsgemeinshaft (DFG) within the CRC 1319
"Extreme light for analysis and control of molecular chirality" for their stay in Germany.
I.D.P., B.M.L., and V.L.S. would like to thank the Department of Physics,
Technische Universit\"{a}t Kaiserslautern, and the Institute of Physics,
University of Kassel, for the hospitality extended to them. V.L.S. appreciates
the support from Southern Federal University within the inner project No.
3.6105.2017/8.9. I.D.P. and B.M.L. appreciate the support from Russian
Foundation for Basic Research (RFBR (Russia) Grant No. 16-02-00666A) during the work on
this review.

\section*{References}

\end{document}